\documentclass[modern]{aastex61}
\usepackage{graphicx}
\usepackage{psfig}
\usepackage{natbib}
\usepackage{subfigure} 

\newcommand{\gsim}{\mbox{\hspace{.2em}\raisebox{.5ex}{$>$}\hspace{-.8em}\raisebox{-.5ex}{$\sim$}\hspace{.2em}}}
\newcommand{\lsim}{\mbox{\hspace{.2em}\raisebox{.5ex}{$<$}\hspace{-.8em}\raisebox{-.5ex}{$\sim$}\hspace{.2em}}}

\newcommand{\E}[1]{\times 10^{#1}}
\newcommand{\twCO}{$^{12}$CO}  \newcommand{\thCO}{$^{13}$CO}

\newcommand{\HII}{\mbox{H\,\textsc{ii}}}

      \newcommand{\ps}{\,{\rm s}^{-1}}
    \newcommand{\Msun}{M_{\odot}}   
    \newcommand{\km}{\,{\rm km}}

\begin{document}

\title{
Local Molecular Gas toward the Aquila Rift Region
}

\shorttitle{Local CO gas toward the Aquila Rift Region}

\correspondingauthor{Yang Su}
\email{yangsu@pmo.ac.cn}

\author[0000-0002-0197-470X]{Yang Su}
\affil{Purple Mountain Observatory and Key Laboratory of Radio Astronomy,
Chinese Academy of Sciences, Nanjing 210023, China}

\author{Ji Yang}
\affiliation{Purple Mountain Observatory and Key Laboratory of Radio Astronomy,
Chinese Academy of Sciences, Nanjing 210023, China}
\affiliation{School of Astronomy and Space Science, University of Science and
Technology of China, 96 Jinzhai Road, Hefei 230026, China}

\author{Qing-Zeng Yan}
\affiliation{Purple Mountain Observatory and Key Laboratory of Radio Astronomy,
Chinese Academy of Sciences, Nanjing 210023, China}

\author{Yan Gong}
\affiliation{Purple Mountain Observatory and Key Laboratory of Radio Astronomy,
Chinese Academy of Sciences, Nanjing 210023, China}
\affiliation{Max-Planck Institute f{\"u}r Radioastronomy, Auf dem H{\"u}gel 69,
D-53121 Bonn, Germany}

\author{Zhiwei Chen}
\affiliation{Purple Mountain Observatory and Key Laboratory of Radio Astronomy,
Chinese Academy of Sciences, Nanjing 210023, China}

\author{Shaobo Zhang}
\affiliation{Purple Mountain Observatory and Key Laboratory of Radio Astronomy,
Chinese Academy of Sciences, Nanjing 210023, China}

\author{Yan Sun}
\affiliation{Purple Mountain Observatory and Key Laboratory of Radio Astronomy,
Chinese Academy of Sciences, Nanjing 210023, China}

\author{Miaomiao Zhang}
\affiliation{Purple Mountain Observatory and Key Laboratory of Radio Astronomy,
Chinese Academy of Sciences, Nanjing 210023, China}

\author{Xuepeng Chen}
\affiliation{Purple Mountain Observatory and Key Laboratory of Radio Astronomy,
Chinese Academy of Sciences, Nanjing 210023, China}
\affiliation{School of Astronomy and Space Science, University of Science and
Technology of China, 96 Jinzhai Road, Hefei 230026, China}

\author{Xin Zhou}
\affiliation{Purple Mountain Observatory and Key Laboratory of Radio Astronomy,
Chinese Academy of Sciences, Nanjing 210023, China}

\author{Min Wang}
\affiliation{Purple Mountain Observatory and Key Laboratory of Radio Astronomy,
Chinese Academy of Sciences, Nanjing 210023, China}

\author{Hongchi Wang}
\affiliation{Purple Mountain Observatory and Key Laboratory of Radio Astronomy,
Chinese Academy of Sciences, Nanjing 210023, China}
\affiliation{School of Astronomy and Space Science, University of Science and
Technology of China, 96 Jinzhai Road, Hefei 230026, China}

\author{Ye Xu}
\affiliation{Purple Mountain Observatory and Key Laboratory of Radio Astronomy,
Chinese Academy of Sciences, Nanjing 210023, China}

\author{Zhibo Jiang}
\affiliation{Purple Mountain Observatory and Key Laboratory of Radio Astronomy,
	Chinese Academy of Sciences, Nanjing 210023, China}

\begin{abstract}
We present the results of a $\sim$250 square degrees CO mapping 
($+26^{\circ} \lsim l \lsim+50^{\circ}$ and $-5^{\circ}\lsim b \lsim +5^{\circ}$)
toward the Aquila Rift region at a spatial resolution of $\sim 50''$ and
a grid spacing of 30$''$. 
The high dynamic range CO maps with a spectral resolution of $\sim 0.2\km\ps$ 
display highly structured molecular cloud (MC) morphologies with 
valuable velocity information, revealing complex spatial and dynamical 
features of the local molecular gas.
In combination with the MWISP CO data and the Gaia DR2, 
distances of the main MC structures in the local ISM
are well determined toward the Aquila Rift region.
We find that the total MC mass within 1 kpc is about 
$\gsim 4.1\times10^{5} \Msun$ in the whole region.
In fact, the mass of the molecular gas is dominated by the 
W40 giant molecular cloud (GMC) at $\sim$~474~pc ($\sim 1.4\times10^{5} \Msun$) 
and the GMC complex G036.0$+$01.0 at $\sim$~560--670~pc ($\sim 2.0\times10^{5} \Msun$), 
while the MCs at $\sim$~220--260~pc have gas masses of $\sim 10^{2}-10^{3} \Msun$.
Interestingly, an $\sim$~80~pc long filamentary 
MC G044.0$-$02.5 at a distance of $\sim 404$~pc
shows a systematic velocity gradient along and
perpendicular to the major axis of the filament.
The HI gas with the enhanced emission has the similar spatial morphologies 
and velocity features compared to the corresponding CO structure, 
indicating that the large-scale converging HI flows are probably 
responsible for the formation of the MC.
Meanwhile, the long filamentary MC consists of many sub-filaments 
with the lengths ranging from $\sim$~0.5~pc to several pc, as well as prevalent 
networks of filaments in other large-scale local MCs.
\end{abstract}


\section{Introduction}
A substantial amount of interstellar gas in the Milky Way is in
molecular clouds (MCs) and star formation is believed to
occur in dense regions of MCs.
MCs are therefore important in studying the nature of the 
cold and dense molecular gas in the interstellar medium (ISM). 
Detailed studies can provide us with the solid knowledge of 
the properties of MCs, such as the morphology,
distribution, and dynamics of the molecular gas.
In observations, MCs are generally traced by CO emission because of 
its low excitation energy, low critical density, and easy detection 
over other gas tracers \citep{2013ARA&A..51..207B,2015ARA&A..53..583H}.
Actually, a lot of CO surveys have been done to study the
molecular gas in the Galaxy 
\citep[e.g.,][]{1998ApJS..115..241H,2001ApJ...547..792D,
2013PASA...30...44B,2015ApJ...812....6B,2015ARA&A..53..583H}.

Among the most important surveys is the Milky Way Imaging Scroll Painting 
(MWISP\footnote{http://english.dlh.pmo.cas.cn/ic/}) project,
which provides the unbiased \twCO, \thCO, and C$^{18}$O~($J$=1--0) database
toward the northern Galactic plane for regions of  
$l=-10^{\circ}$ to $+250^{\circ}$ and $|b|\lsim$~5\fdg2.
The survey data sets of the MWISP have a large spatial dynamic range,
high sensitivity, and high-velocity resolution for CO and its isotopes
\citep[see Table~1 and Section 2.3 in][]{2019ApJS..240....9S}.
Due to these characteristics and advantages, the MWISP CO data
provided us with a good opportunity to unveil the nature of the MCs
as well as the structures of the Milky Way in our previous studies
\citep[i.e.,][]{2016ApJ...828...59S,2017ApJS..230...17S}.
These large-scale CO maps with the high sensitivity on a fully sampled 
angular grid of $30''$ also allow us to investigate details of MCs' structures and 
dynamics, especially for local MCs due to their small distances.

One can use the optically thick \twCO\ emission to trace the enveloping layer 
(i.e., $\sim 10^{2}$~cm$^{-3}$) of extended MCs and to reveal 
the dynamical features of the molecular gas of low surface brightness. 
On the other hand, the optically thin \thCO\ and C$^{18}$O emission can 
trace the denser region 
(i.e., $\sim 10^{3}-10^{4}$~cm$^{-3}$) in local MCs.
In particular, the enhanced concentration of C$^{18}$O emission seems to 
be associated with star-forming regions. 
Using the abundant data of \twCO, \thCO, and C$^{18}$O,
we can obtain valuable information on the physical properties of 
the local MCs and star formation activities in MCs.

So far, an excellent case is the 100 deg$^2$ CO mapping toward the nearby Taurus MC,
in which the researchers reveal a very complex and highly structured molecular-gas
morphology using the FCRAO 13.7~m telescope \citep{2008ApJ...680..428G,2008ApJS..177..341N}.
Thanks to the high sensitivities of the MWISP, the large-scale CO survey 
can be extended to other nearby MCs, which will shed light on detailed structures, 
distributions, and dynamics of the local molecular gas.

In this paper, we focus on the local MCs toward the Aquila Rift region near the 
Galactic plane, i.e., the region in the first quadrant of $l = +$25\fdg8 to $+$49\fdg7 
and $|b|\lsim$~5\fdg2. This region has been
fully covered by observations of the MWISP over the past seven years. 
In section 2, we briefly describe the observations and data. Section 3
shows the results and some discussions of the local molecular gas
by taking advantages of the high spatial dynamic range CO images. 
Based on the properties of the local MCs, such as distances,
mass distributions, and velocity structures of the gas,
we construct panoramic views
of the local MCs ($\lsim$~1~kpc) by combining the MWISP CO data 
and the Gaia DR2 data.
Finally, we summarize the new findings and main conclusions in Section 4.

\section{Observations and Data}
The observations and data are part of the MWISP project that is described 
in our previous paper \citep{2019ApJS..240....9S}. 
Briefly, the $\sim$~250~deg$^2$ region
of $l = +$25\fdg8 to $+$49\fdg7 and $|b|\lsim$~5\fdg2 
was covered between 2011 November and 2018 April with the 3$\times$3 multibeam 
sideband-separating Superconducting Spectroscopic Array Receiver
(SSAR) system \citep{Shan} on the Delingha 13.7~m telescope.
The \twCO, \thCO, and C$^{18}$O~($J$=1--0) lines were simultaneously observed
using the position-switch On-The-Fly \citep[OTF; see][]{2018AcASn..59....3S} 
mode. The half-power beam width (HPBW) is $\sim 50''$ at a frequency
of $\sim$~110--115~GHz. The tracking accuracy and the pointing accuracy are 
$\sim 1''$--3$''$ and $\lsim 5''$, respectively.

The whole region was divided into 1001 cells of 30$'\times30'$, each being 
mapped along the Galactic longitude ($l$) and latitude ($b$) at least twice.
After fitting the first-order (or linear) baseline,
the final three-dimensional (3D) FITS data have
typical rms noise levels of $\sim$~0.5 K for
\twCO\ ($J$=1--0) at a channel width of 0.16~$\km\ps$ and $\sim$~0.3~K
for \thCO\ ($J$=1--0) and C$^{18}$O ($J$=1--0) at 0.17~$\km\ps$
with a uniform grid spacing of $30''$.
Here, the main beam temperature is calibrated based on
$T_{\rm MB}=T_{\rm A}^{\star}/(f_{\rm b}\times\eta_{\rm MB})$,
and the filling factor $f_{\rm b}$ was assumed to be 1 for the CO emission.
The data were reduced using the
GILDAS software\footnote{http://ascl.net/1305.010 or
http://www.iram.fr/IRAMFR/GILDAS}.

\section{Results and Discussions}
\subsection{Overall Distribution of the Local CO Gas}
Here, we study the properties of the local MCs
using the MWISP \twCO, \thCO, and C$^{18}$O data.
In the previous study, we have shown that the local MCs 
toward the region of $l= +$25\fdg8 to $+$49\fdg7 and $|b|\lsim$~5\fdg2
are mainly in the velocity range of 0--20~km~s$^{-1}$ 
\citep[see Figures~6, 7, and 9 in][]{2019ApJS..240....9S}.
In this region, the prominent feature is the $\sim$~200 square degrees
dark lane seen in the optical image, which is called the Aquila Rift 
\citep[e.g., see the review in][]{2008hsf1.book...18P}.  
A considerable amount of molecular gas is situated in the Aquila Rift,
which was investigated by some authors based on the previous 1.2~m telescope survey
\citep{1985ApJ...297..751D,1987ApJ...322..706D} and the recent 1.85~m telescope survey
\citep{2017ApJ...837..154N}. The whole Aquila Rift is a large extended region
(i.e., $l\sim +20^{\circ}$ to $+50^{\circ}$ and $b\sim -5^{\circ}$ to $+10^{\circ}$)  
and is not fully covered by the MWISP survey ($|b|\lsim$~5\fdg2).
However, the MWSIP data have the higher angular resolution and 
sensitivity \citep{2019ApJS..240....9S}, which allow us to search for 
new observational features and to study the properties
of the local molecular gas.
 
\begin{figure}[ht]
\includegraphics[trim=70mm 0mm 0mm 0mm,scale=0.4,angle=0]{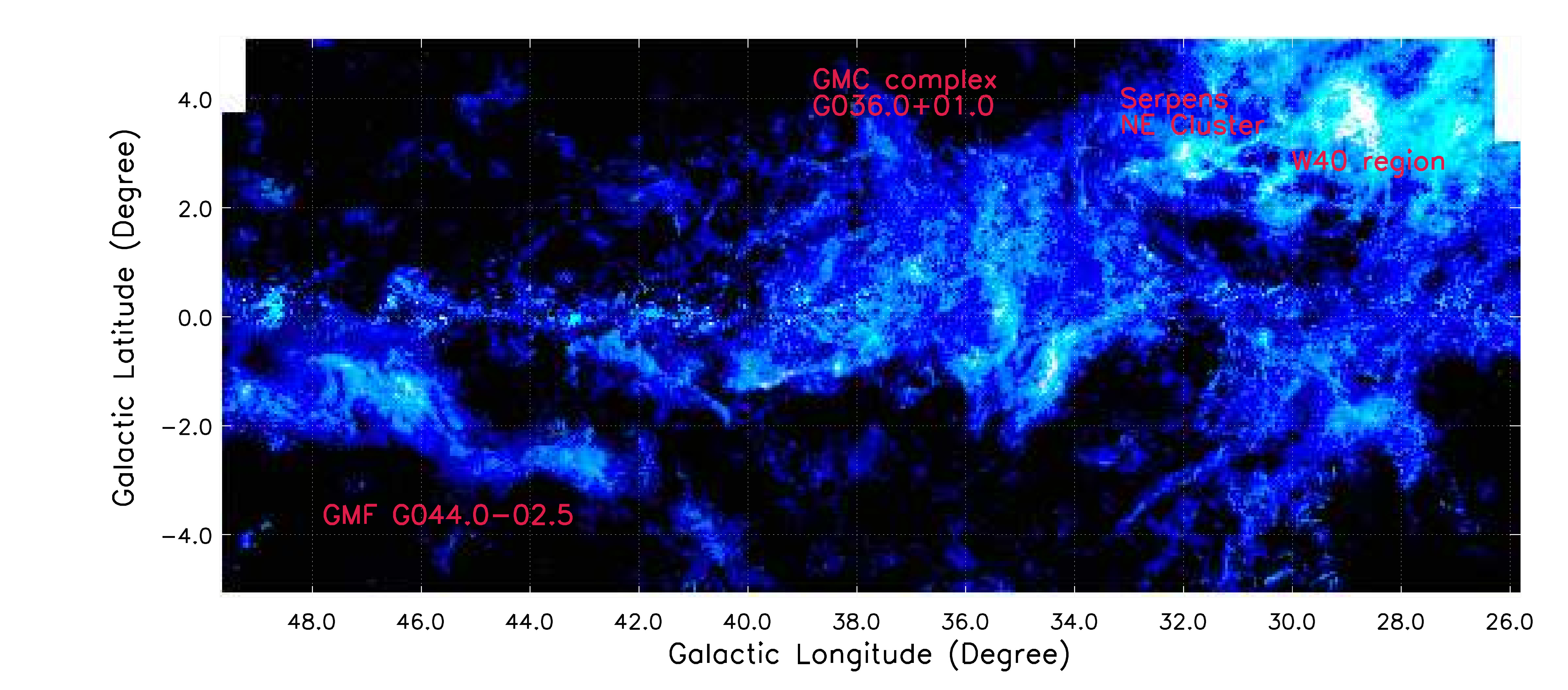}
\caption{
\twCO\ ($J$=1--0, blue), \thCO\ ($J$=1--0, green), and C$^{18}$O ($J$=1--0, red)
intensity map in the velocity range of [-1, 25]~km~s$^{-1}$.
For subsequent studies, several regions (i.e., W40 region, Serpens NE cluster, 
GMC complex G036.0$+$01.0, and GMF G044.0$-$02.5) are labeled on the map.
\label{aquila}}
\end{figure}

Figure~\ref{aquila} displays the \twCO\ (J=1--0, blue), \thCO\ (J=1--0, green), 
and C$^{18}$O (J=1--0, red) intensity map in the [-1, 25]~km~s$^{-1}$.
Note that in this velocity interval, CO emission of the local gas
near regions of $b \sim 0^{\circ}$ suffers some contamination from the Perseus arm
\citep[e.g., see Table~1 in][]{2019ApJ...885..131R}. However, this 
effect is not serious because of the limited extension 
of the Perseus gas at larger distances.
Obviously, the molecular gas traced by the large-scale extended CO emission 
displays complex structures and plentiful details in the mapped region.
Many CO structures, such as filaments, arcs/shells/bubbles, 
finger-like protrusions, and other irregular morphologies, 
are unveiled due to the high dynamic range of our survey.

Some MCs with relatively strong \thCO\ and/or C$^{18}$O 
emission are found to be embedded in the enveloping structures traced
by the faint, diffuse, and extended \twCO\ emission.
The multiple layers traced by the \twCO, \thCO, and C$^{18}$O emission,
together with different velocity components and, therefore, probably 
different distances (see Section 3.2), help us construct a useful 
panoramic 3D picture of the local molecular gas.  

\begin{figure}
\vspace{-3ex}
\gridline{
          \fig{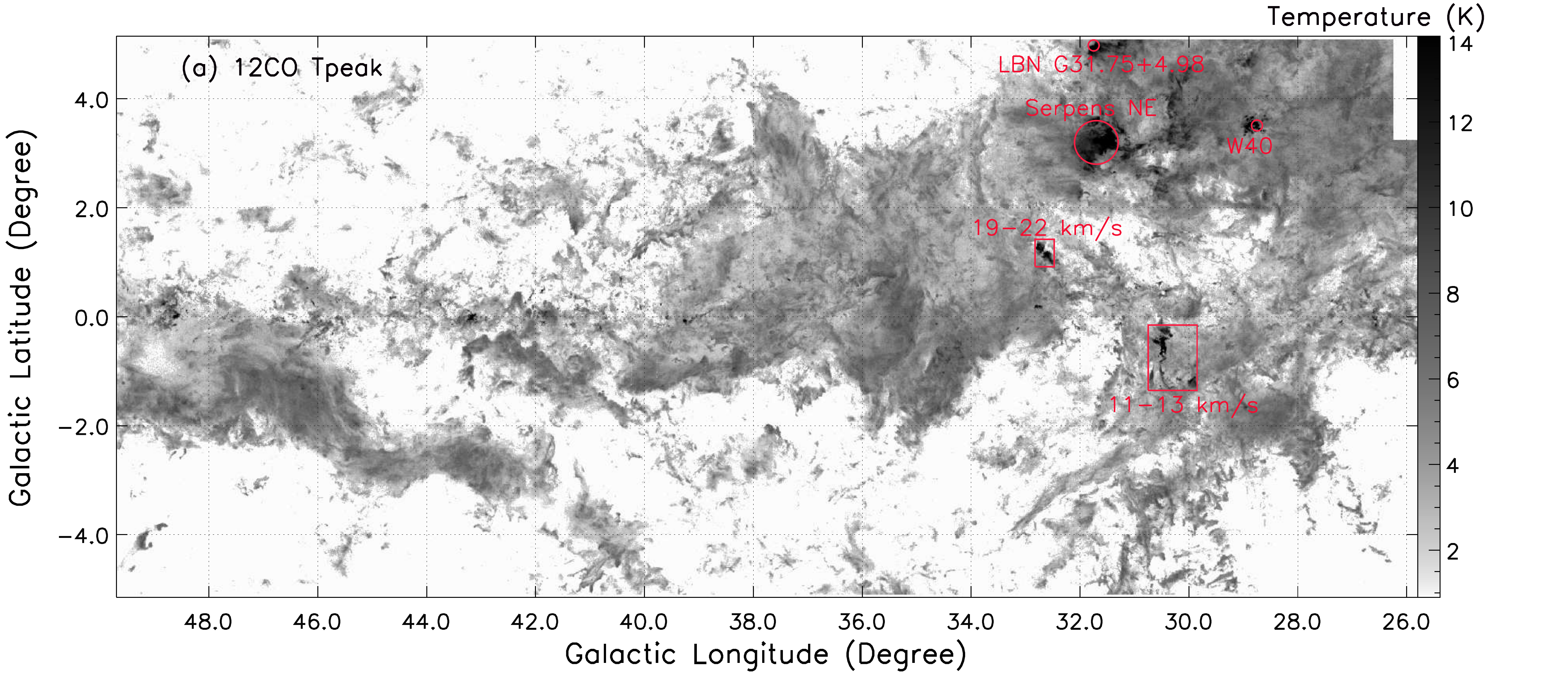}{1.\textwidth}{}
          }
\vspace{-7ex}
\gridline{
          \fig{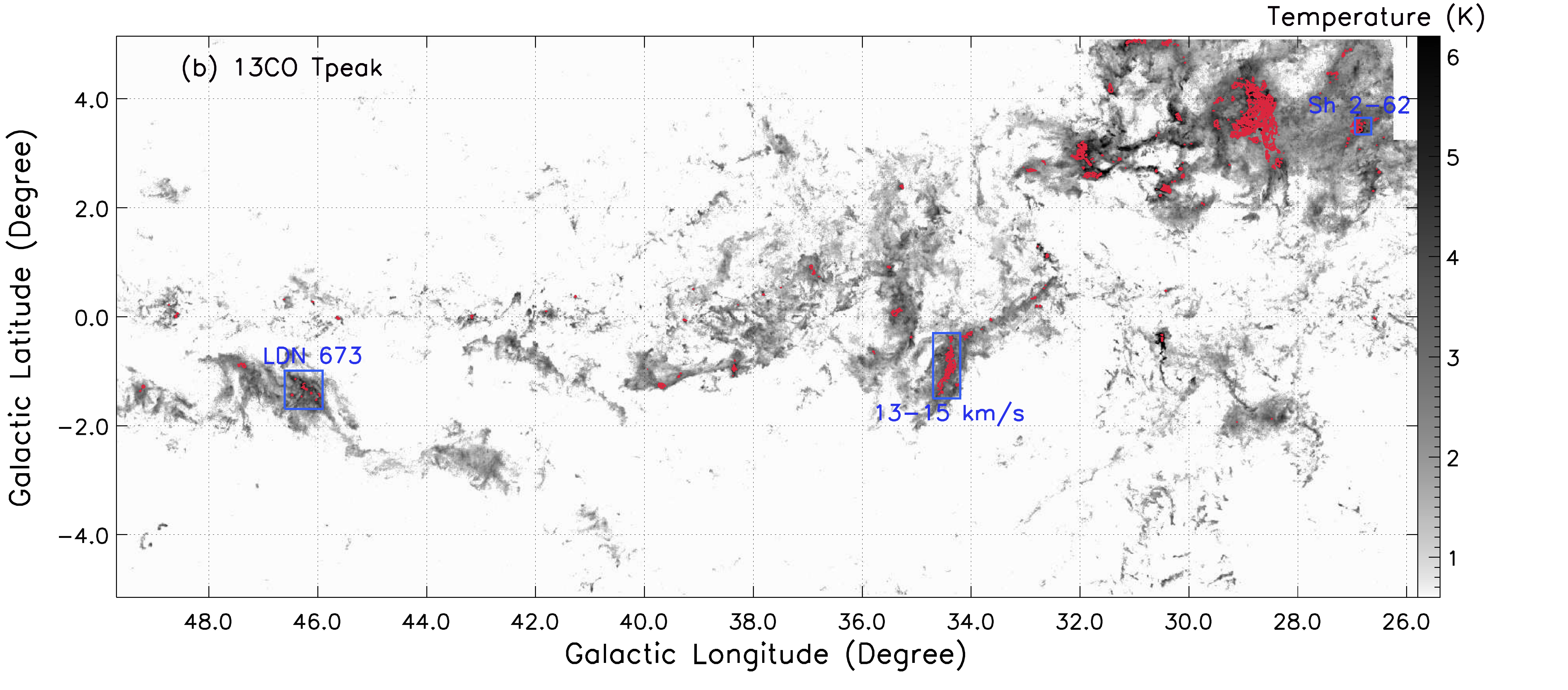}{1.\textwidth}{}
          }
\vspace{-7ex}
\gridline{
          \fig{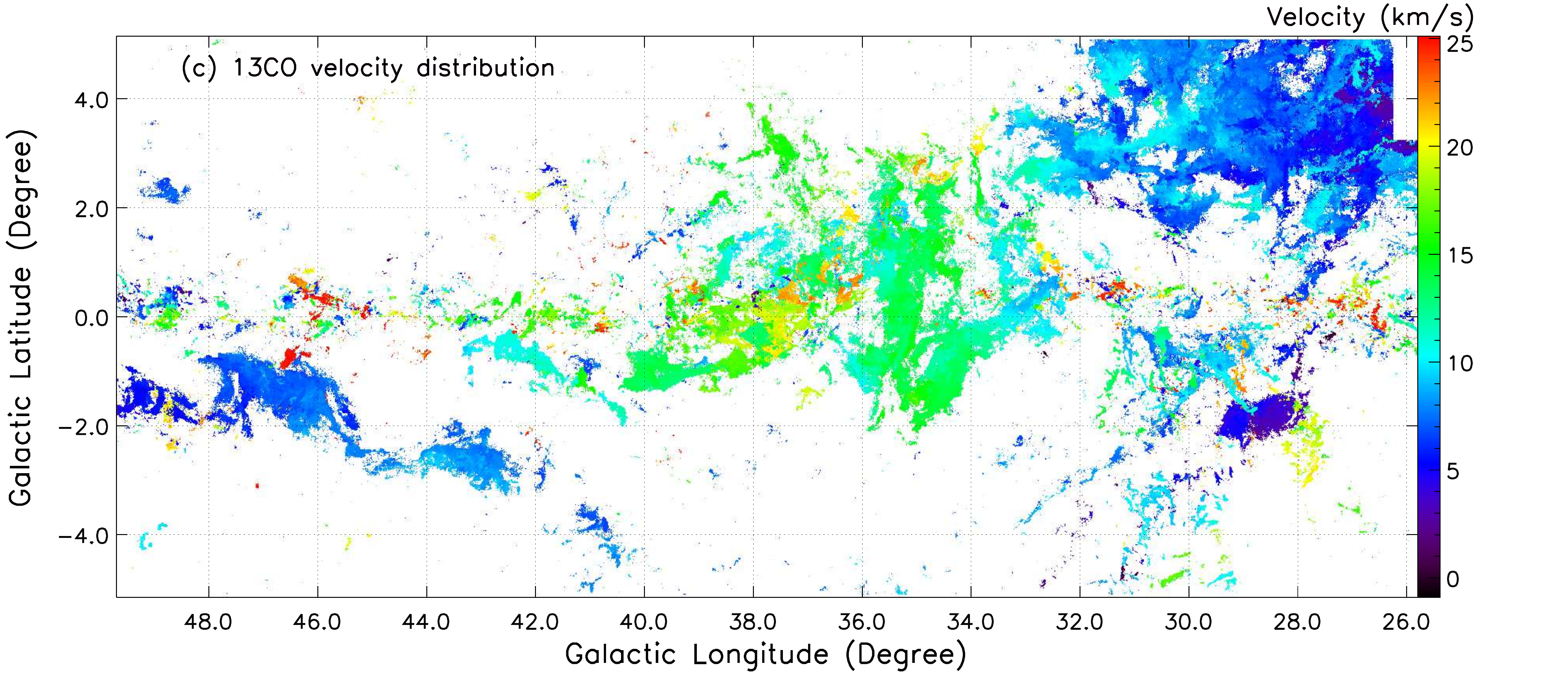}{1.\textwidth}{}
          }
\vspace{-6ex}
\caption{
Panel (a): Peak temperature ($T_{\rm MB}$) of the \twCO\ emission in the velocity 
interval of [-1, 25]~km~s$^{-1}$ in the whole region.
The three red circles show the known active star-forming regions 
with $T_{\rm peak12}\gsim 10$~K, i.e., W40, Serpens NE Cluster, 
and LBN 031.75$+$04.98. 
The two red rectangles show other regions with high peak temperature 
of \twCO\ emission.
Panel (b): Peak temperature of the \thCO\ emission 
overlaid with the C$^{18}$O integrated emission contours of
1, 2, 3, and 4~K~km~s$^{-1}$ (in red). The two blue boxes show the 
active star-forming regions with $T_{\rm peak12}\lsim 10$~K, i.e., Sh 2-62 and LDN 673.
The blue rectangle shows the C$^{18}$O concentration at 13--15~km~s$^{-1}$.
Panel (c): Intensity-weighted \thCO\ mean velocity (first moment) map
(or the velocity field).
\label{t1213}}
\end{figure}

\begin{figure}
\gridline{
          \hspace{-5ex}
          \fig{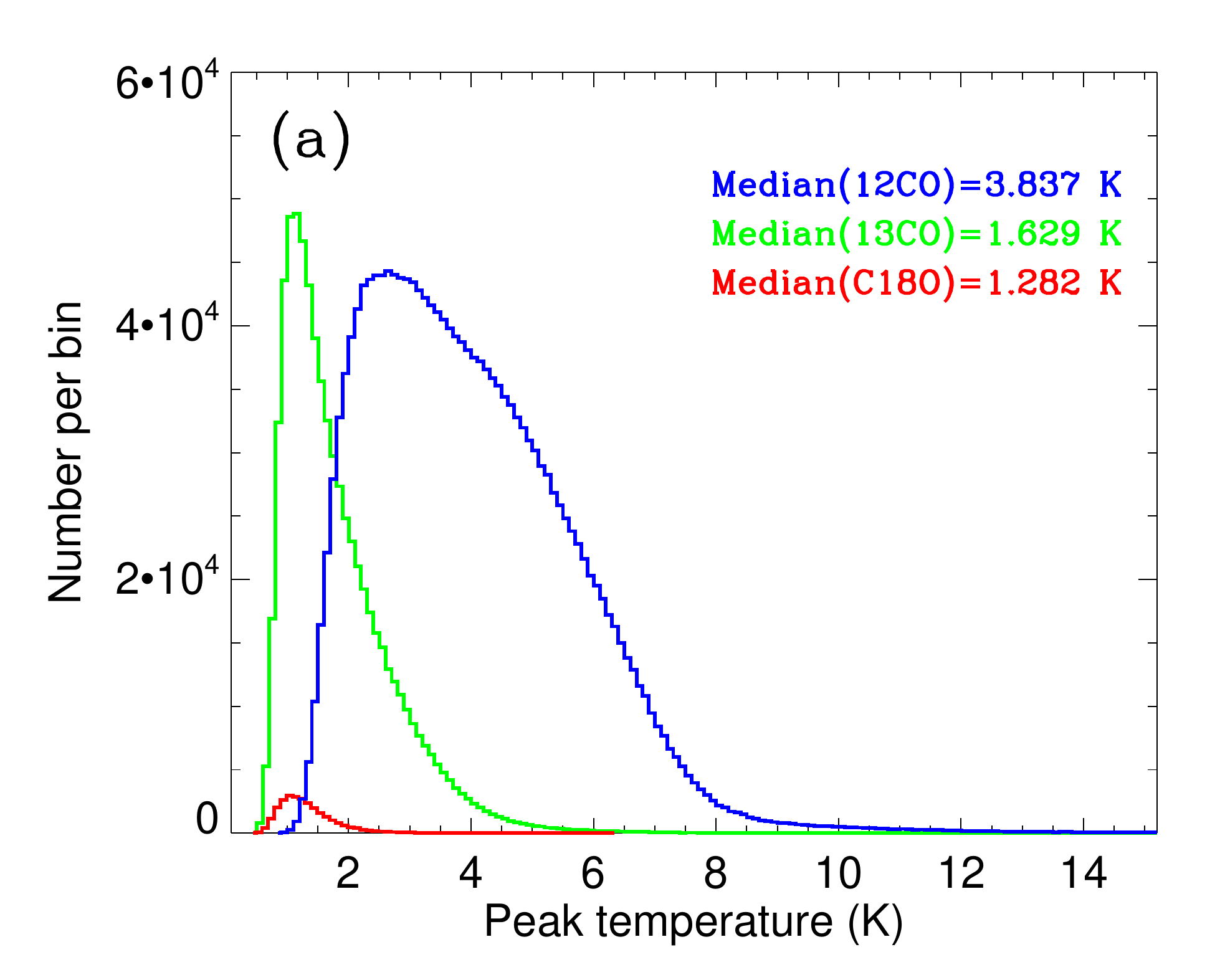}{0.55\textwidth}{}
          \hspace{-5ex}
          \fig{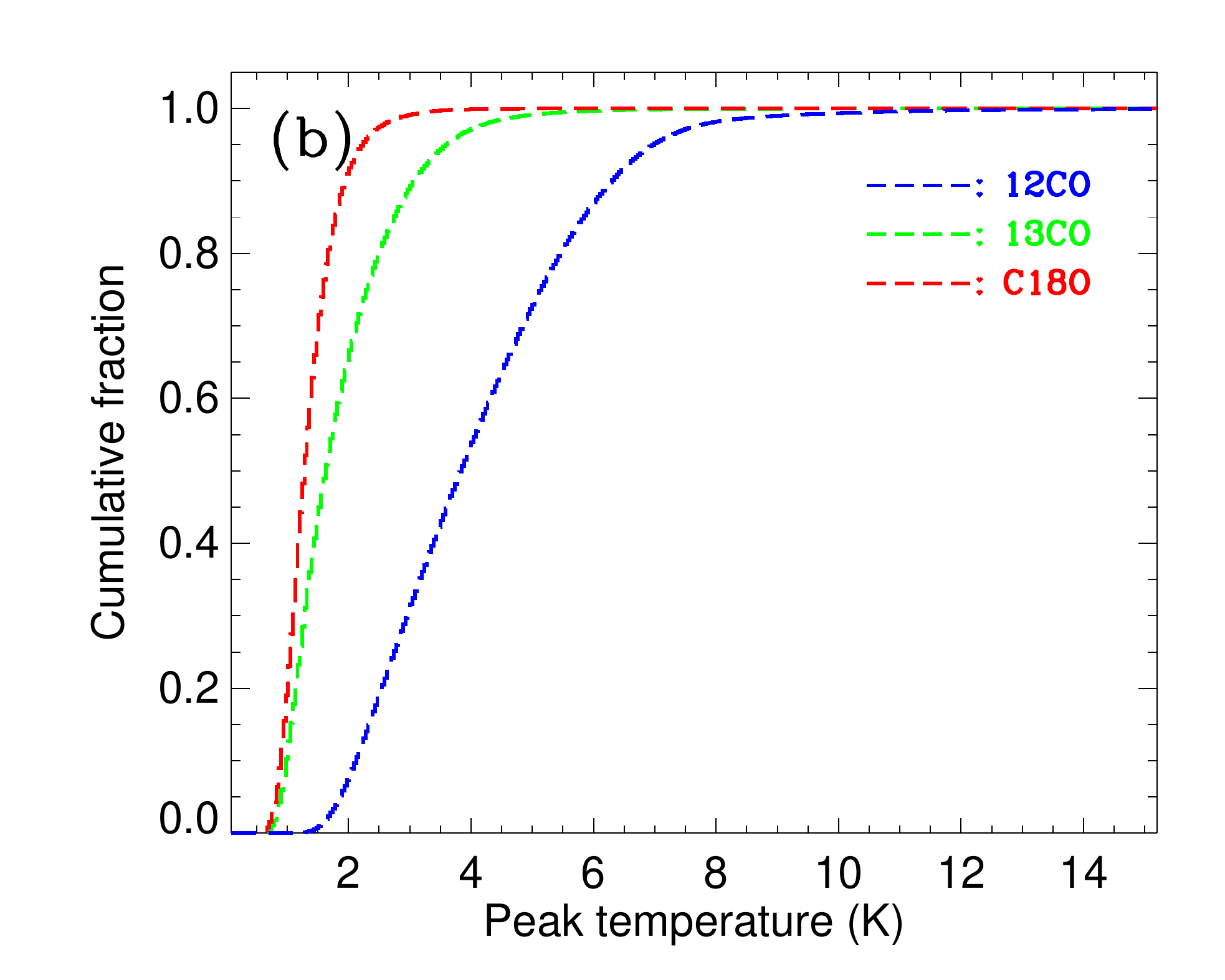}{0.55\textwidth}{}
          }
\caption{
Panel (a): Distribution of the peak temperatures of the three lines 
for the local molecular gas in the velocity range of [-1, 25]~km~s$^{-1}$.
The blue, green, and red lines indicate the 
1.8$\times10^6$, 7.0$\times10^5$, and 2.6$\times10^4$ samples for
the \twCO, \thCO, and C$^{18}$O emission, respectively.
Panel (b): Cumulative distribution of peak temperatures for the three CO lines.
Note that points with three consecutive channels greater than
3$\sigma$ are considered to be the valid samples.
The total number of pixels in the whole observed region
is $\sim3.5\times10^6$.
\label{tacc}}
\end{figure}

Figure~\ref{t1213} displays the peak temperature of the \twCO\ emission,
the peak temperature of the \thCO\ emission, and the velocity distribution
of the \thCO\ emission for the local molecular gas in the covered region.
For a total of $\sim 3.5\times10^6$ pixels on a 30$''$ grid, 
about a half of the samples in the coverage have detected \twCO\ emission,
while only $\sim$~20\% ($\sim$~0.7\%) pixels have \thCO\ (C$^{18}$O) emission 
(Figure~\ref{tacc}).
For the detectable \twCO\ emission, over 50\% of \twCO\ samples have a 
peak temperature less than 4~K, indicating subthermal excitation of 
CO emission for the low-density molecular gas in the local ISM.

The \twCO\ gas with higher peak temperature (i.e., $T_{\rm peak12}\gsim$~10~K)
is roughly associated with the nearby star formation activities, 
i.e., the \HII\ region W40 at ($l=$~28\fdg75, $b=$~3\fdg51),
the Serpens NE Cluster at ($l=$~31\fdg74, $b=$~3\fdg12) \citep{2019ApJ...878..111H}, 
and the LBN 031.75$+$04.98 (see red circles in Figure~\ref{t1213}).
This result is consistent with the work of \citet{2019ApJS..243...25W}
for the Gemini OB1 MCs, where the authors show that the high-temperature
CO gas is related to nearby massive star-forming regions.
The two red rectangles near the Galactic plane show other regions 
with high \twCO\ temperature but lacking distance information.
We speculate that the surrounding star formation activities 
are probably responsible for the higher CO temperatures there.

The optical depths of the \thCO\ and C$^{18}$O lines can be estimated
using equations described in \cite{2015ApJ...811..134S}.
In brief, we assume that the excitation temperatures ($T_{\rm ex}$) of the \thCO\ 
and C$^{18}$O lines have the same value as that of the optically thick \twCO\ line,
and the beam filling factor of the gas, $f_{\rm b}$, is assumed to be 1.
Note that the estimated optical depth of $\tau_{13}$ ($\tau_{18}$)
is related to the values of $T_{\rm ex}$, $f_{\rm b}$, and $T_{\rm mb13}$ ($T_{\rm mb18}$).
Therefore, the estimated $\tau_{13}$ is likely to be a lower limit 
because of the possible underestimated $T_{\rm mb13}$ (i.e., self absorption 
of \thCO\ emission). On the other hand, $\tau_{18}$ is likely overestimated 
due to the underestimated $T_{\rm ex}$ from \twCO\ emission
(i.e., self absorption of \twCO\ emission).

For the entire coverage, about 82\% of detected \thCO\ gas shows a
low optical depth, $\lsim$0.6, indicating the dominant optically thin \thCO\ emission
in the local MCs. However, the optical depth of the 
\thCO\ line is not always below one.  A few \thCO\ points ($\lsim$4\%) have
$\tau_{13}\gsim 1$, especially in regions near the \HII\ region W40.
The high-density gas traced by the strong C$^{18}$O emission is also mainly
located near such regions \citep[e.g., see red contours in Figure~\ref{t1213}, 
also see Section 3.3.2 in][]{2019ApJS..240....9S}.
In fact, $\tau_{18}$ is always $<1$, confirming that C$^{18}$O emission
is optically thin in the Aquila Rift region.

Compared to the optically thick \twCO\ emission, the distribution of 
the optically thin \thCO\ and C$^{18}$O emission is obviously diminished 
and is limited in the interior of the envelope traced by the extended
\twCO\ emission (Figures~\ref{aquila} and \ref{t1213}).
Despite this, the \thCO\ emission successfully traces the main body of 
the local MCs because of the high dynamic range and sensitive observations. 
The velocity field of the dense molecular gas is also well described
by the first moment of the \thCO\ emission.
Moreover, the concentration of the C$^{18}$O gas represents the 
local dense gas region, where star formation activities are ongoing 
(i.e., many HH objects in LDN 673 discovered by \citealp{2018ApJ...852...13R}; 
thousands of young stars and protostars in Serpens MC, W40 region, and Sh 2-62
studied by \citealp{2019ApJ...878..111H}).

On the other hand, there are some C$^{18}$O regions with relatively low \twCO\
temperatures (e.g., the blue rectangle region for the 13--15$\km\ps$ gas 
in Figure~\ref{t1213}). The information
of star formation near these regions is unclear. We speculate that
these regions are probably in the early evolutionary stage of star formation
\citep[e.g., see the case of the Serpens filament investigated by][]{2018A&A...620A..62G}.
The Serpens filament is believed to be accreting material in the very early 
evolutionary stage of star formation, leading to a lower \twCO\ temperature
(self absorption) and strong C$^{18}$O emission. 
In fact, \thCO\ emission in some positions also displays a
self-absorption feature compared to the optically thin C$^{18}$O emission.
We will study these samples in the future
based on multi-wavelength data and statistical methods
\citep[e.g., see studies in][]{2019ApJS..243...25W}.

\begin{figure}
\vspace{-3ex}
\gridline{
          \hspace{1ex}
	  \fig{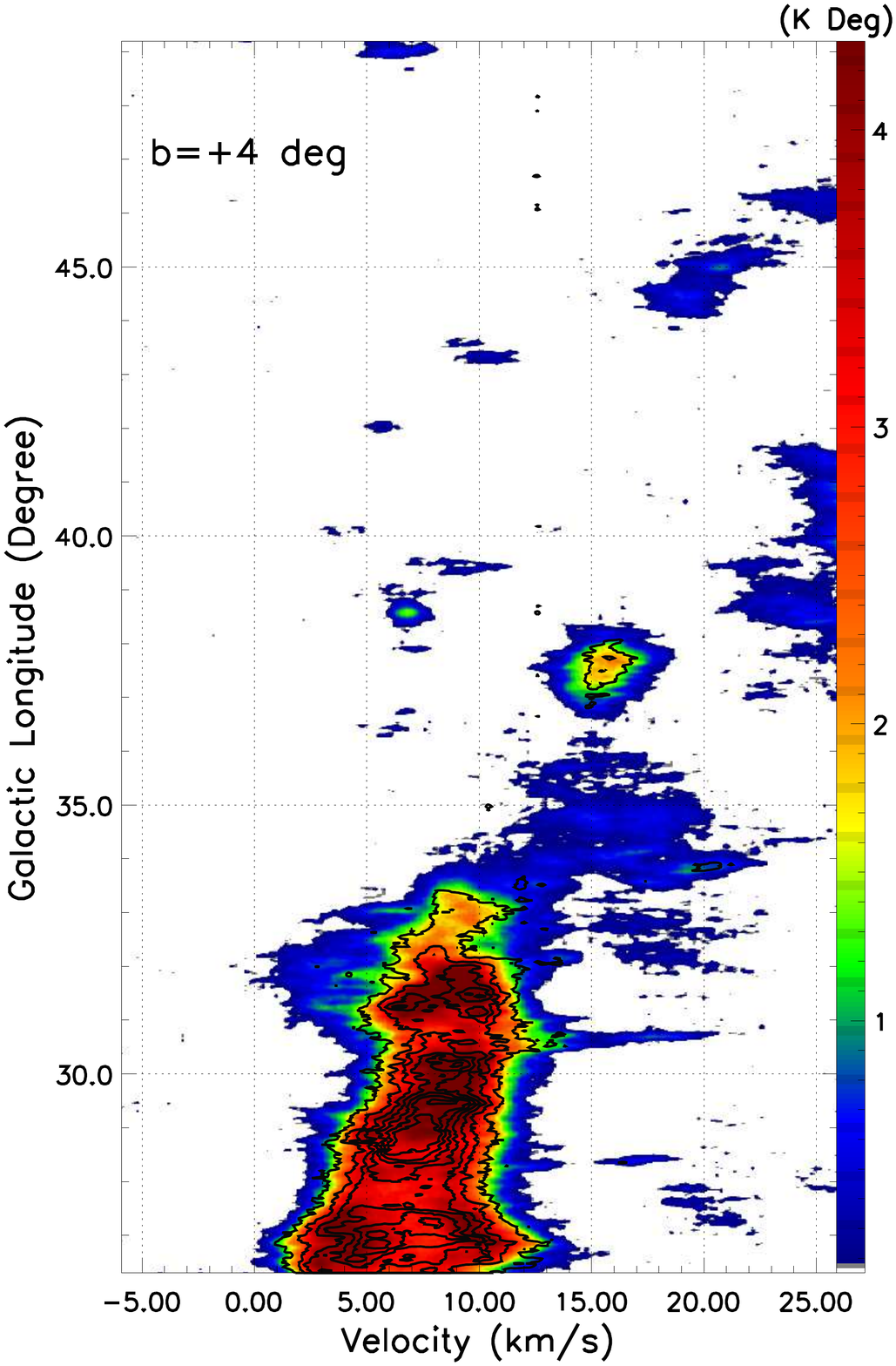}{0.5\textwidth}{}
	  \hspace{-1ex}
          \fig{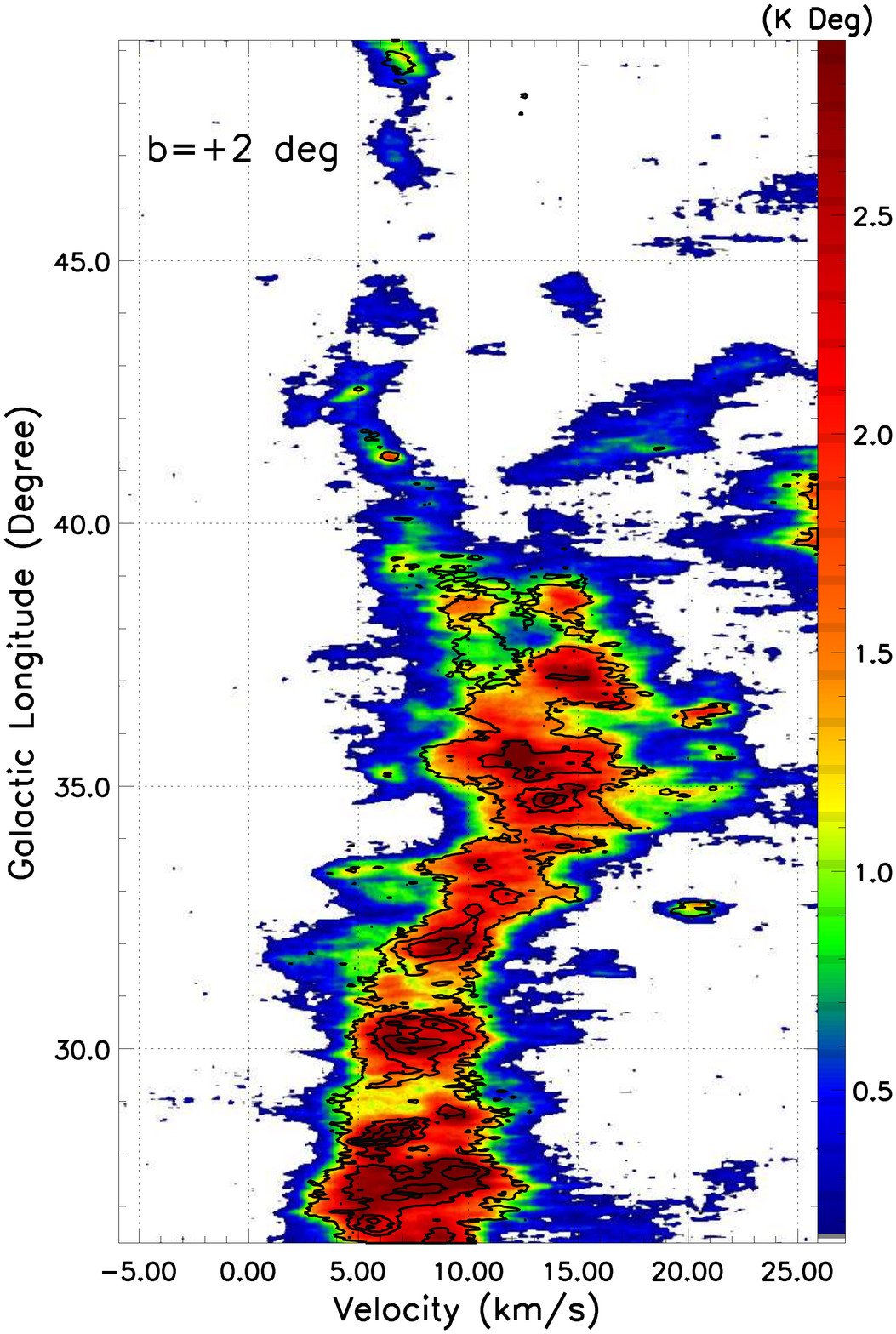}{0.5\textwidth}{}
          }
\vspace{-7ex}
\gridline{
          \hspace{1ex}
	  \fig{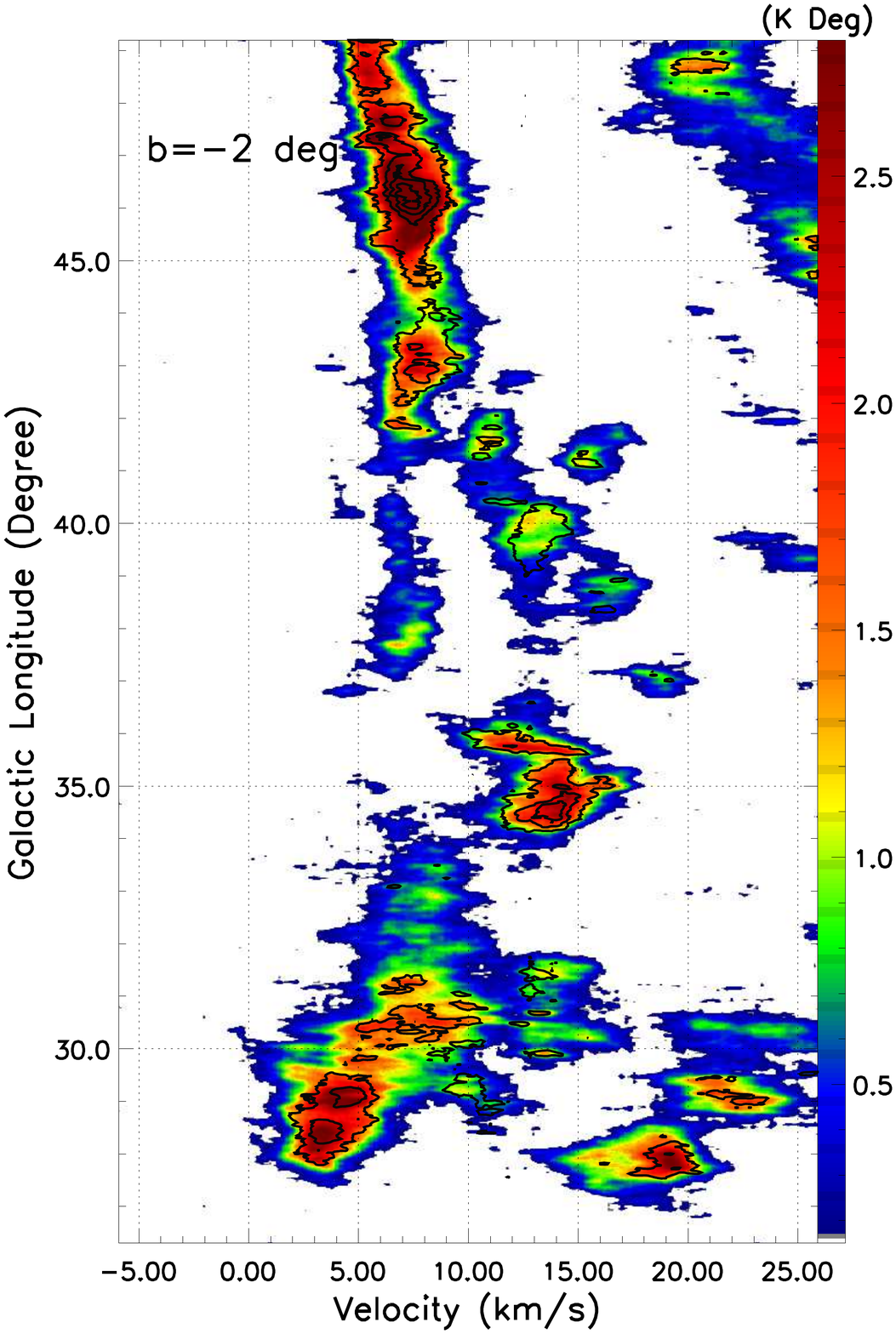}{0.5\textwidth}{}
          \hspace{-1ex}
          \fig{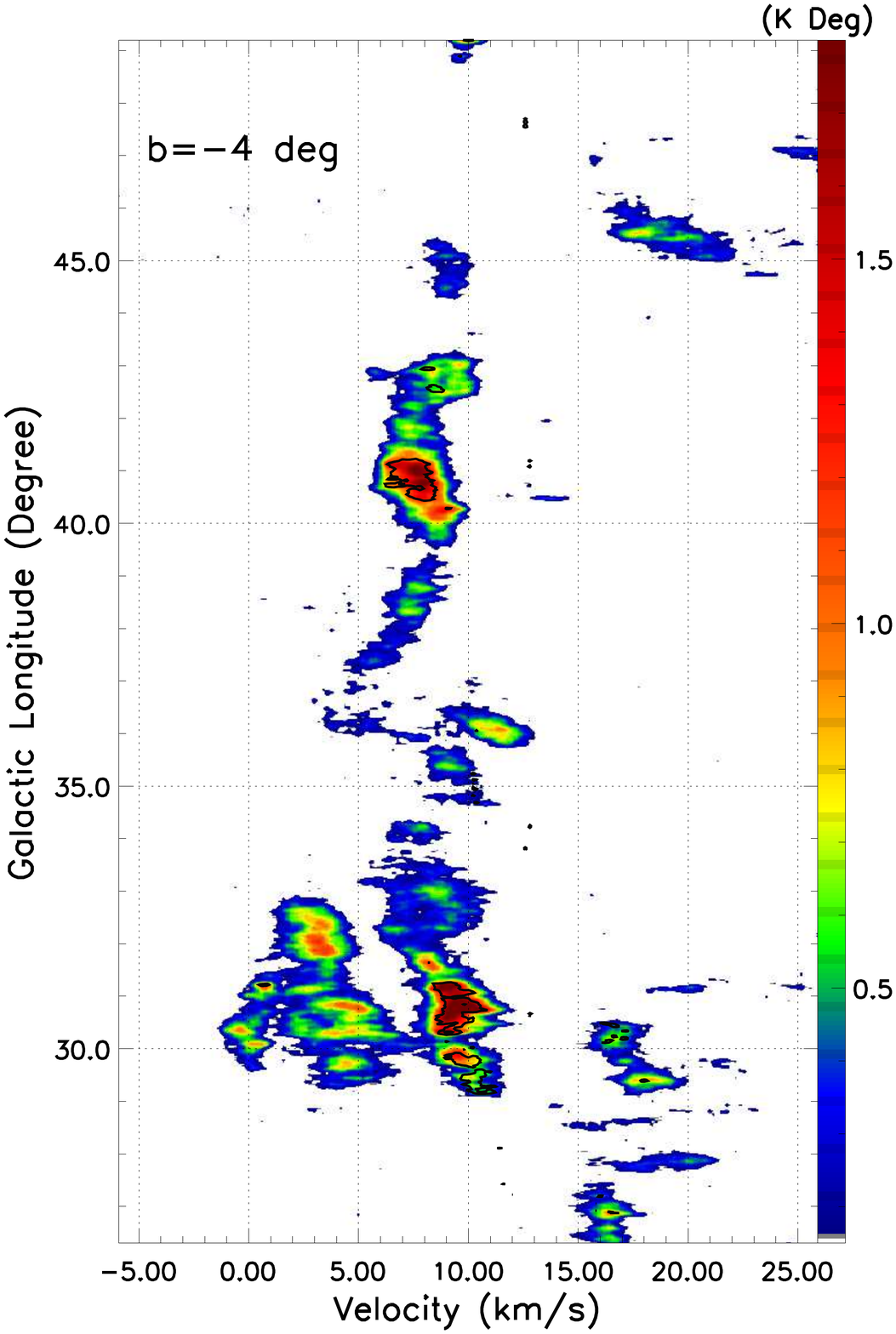}{0.5\textwidth}{}
          }
\vspace{-5ex}
\caption{
Longitude--velocity diagrams of the \twCO\ (color) and \thCO\ (contours) emission for 
the local molecular gas along $b=+4^{\circ}, +2^{\circ}, -2^{\circ},$ and $-4^{\circ}$, 
respectively. The slices have a length of 23\fdg9 and a width of 2\fdg0.
The contours of \thCO\ emission start from 0.15~K~Deg with a step of 0.3~K~Deg.
\label{pv}}
\end{figure}

Figure~\ref{pv} describes the molecular-gas distribution via 
the traditional position$-$velocity (PV) diagrams along
$b= +4^{\circ}$, $+2^{\circ}$, $-2^{\circ}$, and $-4^{\circ}$, respectively.
Roughly, the CO emission in the observed area can be divided into three main parts:
the molecular gas at 
(1) $l\sim +26^{\circ}$ to $+33^{\circ}$ with $V_{\rm LSR} \sim$~0--15~km~s$^{-1}$,
(2) $l\sim +32^{\circ}$ to $+41^{\circ}$ with $V_{\rm LSR} \sim$~10--20~km~s$^{-1}$,
and (3) $l\sim +40^{\circ}$ to $+50^{\circ}$ with $V_{\rm LSR} \sim$~3--11~km~s$^{-1}$. 
Again, the main body of local MCs is well described by the \thCO\
emission (contours in the figure), which is similar to results
shown in Figure~\ref{aquila}.
For comparison, the peak-temperature and velocity distributions of \thCO\ emission
in Figure~\ref{t1213} also show similar results: MCs are mainly 
concentrated in well-defined space and velocity ranges.
We investigate the properties of the molecular gas 
via these defined sub-regions, as well as the coherent spatial 
and velocity structures of the local MCs in Section 3.4.

\subsection{Distances of the Local MCs}
MCs with similar LSR velocities can be located at quite different distances.
The situation is even more serious for the complex MCs with
multi-velocity components in the first quadrant.
Therefore, a more accurate distance is very important 
in studying the physical properties of the local MCs.
In this work, distances of local MCs are estimated based on
the Gaia DR2 data \citep{2016A&A...595A...1G,2018A&A...616A...1G}.
The method is simple and straightforward since the extinction is
determined by the amount of dust along the
line of sight (LOS). Assuming that the extinction
of a local MC is mainly determined by the dust in the cloud, 
the extinction ($A_{\rm G}$) of stars
behind the MC will be obviously larger than that of stars in front of the MC.
The key point is how to pick up the jump correctly in the parallax--$A_{\rm G}$ space.
Here, we employ the Markov Chain Monte Carlo (MCMC) algorithm
discussed very recently by \cite{2019A&A...624A...6Y}.

To obtain a more accurate distance estimation of the local
molecular gas, we must define the structure of the MCs first.
Generally, the CO data are viewed channel by channel to
search for coherent MC structures in the velocity field 
on a scale of several degrees. 
The selected MC structures with coherent velocity features 
are also checked in the corresponding PV diagrams.
The boundary of an MC is then defined by the selected velocity 
interval with the selected threshold of the integrated CO intensity, 
which is crucial for the subsequent analysis.

Briefly, there are two cases in the definition of the
CO structure. 
One case is the overlapping MCs with adjacent LSR velocities.
The non-overlapping velocity intervals are integrated to show
the distribution of the different MCs. 
That is, for individual examples, the overlapping MCs are divided into 
different parts based on their LSR velocities 
(e.g., MCs G030.5$-$01.5 and MCs G029.0$-$2.0a in Figure~\ref{aqu_03}) 
or spatial distributions (e.g., GMC G036.0$+$01.0 in Figure~\ref{aqu_1319}).
And then the individual CO
structures can be used to calculate the corresponding distance of the MCs.
The overlapping parts of MCs in an area are usually avoided in the 
estimation of the distance (e.g., see W40a and W40b in Figure~\ref{aqu_03}).
Regions near the Galactic plane are also avoided due to the complex
gas environments there. The other case is the MCs with little
velocity confusion. We used all CO emission in the velocity interval 
to define the MC structure, e.g., MC G044.0$-$02.5. 
The distance of the MC is estimated by using the whole CO structure.
We also divided the single MC into several sub-regions to check 
the consistency of the distance estimation (see Figure~\ref{aqu_312}).

The integrated-intensity threshold of a defined MC
is roughly $\gsim$~3--5~K$\km\ps$, depending on the 
cloud properties (i.e., sizes, environments, and locations). 
The threshold of the CO intensity is different for 
different MC structures. For example,
the thresholds of MCs are $\sim$~3--5~K$\km\ps$ for the W40a region 
and $\sim$~10--15~K$\km\ps$ for the W40b region, respectively.
Based on the CO structure from the selected velocity interval, 
the $A_{\rm G}$ of stars within the MC are then sorted 
according to the distances of stars.
Using the MCMC algorithm, the jump point of on-cloud $A_{\rm G}$ 
in the distance--$A_{\rm G}$ space is picked up to estimate the distance 
of the CO cloud. We also check the validity of the distance estimation 
by comparing the $A_{\rm G}$ 
distribution of the nearby off-cloud samples with that of on-cloud samples.
Results with large uncertainties are excluded according to
the comparison between the on-cloud $A_{\rm G}$ and off-cloud $A_{\rm G}$.

Generally, the estimated distance error of local MCs is less than 5\%, 
which is mainly from the uncertainties in $A_{\rm G}$, the different 
properties of the MCs (i.e., size, structure/morphology, and column 
density, etc.), and the contamination of other gas components along the LOS.
As a result, the distance error is relatively small 
for large-scale MCs with the strong CO emission 
(i.e., MC structure with a sky coverage of several square degrees and CO integrated 
intensity of $\gsim$~10~K~km~s$^{-1}$). 
On the other hand, the dust in some
extended foreground cloud leads to the complicated distribution of 
$A_{\rm G}$ along the LOS. It is thus hard to calculate the distance
of individual MCs in the region with multi-velocity components,
e.g., MCs near the Galactic plane.
In some cases, the distance of small CO patches cannot be determined 
due to the limited on-cloud Gaia samples.
Despite this, some small MCs with coherent velocity and
spatial structures can be considered together to estimate the distance
of the molecular gas on a large scale (see rectangles in Figure~\ref{aqu_29}).
Meanwhile, distances of some CO structures with weak emission 
(i.e., $\lsim$~3--5~K~km~s$^{-1}$) cannot be obtained 
because of the low column density of MCs and the large $A_{\rm G}$ error of stars.

With the method discussed above, distances of some large-scale 
CO structures are successfully determined for the local MCs 
based on the MWISP CO data and the Gaia DR2 (see Table 1).
All selected regions with estimated distance
are labeled in Figures~\ref{aqu_03}--\ref{aqu_312}.
These MCs are distributed in a large distance range of 
$\sim$~220--770~pc in the whole region.
In the region of $l\sim +26^{\circ}$ to $+33^{\circ}$ (Figure~\ref{aqu_03}),
the dominant CO gas, which is mainly associated with the Serpens
clusters and the \HII\ W40 region, is located at $\sim$~470~pc.
Our results roughly agree with the previous studies based on 
VLBA measurements \citep[i.e., 436.0$\pm$9.2 pc in][]{2017ApJ...834..143O}
and the independent analysis from Gaia DR2 
\citep[i.e., $\sim$~400--500~pc in][]{2018ApJ...869L..33O,2019ApJ...879..125Z}.
In regions of $b\lsim 0^{\circ}$, some interesting MC structures 
are located at $\sim$~400--550~pc. 
Interestingly, toward the region of $l\sim +26^{\circ}$ to $+33^{\circ}$,
considerable molecular gas at $\sim$~240--260~pc is situated in front of the 
$\sim$~400--550~pc MCs (see the left panel of Figure~\ref{aqu_03}), 
which leads to some controversy over MC distances in previous studies
\citep[e.g., see discussions and references in][]{2017ApJ...837..154N}.
Additionally, the nearer MCs at $\sim$~220--240~pc also extend from the
$l\sim +36^{\circ}$ to $+44^{\circ}$ region with larger LSR velocities 
(Figure~\ref{aqu_29}).

Toward the region of $l\sim +32^{\circ}$ to $+41^{\circ}$, the distribution 
of the local molecular gas is quite complicated.
Many small MCs are estimated to be at $\sim$~220--240~pc (Figure~\ref{aqu_29}),
and some CO clouds are located at $\sim$~610~pc; 
although, these MCs cannot be easily identified due to their complex and intricate
velocity components. Nevertheless, we suggest that 
most of the molecular gas in the region is
located at a distance range of $\sim$~560--670~pc based on
Figures~\ref{aqu_29} and \ref{aqu_1319}. By considering the mass, the extended size, and 
the distance error (i.e., $\lsim$~5\%, 20--30~pc) 
of the molecular gas, we defined a GMC named as G036.0$+$01.0 
(the nickname of ``Phoenix cloud").
This GMC is located at about $\sim$~610~pc and  
is less studied in the previous literature.

Toward $l\sim +40^{\circ}$ to $+50^{\circ}$,
the molecular gas is dominated in the $b\lsim 0^{\circ}$ region.
Two local MCs with little contamination from other velocity components 
are estimated to be located at $\sim$~400~pc and $\sim$~770~pc, respectively. 
We identified the prominent $\sim$~404~pc MC feature as MC G044.0$-$02.5
(the nickname of ``River cloud"; Figure~\ref{aqu_312}), 
which was named as Cloud B in previous studies 
\citep[e.g., see][]{1985ApJ...297..751D,1987ApJ...322..706D}.
We also note that the Cloud A at $\sim$~25--27$\km\ps$ 
\citep{1985ApJ...297..751D,1987ApJ...322..706D} probably has a distance of 
$\gsim$~1.1~kpc, which is thus not discussed in the present paper
\citep[see the 20--30$\km\ps$ maps in ranges of 
$l\sim +44^{\circ}$ to $+47^{\circ}$ in Figures~6 and 7 of][]{2019ApJS..240....9S}.
Figure~\ref{cone_dis} displays an example for the distance estimation
of a part of MC G044.0$-$02.5, which is consistent with the distance 
of the whole filamentary MC. 
Details of the method and an explanation of the figure 
can be found in \cite{2019A&A...624A...6Y}.

All of these new results, together with other similar studies
\citep[e.g., see Table~3 and Figure~7 in][]{2019ApJ...878..111H},
remarkably refine our knowledge on the local MCs. More recently,
a narrow and elongated arrangement of dense gas 
in the local ISM is well unveiled based on the accurate distances 
of star forming regions 
toward the anticenter of the Milky Way
\citep{2020Natur.578..237A}. We hope that our study will be helpful in 
revealing possible large-scale structures toward the inner Galaxy in the future.

\subsection{The Mass Distribution of the Local Molecular Gas}
After combining the MWISP CO data and more accurate distances
of the MCs within $\lsim$~1~kpc, we can discuss the mass distribution 
of the local molecular gas toward the Aquila Rift region.

For CO observations, the column density of MCs can be estimated from 
the \twCO\ luminosity and the CO-to-H$_2$ conversion factor, $X_{\rm CO}$.
The total mass of the molecular gas can be calculated 
if we know the MC's distance.
Usually, the value of $2\E{20}$~cm$^{-2}$(K~km~s$^{-1})^{-1}$ is 
recommended by many studies \citep[e.g., ][]{2001ApJ...547..792D,2013ARA&A..51..207B}.
Based on the MWISP data, in fact, the conversion factor 
can be estimated from the \twCO\ and \thCO\ emission.
And then we can use the estimated value of $X_{\rm CO}$ 
to calculate the mass of the local MCs.
 
Here, we use samples of MC G044.5$-$02.0 to estimate the value
of $X_{\rm CO}$ due to little velocity confusion in the
direction (see Section 3.4.3). 
First, assuming that the \twCO\ emission
is optically thick and the beam filling factor of the extended gas is 1, 
we can estimate the excitation temperature from the peak
temperature of \twCO\ emission. Second, the column density
of \thCO\ is calculated pixel by pixel 
in the local thermodynamic equilibrium (LTE) assumption
\citep[see discussions and equations in Section 3.1.2 of][]{2015ApJ...811..134S}.
Third, the column density of H$_2$ can be obtained 
by adopting [\twCO]/[\thCO]=66 \citep{2014A&A...570A..65G} 
and [H$_2$]/[\twCO]$\sim 1.7\times10^4$.
The value of $X_{\rm CO}$ is thus calculated from the resulting ratio between
$N({\rm H_2})$ from the \thCO\ emission and $I_{\rm CO}$ from the
corresponding \twCO\ data.
The mean value of the X-factor is $\sim2\E{20}$~cm$^{-2}$(K~km~s$^{-1})^{-1}$
under the above assumptions.
We find that the $N({\rm H_2})$/$I_{\rm CO}$ vs. $I_{\rm CO}$ trend
from MC G044.5$-$02.5 is similar to the recent result from the
Galactic cluster-forming molecular clumps
\citep[i.e., see details in][]{2018ApJ...866...19B}.
We emphasize, however, that the derived value of $X_{\rm CO}$ 
largely depends on the uncertainties of the abundance ratios 
of [\twCO]/[\thCO] and [H$_2$]/[\twCO], which are difficult to 
estimate just based on the MWISP CO data.

The MC distances and masses, together with other physical parameters,
are briefly summarized in Table~1.
The two largest concentrations are W40b and the GMC complex G036.0$+$01.0,
which have a total mass of $\gsim 3.4\times10^{5} \Msun$.
The former is related to the embedded star-forming regions 
of the \HII\ region W40 and Serpens NE cluster,
while the latter region is less studied in the literature.
The distance separation between the two main mass concentrations 
is about 130~pc, which is roughly twice the extension of 
the GMC complex 
G036.0$+$01.0 (i.e., 6\fdg8~$\approx$~70~pc at a mean distance of $\sim$~610~pc).
Other local MCs with $\sim 10^{3}-10^{4}\Msun$ are 
either in distinct groups (Figures~\ref{aqu_03} and \ref{aqu_312})
or in a single concentration (Figure~\ref{aqu_29}).

All of these show that molecular gas toward the Aquila Rift region
is distributed in different distances, which agrees well with the new
results from the very recent study \citep[see details in][]{2019ApJ...878..111H}.
As shown in Figure~7 of \cite{2019ApJ...878..111H}, the $\sim 350$--480~pc
for the cyan circles matches our value of $\sim 474$~pc for 
the molecular gas toward the W40b region (the middle panel in Figure~\ref{aqu_03}) 
and the $\gsim 550$~pc for the red circles agrees with
the value of $\sim 560$--670~pc toward the GMC complex G036.0$+$01.0 (Figure~\ref{aqu_1319}).
The distance of $\sim 235$~pc for the molecular gas at the W40a region 
(left panel in Figure~\ref{aqu_03}) is also consistent with 
the value of $\sim 250$~pc for the extended dust cloud 
\citep[i.e., the Serpens Cirrus, see Table 3 and 
the yellow contour in Figure~7 of][]{2019ApJ...878..111H}.  
Moreover, the distance of $\sim 404$~pc toward the MC 
G044.0$-$02.5 (Figure~\ref{aqu_312}) is also consistent with
their estimated value of $\sim 407$~pc for the associated 
cloud LDN 673 \citep[see Table 1 in][]{2019ApJ...878..111H}.

Toward the Aquila Rift, we conclude that the dominant mass (i.e., $\gsim$~95\%) 
of the local MCs in the coverage is in regions $\gsim 400$~pc,
while only several $10^{3}\Msun$ MCs are located in regions of $\lsim 300$~pc.
We plan to build a sample database of local MCs with more reliable
properties in the next study (e.g., MC identification, distance estimation, 
and mass distribution, etc.). 
The MWISP CO data from further accumulation 
(e.g., $l \sim 20^{\circ}-26^{\circ}$), together with the future releases 
from the Gaia data,
are helpful in studying the properties of the whole local molecular gas.

\subsection{Individual Regions}
In this Section, we discuss the physical properties of 
the local MCs with three regions as examples:  
molecular gas in $l\sim +26^{\circ}$ to $+33^{\circ}$,
GMC complex G036.0$+$01.0 in $l\sim +32^{\circ}$ to $+41^{\circ}$, 
and MC G044.0$-$02.5 in $l\sim +40^{\circ}$ to $+50^{\circ}$.
We mainly focus on the molecular-gas distribution with different LSR
velocities and the dynamical features of the local molecular gas.

\subsubsection{MCs toward $l\sim +26^{\circ}$ to $+33^{\circ}$ Regions}
Figure~\ref{W40_45115_rgb} displays the \thCO\ emission 
in the velocity intervals of [4.5, 7.0]~km~s$^{-1}$ (blue),
[7.0, 9.5]~km~s$^{-1}$ (green), and [9.5, 12.0]~km~s$^{-1}$ (red). 
The molecular gas in the velocity range of 4.5--12.0~km~s$^{-1}$ is mainly located 
at $\sim 470$~pc, excluding some CO gas at $V_{\rm peak}\sim$~2--4~km~s$^{-1}$ 
(or $\sim 240$--260~pc in the left panel of Figure~\ref{aqu_03}).

Toward this region, we find that \thCO\ emission is a good
tracer for identifying MC structures due to its relatively larger extension
compared to the C$^{18}$O emission and its smaller optical depth (Section 3.1)
compared to the \twCO\ emission.
Interestingly, at least tens of filaments (the elongated 
structure with an aspect ratio larger than $\sim$5) with different velocity 
structures are found to be located in the complex region
of \thCO\ emission (see some typical filaments indicated 
by arrows in Figure~\ref{W40_45115_rgb}).
The typical length and width of these filamentary structures
are tens of arcminutes and several arcminutes, respectively.
Additionally, the typical column density of these filaments 
is in the range of $\sim 10^{21}-10^{22}$~cm$^{-2}$ estimated from the \thCO\ gas.
Several filamentary structures with
somewhat different velocities seem to just overlap on the
Serpens NE region, which is also seen by \cite{2017ApJ...837..154N} 
according to \twCO\ (J=2--1) and \thCO\ (J=2--1) data.
Systematic analysis of these filaments will be presented in
future works by using some algorithms on the 3D \thCO\ and C$^{18}$O data cubes
\citep[e.g., see discussions in][]{2017ApJ...838...49X,2019ApJ...880...88X}.

We note that the C$^{18}$O emission 
is roughly concentrated in these filamentary structures
(white contours associated with the filamentary structures 
in Figure~\ref{W40_45115_rgb}).
Here, we use PV diagrams (Figure~\ref{W40pv}) to show the
detailed velocity structures of \thCO\ and C$^{18}$O emission
along two filaments (see red arrows in Figure~\ref{W40_45115_rgb}).
Including the main velocity structure
seen in \thCO\ and C$^{18}$O emission (i.e., V$_{\rm LSR} \sim 6-8\km\ps$),
the \thCO\ gas also displays additional velocity components 
(i.e., V$_{\rm LSR} \sim 3-6\km\ps$ and V$_{\rm LSR} \sim 8-11\km\ps$)
near the main LSR velocity structure.
In fact, very weak C$^{18}$O emission (e.g., T$_{\rm peak18}\lsim$~0.4--0.5~K 
at $\sim8.5\km\ps$ in the right panel of Figure~\ref{W40pv}) 
is also found to be associated with this
additional velocity structure seen in \thCO\ emission.

The star formation activities \citep[e.g.,][]{2013ApJ...779..113M,2014ApJ...791L..23N,
2015ApJS..220...11D,2019ApJ...878..111H}
are strong in regions with enhanced C$^{18}$O emission
and the high peak temperature of \twCO\ gas, e.g., 
the Serpens NE cluster, the \HII\ region W40, and
the \HII\ region Sh 2-62 (or the MWC 297 region at $l= $26\fdg8 and $b= +$3\fdg5).
The most powerful source in this general region is the \HII\ region W40, 
which is responsible
for the spatial and velocity distributions of the surrounding molecular gas
\citep[e.g., see Figures~15--16 and Section 3.3.2 in][]{2019ApJS..240....9S}.
On the other hand, the gas in regions with strong C$^{18}$O emission
but relatively low \twCO\ temperature is probably in such
an early stage that star formation is ongoing \citep[e.g., see the Serpens 
filament and embedded young stellar object (YSO) candidates in][]{2018A&A...620A..62G}.

Figure~\ref{aqu_17_612_rgb} displays the gas distribution at
$\sim 260$~pc (the upper panel for MCs G029.0$-$02.0a) and 
$\sim 400$~pc (the lower panel for MCs G030.5$-$01.5).
The CO gas with somewhat different LSR velocities (or distances) has
a similar northwest-southeastern extension from 
$b\sim +$0\fdg5 to $b\sim -$5\fdg0 (or with a length of $\sim$30--50~pc).
This finding shows that at least two distinct gas components 
with different distances 
are located in the region, which is similar to the W40 region
with $b\gsim 0^{\circ}$ (the left and middle panels 
in Figure~\ref{aqu_03}). 

Additionally, many protruding CO structures (see arrows in Figure~\ref{aqu_17_612_rgb}) 
are located toward a large-scale bright optical nebula centered at 
($l \sim 26^{\circ}$, $b \sim -3^{\circ}$) with a diameter of several degrees
\citep[e.g., see the Southern H$\alpha$ Sky Survey by][]{2005MNRAS.362..689P}.
Some of the CO protrusions, which are perpendicular to the above 
elongated northwest-southeastern molecular-gas structures, 
also have remarkable velocity gradients
of $\sim0.7-1.3\km\ps$pc$^{-1}$ at a distance of $\sim 400$~pc 
(e.g., see the black arrow in Figure~\ref{aqu_17_612}, also
see solid lines in Figure~\ref{fingerpv}).
Comparing to the quiescent gas at $V_{\rm LSR} \sim$~9--11~km~s$^{-1}$, 
the \twCO\ gas near the tip of the protrusion also displays 
redshifted velocity up to $\gsim15\km\ps$, indicating interactions 
between the gas and the shock there.
We suggest that these features are probably related to nearby
stellar feedbacks (e.g., winds and radiation from nearby early-type
OB stars or shocks from supernova remnants).

\subsubsection{GMC Complex G036.0$+$01.0}
We have shown that the most striking feature in $l\sim +32^{\circ}$ to $+41^{\circ}$
is the GMC complex G036.0$+$01.0 at a distance of $\sim$~610~pc.
The GMC complex has a total mass of $\sim 2.0\times 10^{5} \Msun$ 
(Table~1). 
Similar to other MCs discussed above, filamentary structures and the
complex velocity field are also prevalent throughout the GMC complex
(Figures~\ref{b_1117_rgb} and \ref{b_1117}).
The whole GMC complex, which extends about $\sim$~35 
square degrees in the map, consists of three main parts: 
the $l\lsim +36^{\circ}$ region, 
the $+36^{\circ} \lsim l \lsim +38^{\circ}$ region,
and the $l\gsim +38^{\circ}$ region (also see Figure~\ref{aqu_1319}). 
These CO clouds, which consist of many arc-like or sub-filamentary 
structures, display different morphologies in the 
individual sub-regions of size of $\sim$~30--50~pc.

We made channel maps to show the spatial
and velocity distributions of the molecular gas for two sub-regions
in the GMC complex (Figures~\ref{b12channel} and \ref{b13channel}). 
According to the \twCO\ channel maps shown in Figure~\ref{b12channel}, 
we find that many elongated structures with a length of
several tens of arcminutes (or several pc at a distance of $\sim$~555~pc) 
and somewhat different LSR velocities (i.e., $\sim13-18\km\ps$) 
are roughly along the southwest-northeast direction, 
which is similar to the whole large-scale CO extension in 
this region (see the arrow in Figure~\ref{b12channel}).

In the densest region of the GMC complex,
the \thCO\ channel maps reveal several arc-like structures on a scale of 
$\sim10-20$~pc (e.g., see dashed lines in Figure~\ref{b13channel}).
The large-scale arc-like structures consist of many
curved structures with a smaller length of several arcminutes.
In fact, these small curved structures can be clearly seen
in the channel maps at a velocity separation of $0.2\km\ps$
due to their coherent spatial and velocity features.
C$^{18}$O emission is concentrated in the
large-scale arc-like structures, in which molecular gas 
displays multi-velocity components revealed by \thCO\ emission
(Figures~\ref{b_1117} and \ref{b13channel}).


Based on the recent study by \cite{2019ApJS..243...25W},
the relatively higher \twCO\ peak temperature of $\gsim$10-15~K is usually 
related to processes of the nearby star formation activities.
The peak temperature of the \twCO\ emission toward the GMC complex
is $\lsim$~10~K (Figure~\ref{t1213}).
We thus suggest that star formation activities are probably
not strong in the surrounding of the GMC complex.
We speculate that the GMC complex is likely at 
an early evolutionary stage when the gas is just accumulating
due to the concentrations of the C$^{18}$O gas
in the arc-like structures (see dashed lines in Figure~\ref{b13channel}).
The densest regions traced by the enhanced C$^{18}$O emission in the GMC
(i.e., red contours in Figure~\ref{b_1117}) are probably
good places to investigate the possible 
star formation activities at the early evolutionary stage.

\subsubsection{MC G044.0$-$02.5}
Toward the $l\sim +40^{\circ}$ to $+50^{\circ}$ region,
the prominent feature of the local molecular gas is the MC G044.0$-$02.5.
The intriguing MC G044.0$-$02.5 displays a long filamentary structure from 
($l\sim +$49\fdg7, $b\sim -$0\fdg9, $V_{\rm LSR}\sim 5.5\km\ps$) to 
($l\sim +$39\fdg5, $b\sim -$5\fdg1, $V_{\rm LSR}\sim 9.5\km\ps$)
based on the \twCO\ emission seen in Figure~\ref{aqu_312_rgb}. 
Roughly, the elongated MC displays some
regular CO morphologies: the vortex-like structure in $l\sim$~[46\fdg7, 49\fdg5],
the plateau structure in $l\sim$~[45\fdg7, 47\fdg7], the concave structure in
$l\sim$~[43\fdg4, 46\fdg1], and the cone structure in $l\sim$~[39\fdg5, 43\fdg9]
(see labels in Figure~\ref{aqu_312_rgb}).
The interesting structure has a length of $\sim$~80~pc and 
a width of $~\sim 7-14$~pc in \twCO\ emission.
We stress that the filamentary MC G044.0$-$02.5 is a single 
giant molecular filament (GMF) based on its elongated spatial structure 
and coherent velocity distribution. 
This is also supported by the distance estimation for 
the four individual parts of the long filamentary MC 
(i.e., $\sim$~400~pc, see Figures~\ref{aqu_312} and \ref{cone_dis}).

As seen in Figure~\ref{c13channel}, the channel maps of \twCO\
and \thCO\ emission show that the long MC structure is composed of
many sub-filaments with lengths of several arcminutes to tens 
of arcminutes. Most of the sub-filaments have similar alignments with the 
long extension of MC G044.0$-$02.5;
although, some tiny elongated structures are also found to
be perpendicular to the major axis of the long filamentary cloud
(see box regions in the \thCO\ maps).
These velocity-coherent fiber-like structures, 
which are very thin (i.e., the observed size of $\sim 1'-2'$
or $\sim$~0.1--0.2~pc at a distance of 404~pc), tend to converge toward 
the trunk of the MC traced by the relatively bright 
\thCO\ emission along the GMF (e.g., see the box regions in the figure).
We named these structures as ``networks of tiny filaments or fibers" 
that are associated with the main trunk of the GMF.

Based on Figure~\ref{aqu_312_rgb},
the vortex-like structure centered at ($l\sim +$48\fdg1, $b\sim -$1\fdg1, 
radius~$\sim$~1\fdg3) is prominent at $V_{\rm LSR}\sim 5-6 \km\ps$.  
At the southwest of MC G044.0$-$02.5, a dendritic structure is
just across the filamentary MC
($l\sim +$40\fdg6, $b\sim -$4\fdg4, $V_{\rm LSR}\sim 8-9 \km\ps$).
The velocity variation of the molecular gas can be clearly 
seen in the channel maps of CO emission (Figure~\ref{c13channel}), indicating 
a systematic velocity gradient of $\sim0.05\km\ps$pc$^{-1}$ along the GMF.
This value agrees with the result from analysis of the GMFs 
in the Milky Way \citep[i.e., $0-0.12\km\ps$pc$^{-1}$,]
[]{2014A&A...568A..73R,2016A&A...590A.131A}.

In addition to the velocity gradient along the major axis of 
the filamentary MC (i.e., northeast-southwest),
G044.0$-$02.5 also shows a velocity gradient perpendicular 
to the major axis of the filament (e.g., $V_{\rm LSR}\sim 6.0-9.5\km\ps$ 
in the $l\sim 42^{\circ}-47^{\circ}$ region; see the red parts of MC 
G044.0$-$02.5 in Figure~\ref{aqu_312_rgb}).
We made PV diagrams across the trunk of the GMF 
(see arrows in Figure~\ref{c13channel}).
In Figure~\ref{conepv}, the PV diagrams show that the velocity gradient 
perpendicular to the trunk of the GMF is $\sim0.3-0.8\km\ps$pc$^{-1}$.
The above intriguing scenario is firstly revealed by the 
large-scale MWISP CO data.

For comparison with other distant MC filaments
that are associated with high-mass star-forming regions
\citep[e.g.,][]{2014A&A...565A.101T,2018ApJ...864...54D},
the $\sim$~80~pc long GMF G044.0$-$02.5 is situated at the local region and
displays abundant details in the spatial and velocity observations 
discussed above.
The relatively low temperature of the molecular gas 
(i.e., T$_{\rm 12COpeak}\lsim 10$~K) shows that 
massive star formation activities are probably lacking
in the area surrounding the GMF. Many questions need to be investigated for the filamentary
cloud, such as the networks of sub-filaments in the GMF, the stability
and fragmentation of the long filament, and the star formation
in the region.

Figure~\ref{cone_312} shows the velocity field of the whole MC.
The different spatial features
(i.e., vortex, plateau, concave, and cone),
as well as the systematic velocity gradients along and 
perpendicular to the trunk of the GMF, are well revealed
by the \thCO\ emission.
According to Figure~\ref{cone_1213}, we find that the dense gas traced 
by \thCO\ emission is concentrated in the region with larger 
\twCO\ linewidths, indicating high turbulence there 
(see the red circle in the figure).
C$^{18}$O emission is also enhanced in the region of
LDN 673, where many fiber-like structures seem to gather together
(see networks of sub-filaments in the upper box of 
Figure~\ref{c13channel}). 

In fact, the overdensity of YSO candidates from IR data
is related to the dense gas region of LDN 673.
We also note that the ongoing star formation in their early phase occurs
in the densest part of the LDN 673 cloud \citep[e.g., the discovery of many
Herbig-Haro objects in LDN 673;][]{2018ApJ...852...13R}.
Another box region in Figure~\ref{c13channel} displays the similar
filamentary networks but where little C$^{18}$O emission is detected
(at $l\sim$~42\fdg2 to 43\fdg9, $b\sim -$3\fdg3 to $-$2\fdg1, i.e.,
LDN 645, LDN 647, and LDN 651).

The interesting velocity features of MC G044.0$-$02.5,
as well as the regular molecular-gas structures,
are clearly seen in the CO channel maps with the high spatial
dynamic range.
We find that filamentary structures are prevalent throughout
the MC from scales of several arcminutes to $\gsim 10^{\circ}$.
These interesting elongated features are present in particular integrated maps
with velocity intervals of $\sim$~1--2~km~s$^{-1}$, suggesting the rapid velocity
change on a small scale of several arcminutes (Figure~\ref{c13channel}).
For both the \twCO\ emission and the \thCO\ emission, in fact, the velocity channel
maps show more filamentary structures than 
the integrated emission maps. 
Some striation-like features can also be discerned in the channel maps,
which is similar to the cases of other nearby MCs
\citep[e.g.,][]{2008ApJ...680..428G,2016A&A...590A.110C,2016MNRAS.461.3918H}.

The dust emission (e.g., see OBSID of 1342230842 and 1342230843
from the $Herschel$ data) shows similar sub-filaments in the region of
LDN 673 centered at ($l\sim$46\fdg4, $b\sim-$1\fdg3).
Therefore, we suggest that these tiny features seen in CO emission are real
density structures because of the good correlation between the dust emission features
and the enhanced \thCO\ and/or C$^{18}$O emission. Additionally, these sub-structures
in individual velocity channels partly reveal the velocity field of the CO gas, 
which is useful for investigating the dynamic features of the molecular gas.
Moreover, the long MC G044.0$-$02.5 is also an excellent laboratory to
study the large-scale dynamical properties of the molecular gas because of the
relatively regular gas morphology, the systematic CO velocity structure,
and little cloud confusion in the same direction.
We believe that MC G044.0$-$02.5 is accumulating gas
and is in an early evolutionary stage (see Section 3.5).

\subsection{MC Formation by Large-scale Converging Flows?}
The new CO data show that the local molecular gas is widely spread
across the whole map.
However, the dominant molecular gas is concentrated in localized regions
with complicated velocity fields, i.e.,
the W40 region, the GMC complex G036.0$+$01.0, and
MC G044.0$-$02.5 (see Section 3.4). 
Toward some gas concentrations, MCs with somewhat different LSR velocities
often have similar extensions and structures
(e.g., see MCs elongated along the northwest-southeast in Figure~\ref{aqu_17_612_rgb} 
and large-scale arc-like structures in Figure~\ref{b13channel}).
These large-scale MC concentrations are highly structured and organized,
as well as those hierarchical sub-structures within them
(also see filamentary networks in box regions of Figure~\ref{c13channel}).
With these details in mind, we wonder how the MCs are formed in
the local ISM.

In theoretical studies, MCs may form out of large-scale converging
atomic gas flows, which probably originate from shocks by stellar feedback
processes (e.g., expansion of \HII\ regions, winds of early-type stars, 
and explosion of
supernovae), as well as large-scale gravitational instability and shear motions.
These mechanisms can produce compressive flows and drive large-scale turbulence
into the ISM. Turbulence, thermal instability, and local gravitational instability
may be responsible for the density fluctuations, leading to the dense gas formation
out of the diffuse atomic hydrogen. 
Several theoretical views on the topic are discussed by some authors
\citep[e.g.,][]{2012A&ARv..20...55H,2014prpl.conf....3D,2016SAAS...43...85K}.
Additionally, many numerical studies on MC formation have also been undertaken 
recently \citep[e.g.,][]{2000ApJ...532..980K,2006MNRAS.371.1663D,2005ApJ...633L.113H,
2006ApJ...648.1052H,2008ApJ...674..316H,2008A&A...486L..43H,
2006ApJ...643..245V,2007ApJ...657..870V,2010MNRAS.404....2G,2014MNRAS.445.2900S}.

Therefore, large-scale gas flows are ubiquitous in the ISM due to the energy 
input by activities of massive stars or Galactic shearing motions. 
Such large-scale gas flows, which probably play a crucial role in 
accumulating the large amount of gas material, 
may display some observational features in multi-wavelengths.
We speculate that the systematic velocity structure seen in CO emission 
is one piece of the observational evidence for large-scale flows. 

For example, the intriguing case of MC G044.0$-$02.5 displays the systematic velocity 
structure along and perpendicular to the major axis of the $\sim$~80~pc long 
filament (Figures~\ref{aqu_312_rgb}--\ref{cone_312}), 
which probably indicates the large-scale gas flows in the local ISM.
Based on Section 3.4, we note that the dense gas of the MC 
often displays multi-velocity components, which can also be found in other regions 
(e.g., see contours in Figure~\ref{W40pv}
and channel maps of Figure~\ref{b13channel}). 
These features seem to support the idea that the gas 
is converging in the localized region of the MC,
although the correlation between the dense gas concentrations and the 
multi-velocity components needs to be further investigated 
for the whole mapping region.

In the following paragraphs, we focus on the observational points of the
formation of MC G044.0$-$02.5 based on
(1) the relationship between HI atomic gas and CO molecular gas, 
(2) the coherent spatial structures of MCs on various scales, 
and (3) the regular dynamical features of gas from its velocity field.

A large amount of atomic gas traced by enhanced HI emission from the
Arecibo data 
is found to be associated with MC G044.0$-$02.5 because of
the similar spatial morphology and velocity structure between them 
(Figure~\ref{HI_aqu_411_rgb}). 
In fact, the large-scale structure of the atomic gas extends to
($l\sim$~37\fdg5, $b\sim -$6\fdg5), which is not covered by the current MWISP CO map.
Based on Figure~\ref{HI_aqu_411_rgb}, we find that 
the CO emission of the intriguing structure appears in the
overlapping HI gas regions, i.e., the yellow parts of
6--9$\km\ps$ (green) and 9--11$\km\ps$ (red) in the map.
Along the directions indicated by 
cyan arrows in the figure, the atomic gas displays a similar velocity 
gradient perpendicular to the main trunk of the CO structure 
(i.e., the HI gas from the green color of 6--9$\km\ps$ to the red color 
of 9--11$\km\ps$; also see the velocity trend of CO gas in Figure~\ref{conepv}).
On a larger scale, the velocity trend of the atomic gas also shows such the
velocity gradient from the northwest to the southeast, which
is comparable to the velocity structure of the CO gas
(Figure~\ref{aqu_312_rgb}).
This feature probably indicates the transition from the converging atomic gas 
to newborn molecular gas on large scales.

In Section 3.4.3, we show that the long MC G044.0$-$02.5 has different
morphologies on scales of $\sim 2^{\circ}-4^{\circ}$ 
(or $\sim$~10--30~pc; see Figure~\ref{aqu_312_rgb}).
The gas in these sub-regions also exhibits regular networks organized by
tiny CO sub-filaments or fibers, which have somewhat different
but coherent velocity features (Figure~\ref{c13channel}).
The different gas structures in these sub-regions are probably the result of
the localized feedback of star formation
or various instabilities of the gas flow in the ISM
(e.g., gravitation, magnetic field, and thermal pressure).

The denser gas traced by \thCO\ and C$^{18}$O emission
is often located in regions of filamentary networks
(see box regions in Figure~\ref{c13channel}).
CO emission of these coherent structures frequently shows asymmetric line profiles
or multi-velocity components rather than a single Gaussian profile
(e.g., see PV diagrams in Figure~\ref{conepv}),
possibly indicating the assembling gas in certain regions.
Similar features were also revealed by some authors in
simulations \citep[e.g.,][]{2009ApJ...704.1735H,2014MNRAS.445.2900S}
and observations \citep[e.g.,][]{2017ApJ...836..211S,
2017ApJS..230....5W,2018ApJ...853..169D}.
For the particular case of LDN 673, the gas seems to converge 
toward the hub-like region (see the upper box region in Figures~\ref{c13channel}
and \ref{cone_312}),
in which the \thCO\ emission is embedded in the enhanced
\twCO\ emission with broader linewidths (the red circle in Figure~\ref{cone_1213}).
The dense C$^{18}$O gas is spatially associated with the concentration 
of the \thCO\ emission with multiple velocity structures (Figure~\ref{cone_312}).
These characteristics probably indicate that molecular gas 
is also assembling in sub-regions of MC G044.0$-$02.5, 
supporting the idea of the simultaneous piling up of material in the whole cloud of 
$\sim$~100~pc and some localized regions of several pc.

We suggest that MC G044.0$-$02.5 forms from large-scale atomic flows,
which provide a natural mechanism to accumulate material in 
large-scale regions of $\sim$~100~pc. We find that bundles of tiny filaments 
or networks of fiber-like structures 
are ubiquitous in the MWISP CO maps.
These elongated structures can be naturally 
explained by turbulent flows, which will lead to gas concentration 
and subsequent fragmentation on scales of 
$\sim$~0.5~pc (i.e., at least five times of the beamwidth) to several pc.

We also note that the magnetic field from the $Planck$ 353~GHz 
data is roughly aligned with the long filament (Figure~\ref{plank_pol}).
This fact shows that the magnetic field may play an important role in
the formation of the molecular-gas structure, which needs to be investigated
by further observations and simulations 
\citep[e.g., see cases in][]{2017ApJ...846..122P,2019MNRAS.485.4509L}.

Taken as a whole, MCs may form from the pre-existing atomic gas, in which 
large-scale converging flows may be crucial for the accumulation of material.
When atomic gas is assembled by large-scale converging flows, 
collision between these filamentary flows becomes inevitable and will lead to
gas compression. Furthermore, in the environment of pressure enhancement by
colliding flows, MC formation is probably rapid due to dynamical processes
\citep[e.g.,][]{2000ApJ...532..980K,2004ApJ...612..921B,
2007ApJ...659.1317G,2010MNRAS.404....2G,2018ApJ...863..103S} and various instabilities
\citep[e.g.,][]{2012A&ARv..20...55H,2014prpl.conf....3D,2016SAAS...43...85K}.
This scenario indicates that MCs are likely evolving transient objects
from filamentary HI gas to H$_2$ gas and then MCs traced by
observed CO emission \citep[e.g.,][]{2007prpl.conf...81B,2008A&A...486L..43H}.

We find a good coincidence between the emergence of 
the overdensity of YSO candidates and the concentrations of 
enhanced C$^{18}$O emission (see Section 3.4.3 and Figure~\ref{cone_312}),
suggesting that star formation at their early evolutionary stages
occurs in the densest part of the forming cloud.
This is consistent with the suggestion that star formation is probably 
rapid within one or two crossing times 
in the localized dense regions of MCs 
(e.g., \citealp{2000ApJ...530..277E,2008ApJ...689..290H}; 
also see \citealp{2001ApJ...562..852H,2001MNRAS.327..663P,
2004ARA&A..42..211E,2007ApJ...668.1064E}).


\section{Summary}
We have analyzed molecular line observations of \twCO, \thCO, and 
C$^{18}$O ($J$=1--0) emission toward the Aquila Rift region 
by using the Delingha 13.7~m telescope. The whole CO mapping, 
which is obtained as part of the MWISP project, covered 
a $\sim$~250~deg$^2$ areas with a grid spacing of 30$''$ 
toward the Aquila Rift region. 
The sensitive molecular line data, together with their high spatial 
dynamic range mapping, allowed us to investigate the details of the local 
molecular gas traced by CO emission. 
Our main results are summarized below.

1. In the covered region of $l=$25\fdg8--49\fdg7 and $|b|\lsim$~5\fdg2,
CO emission exhibits large-scale extended structures for the local MCs within 
an LSR velocity range of $-5\km\ps \lsim $ V$_{\rm LSR} \lsim +25\km\ps$.
About half of the covered region has \twCO\ emission, while
C$^{18}$O emission is only located in the densest region, which 
is usually associated with star formation activities.
The \thCO\ emission, which is embedded in the extended \twCO\ layer,
is found to be a good tracer of the main structure of the local MCs.

2. Our new data have revealed many molecular-gas structures traced by
weak \twCO\ emission, which cannot be discerned by previous CO surveys
because of their relatively low resolution and sensitivity.
The dominant region of the local MCs is occupied by such the weak 
\twCO\ emission, with a low peak temperature of 
$\lsim$~4.5~K (i.e., $\sim$~7.8~K for excitation temperature), 
indicating subthermal excitation in the low-H$_2$-density environment.

\begin{figure}[ht]
\vspace{-3ex}
\gridline{
          \hspace{-11ex}
          \fig{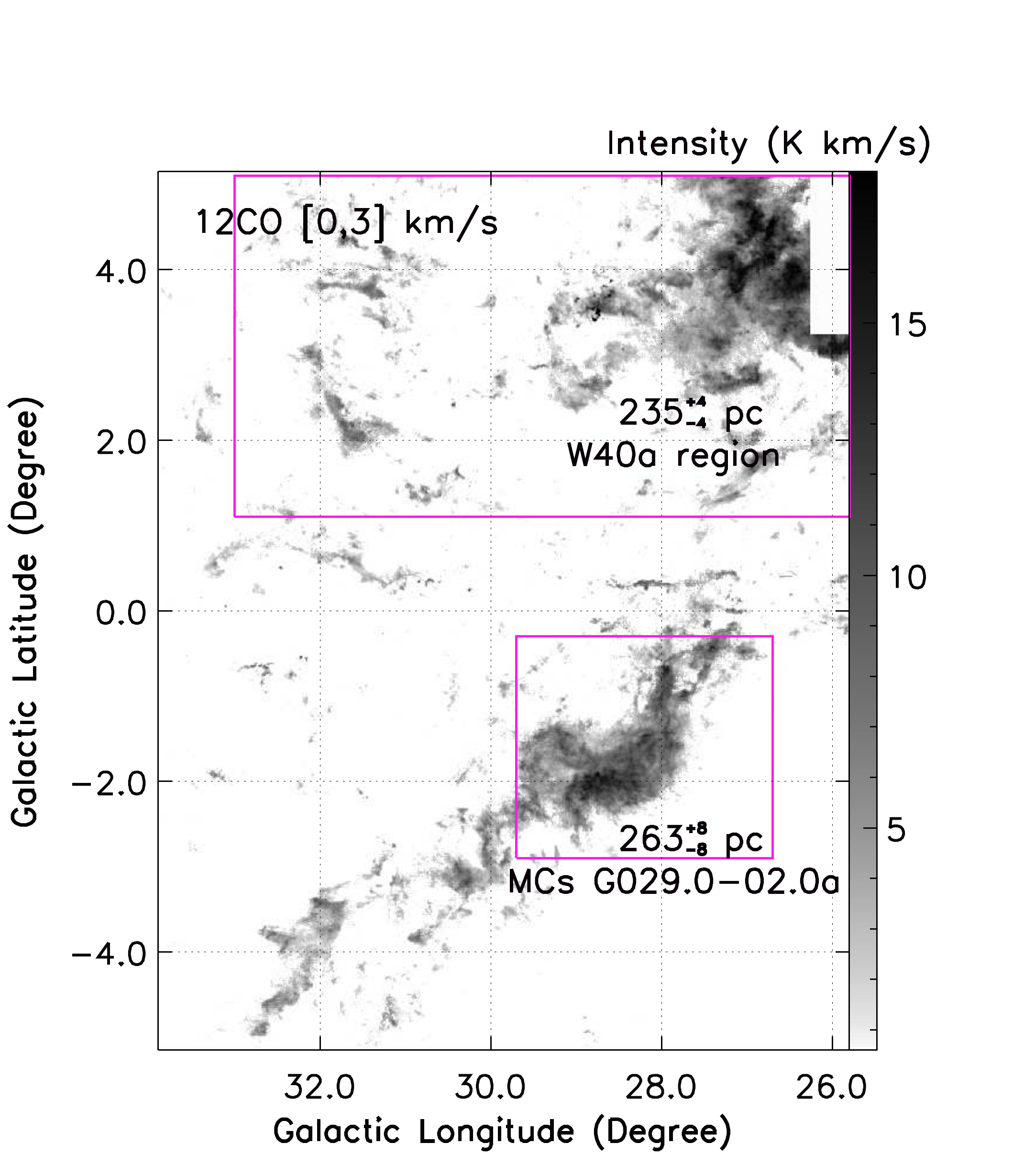}{0.45\textwidth}{}
          \hspace{-4ex}
          \fig{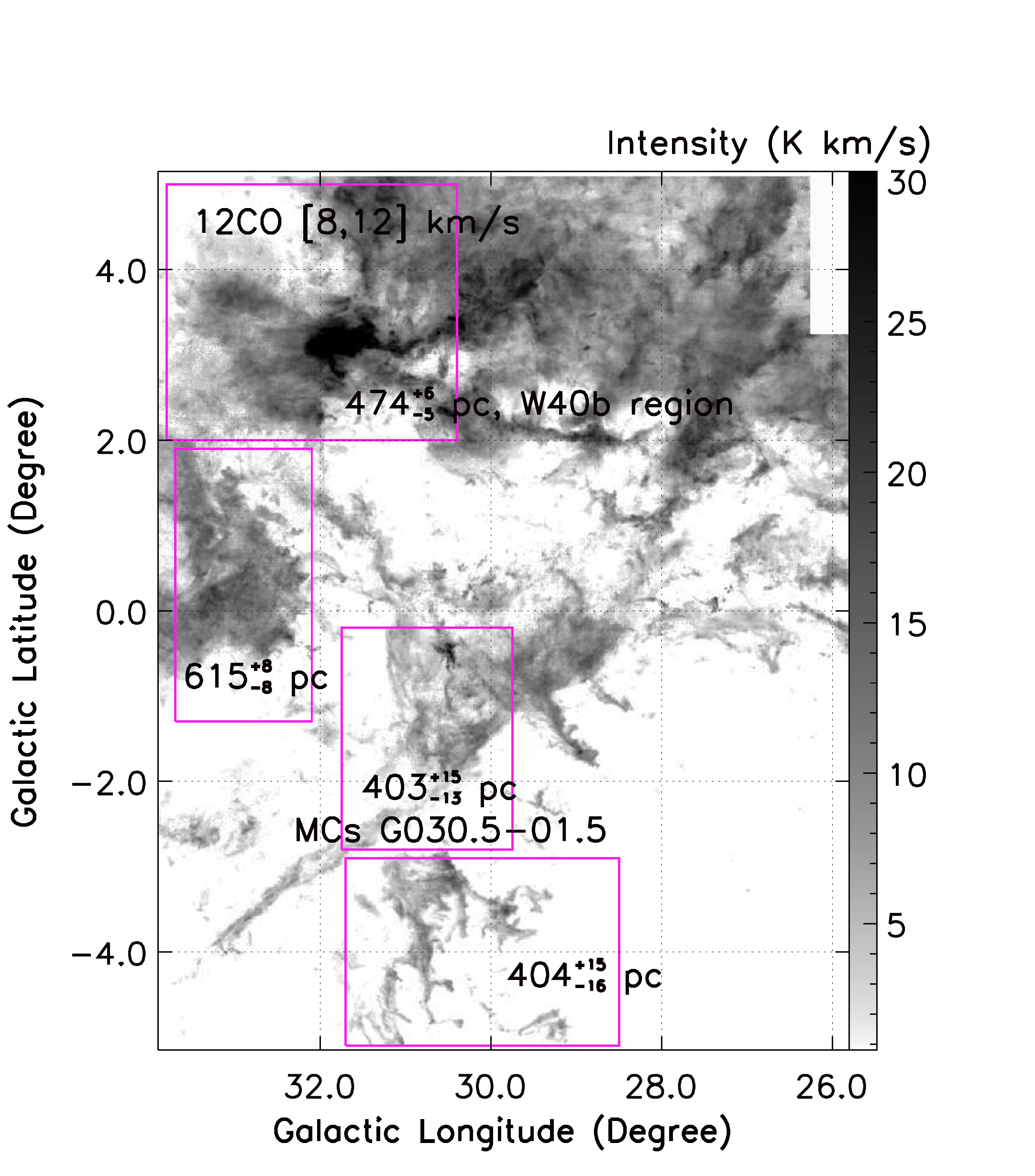}{0.45\textwidth}{}
          \hspace{-4ex}
          \fig{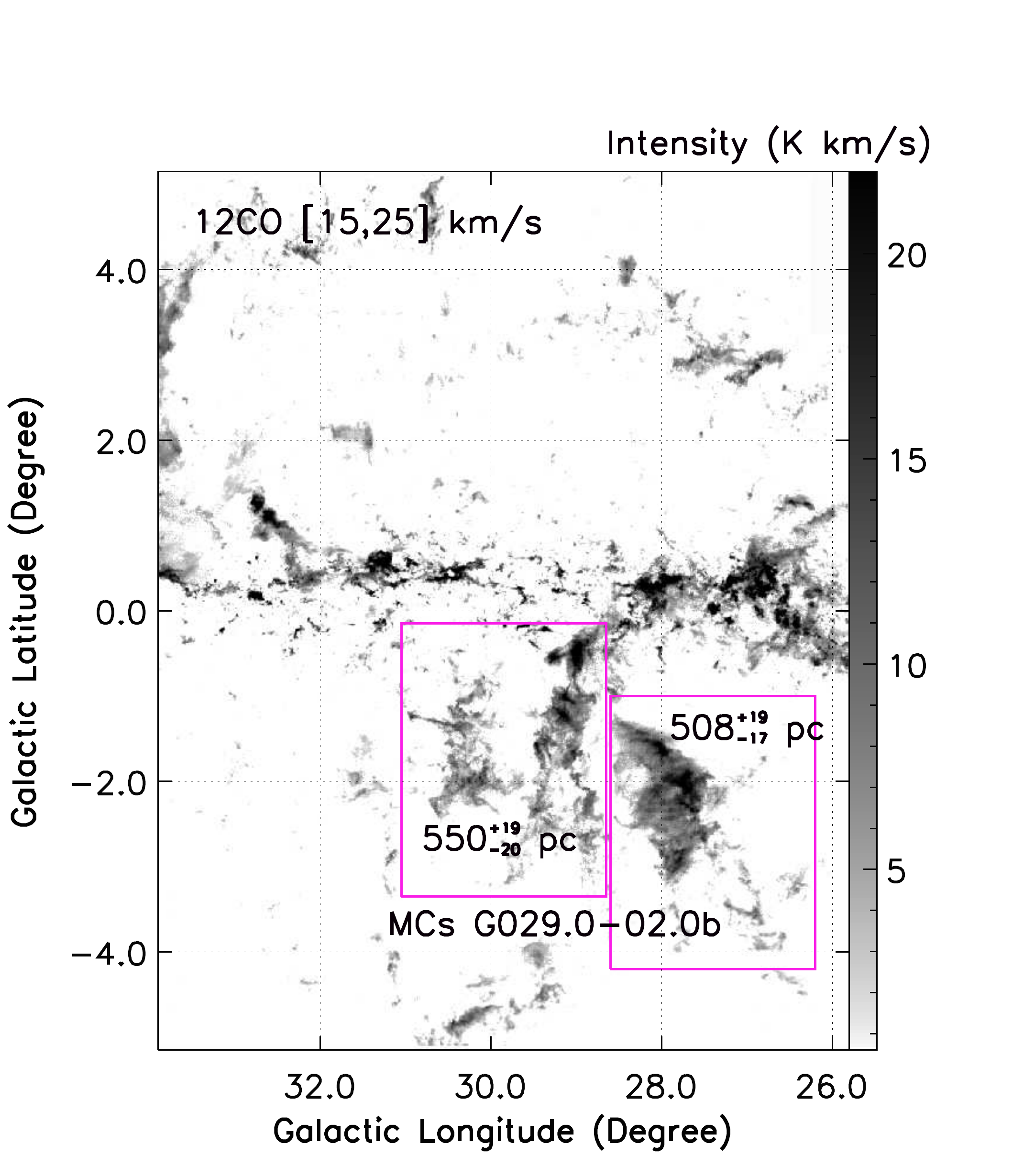}{0.45\textwidth}{}
          }
\vspace{-3ex}
\caption{
\twCO\ ($J$=1--0) intensity maps toward $l\sim +26^{\circ}$ to $+33^{\circ}$ 
in the [0, 3], [8, 12], and [15, 25]~km~s$^{-1}$, respectively. 
The distances of the corresponding molecular gas are labeled on the map.
Note that distances in the map are estimated for the areas
indicated by the rectangles but only using regions with high-integrated
CO intensity (see the threshold in the text).
\label{aqu_03}}
\end{figure}

\begin{figure}[ht]
\includegraphics[trim=0mm 5mm 0mm 30mm,scale=0.6,angle=0]{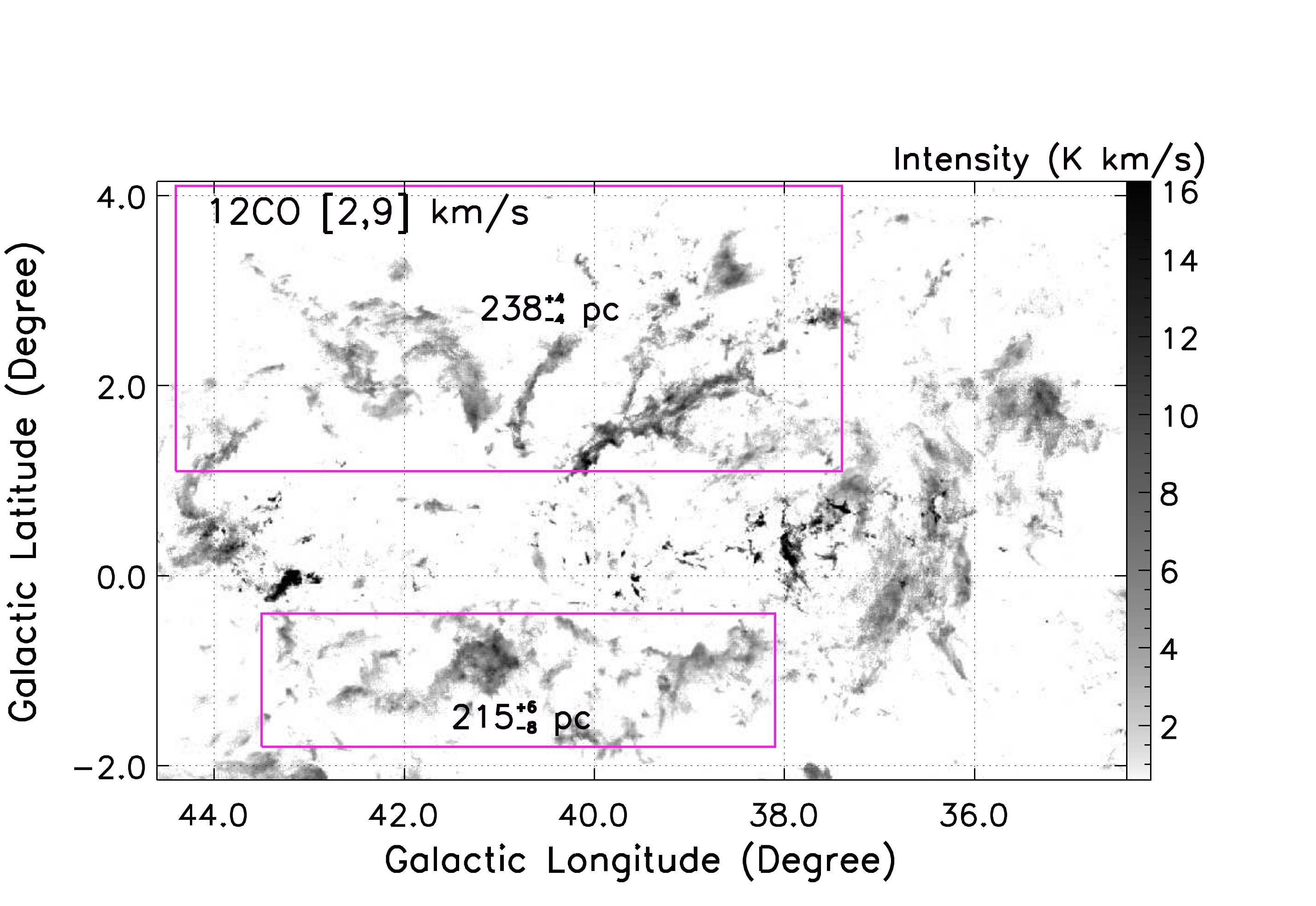}
\caption{
\twCO\ ($J$=1--0) intensity map toward $l\sim +35^{\circ}$ to $+44^{\circ}$
in the [2, 9]~km~s$^{-1}$ interval.
The distances of the corresponding molecular gas are labeled on the map.
Note that distances in the map are estimated for the areas
indicated by the rectangles but only using regions with high-integrated
CO intensity (see the threshold in the text).
The map also displays many shell-like structures with sizes of several pc.
\label{aqu_29}}
\end{figure}

\begin{figure}[ht]
\includegraphics[trim=50mm 0mm 0mm 0mm,scale=0.75,angle=0]{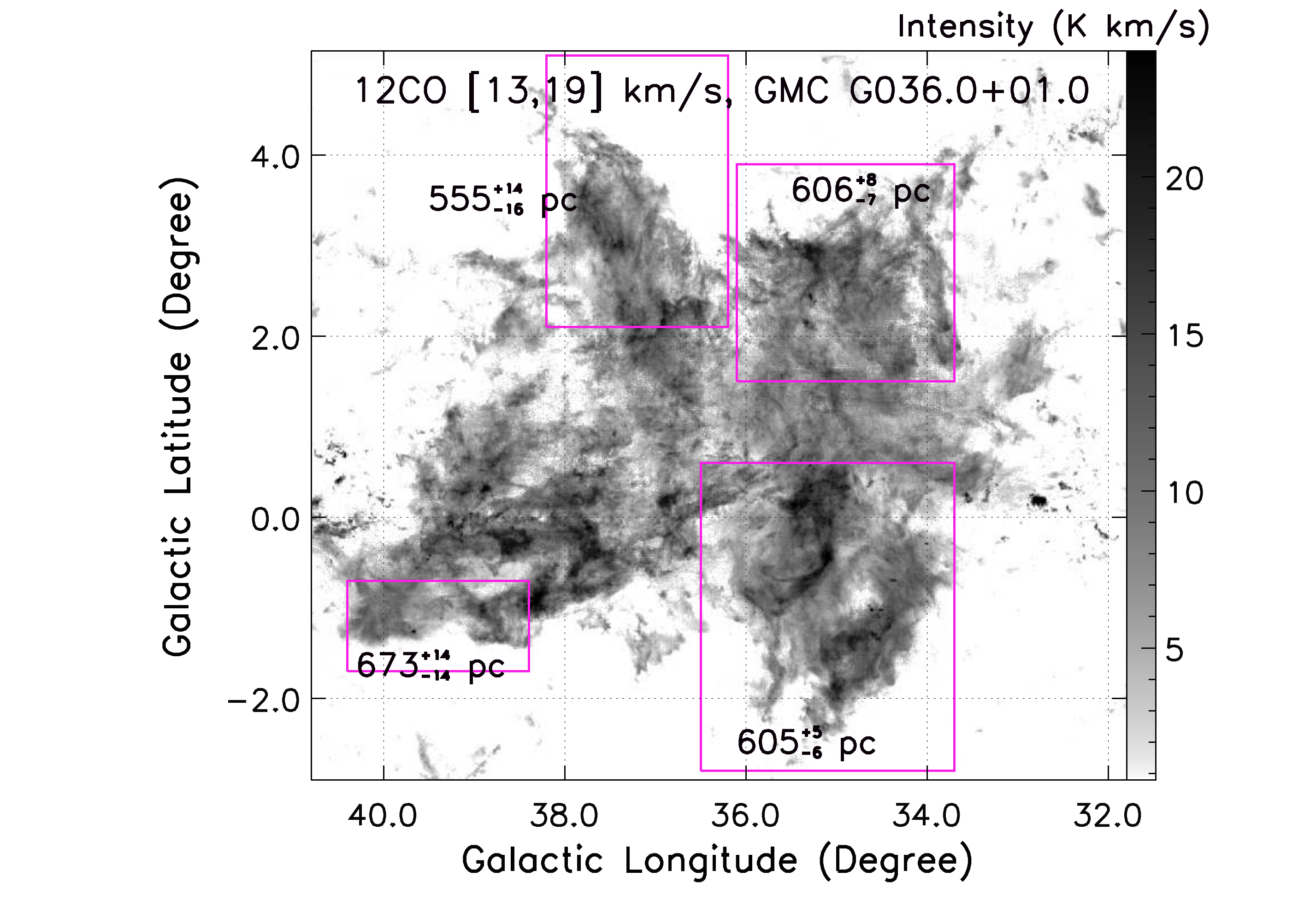}
\caption{
\twCO\ ($J$=1--0) intensity map toward $l\sim +32^{\circ}$ to $+41^{\circ}$
in the [13, 19]~km~s$^{-1}$ interval.
The distances of the corresponding molecular gas are labeled on the map.
Note that distances in the map are estimated for the areas
indicated by the rectangles but only using regions with high-integrated
CO intensity (see the threshold in the text).
\label{aqu_1319}}
\end{figure}

\begin{figure}[ht]
\includegraphics[trim=15mm 10mm 0mm 20mm,scale=0.65,angle=0]{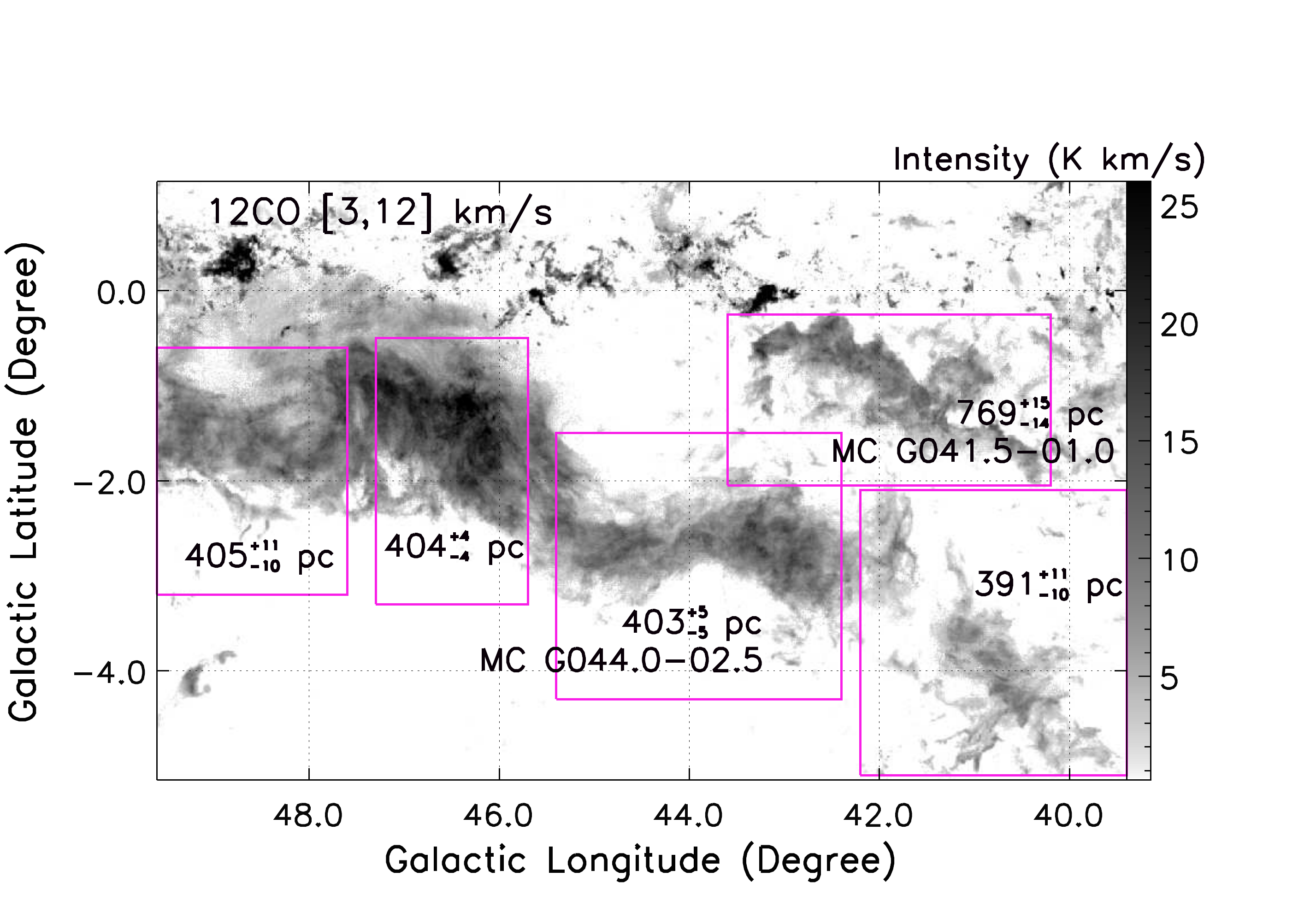}
\caption{
\twCO\ ($J$=1--0) intensity map toward $l\sim +40^{\circ}$ to $+50^{\circ}$
in the [3, 12]~km~s$^{-1}$ interval.
The distances of the corresponding molecular gas are labeled on the map.
Note that distances in the map are estimated for the areas
indicated by the rectangles but only using regions with high-integrated
CO intensity (see the threshold in the text).
\label{aqu_312}}
\end{figure}

\begin{figure}[ht]
\includegraphics[trim=20mm 0mm 0mm 0mm,scale=0.5,angle=0]{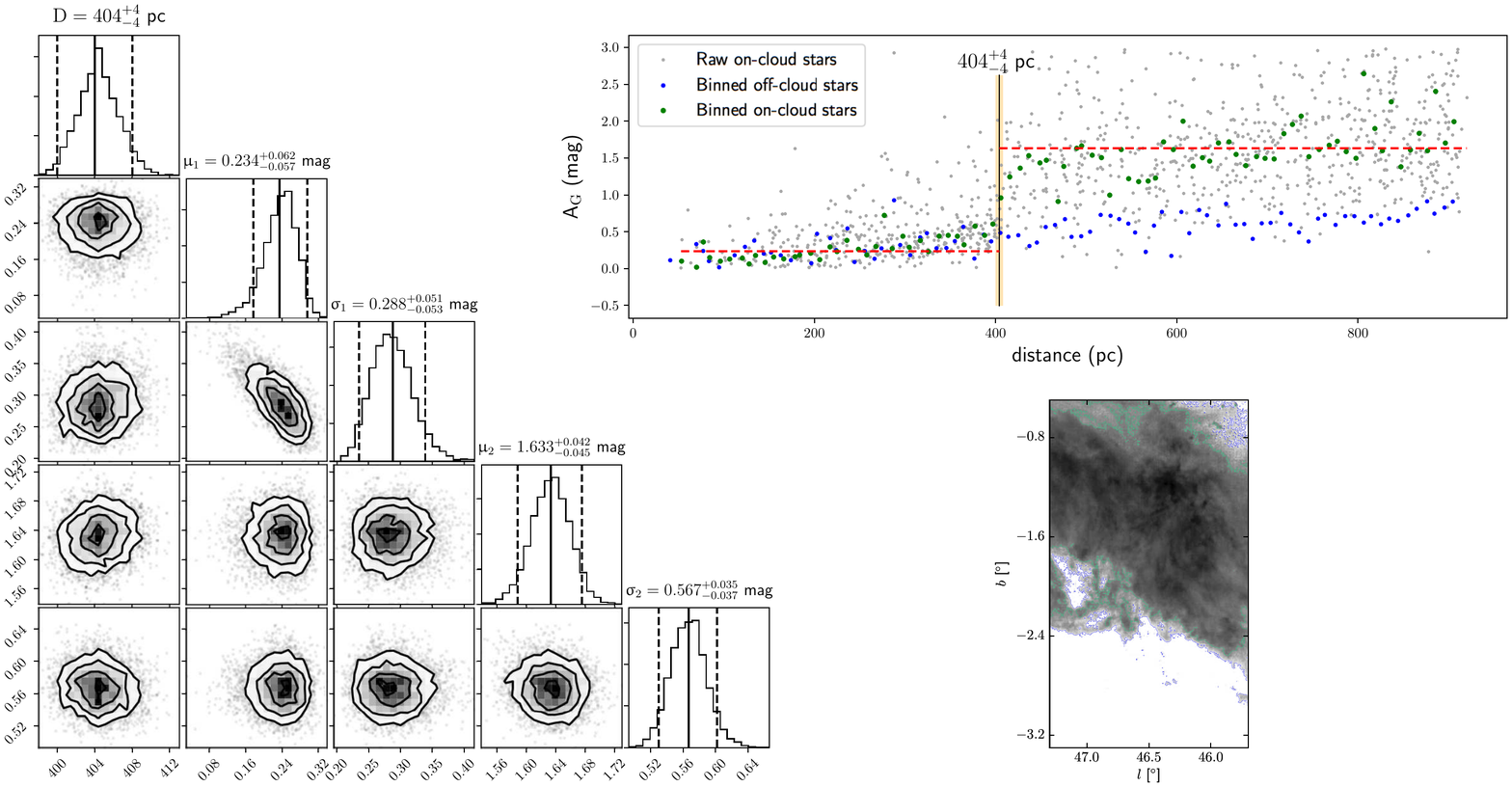}
\caption{
The estimated distance of a part of MC G044.0$-$02.5. 
The green and blue points are 
on- and off-cloud stars, respectively.
The dashed red lines are the modeled extinction $A_G$. 
The vertical line indicates the jump in the distance--$A_{\rm G}$ space.
The distance is estimated from the raw on-cloud Gaia DR2 (gray points). 
The shadow area in the upper right panel is the 95\% HPD distance range. 
The relationships between 
the fitting parameters are shown in the left panel. The integrated CO emission is
shown in the right corner plot. Please see the details in \cite{2019A&A...624A...6Y}.
\label{cone_dis}}
\end{figure}

\begin{figure}[ht]
\includegraphics[trim=20mm 0mm 0mm 0mm,scale=0.6,angle=0]{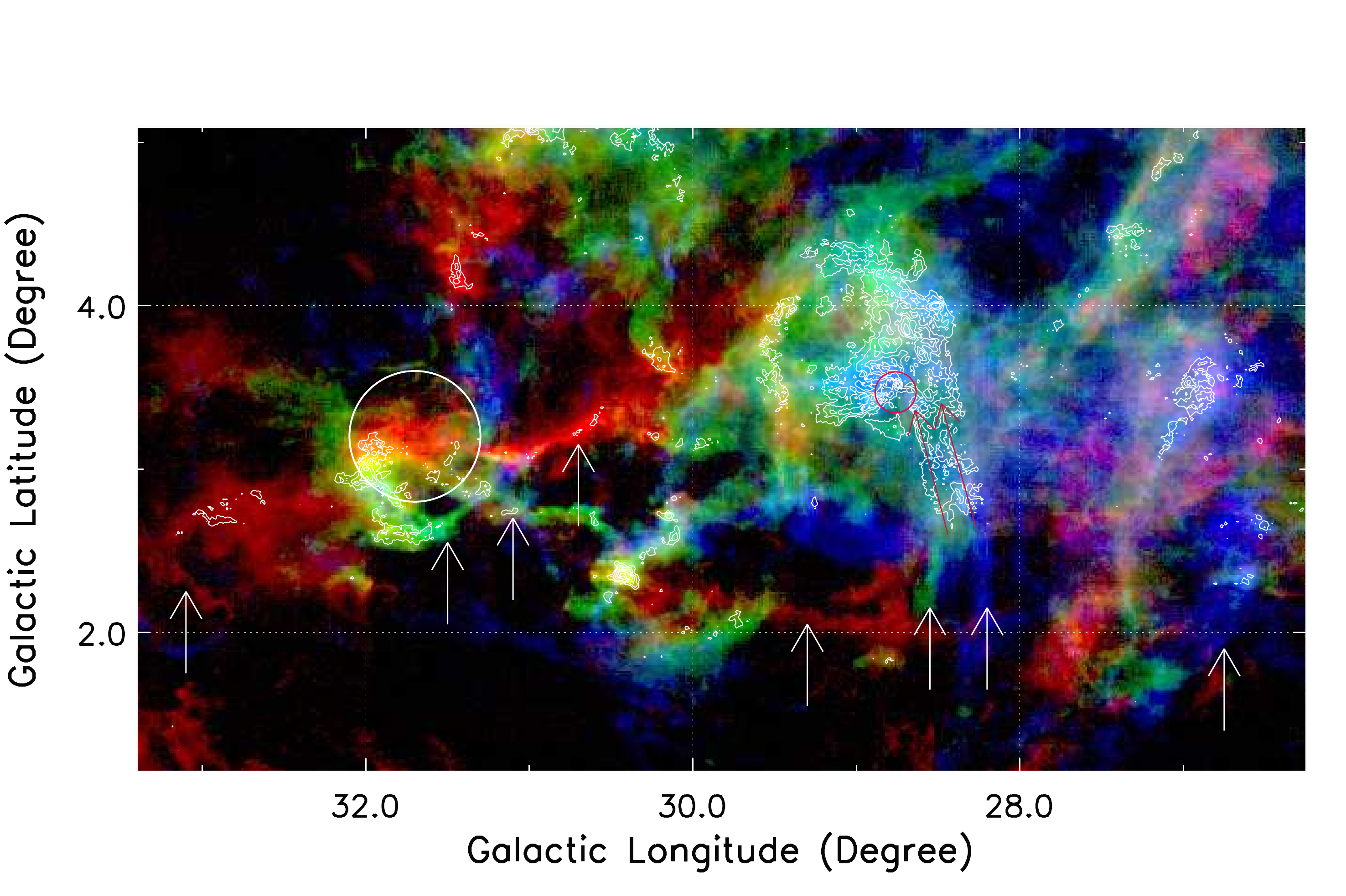}
\caption{
Integrated \thCO\ ($J$=1--0) emission in the interval of [4.5, 7.0]~km~s$^{-1}$ (blue),
[7.0, 9.5]~km~s$^{-1}$ (green), and [9.5, 12.0]~km~s$^{-1}$ (red) toward the
\HII\ region W40 (the red circle at $l=$28\fdg8 and $b=$3\fdg5 with a diameter of 0\fdg2), 
overlaid with the C$^{18}$O ($J$=1--0) integrated emission contours
of 0.5, 1.5, 2.5, 3.5, and 4.5~K~km~s$^{-1}$.
The white circle at $l=$31\fdg7 and $b=$3\fdg2 with a diameter of 0\fdg8
shows the region with higher \twCO\ peak temperatures (see Figure~\ref{t1213}).
The star formation activities (e.g., the Serpens NE cluster and \HII\ region W40) are 
responsible for the relatively high CO temperature in such regions.
The white arrows show some typical filamentary structures seen in \thCO\
emission. The red arrows indicate the direction of PV diagrams 
shown in Figure~\ref{W40pv}.
\label{W40_45115_rgb}}
\end{figure}

\begin{figure}
\gridline{
          \hspace{-5ex}
          \fig{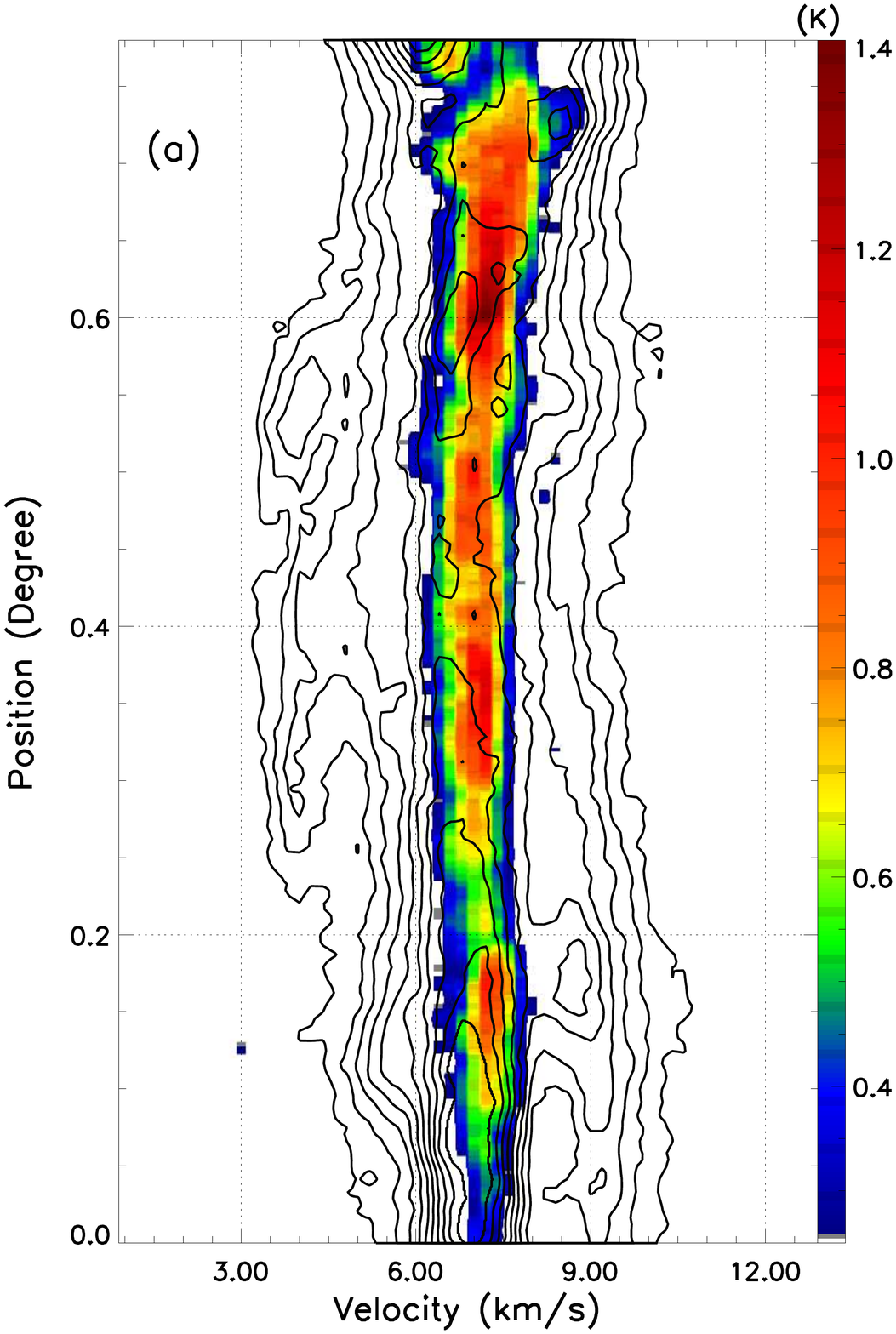}{0.5\textwidth}{}
          \hspace{-5ex}
          \fig{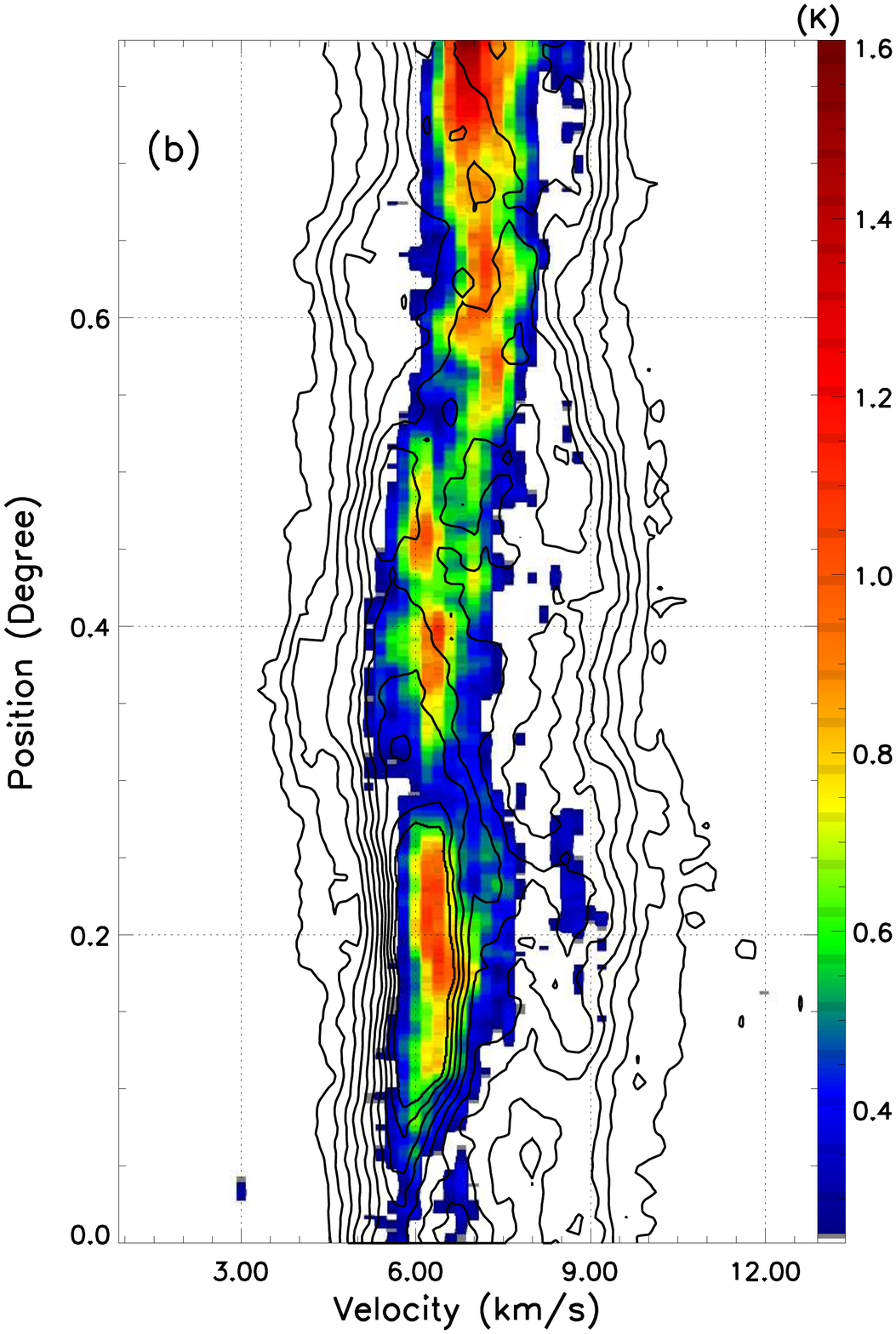}{0.5\textwidth}{}
          }
\caption{
Position--velocity diagrams of the C$^{18}$O (color) and \thCO\ (contours) emission 
of two filamentary structures labeled with red arrows in Figure~\ref{W40_45115_rgb},
i.e., panel (a) from ($l=$28\fdg44, $b=$2\fdg60) to ($l=$28\fdg64, $b=$3\fdg36);
and panel (b) from ($l=$28\fdg27, $b=$2\fdg65) to ($l=$28\fdg48, $b=$3\fdg40).
The two slices have a length of 0\fdg78 and a width of 0\fdg14.
The contours of the \thCO\ emission start from 0.3~K with a step of 0.3~K.
\label{W40pv}}
\end{figure}

\begin{figure}
\gridline{
\hspace{-10ex}
          \fig{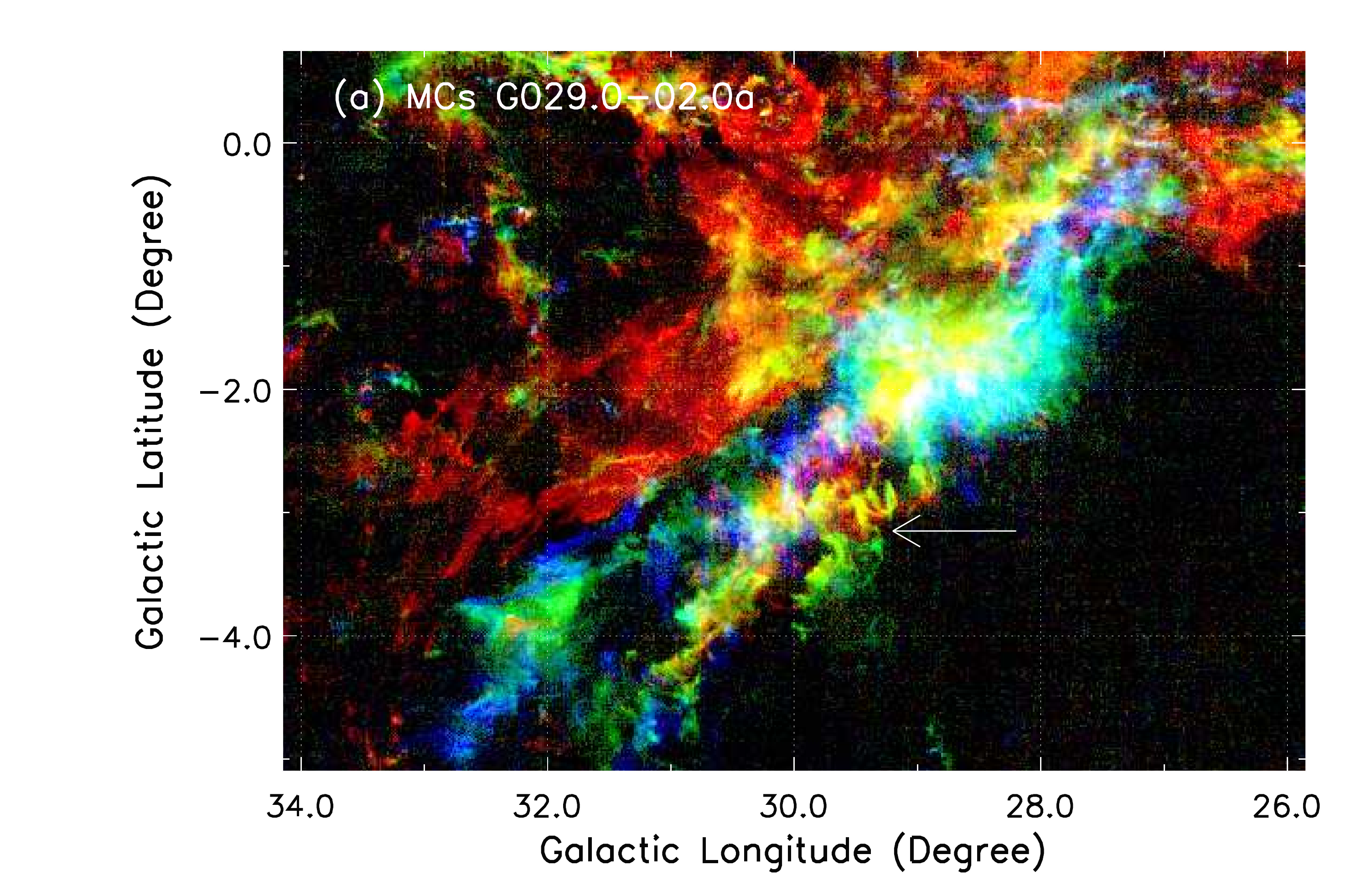}{1.\textwidth}{}
          }
\vspace{-10ex}
\gridline{
\hspace{-10ex}
          \fig{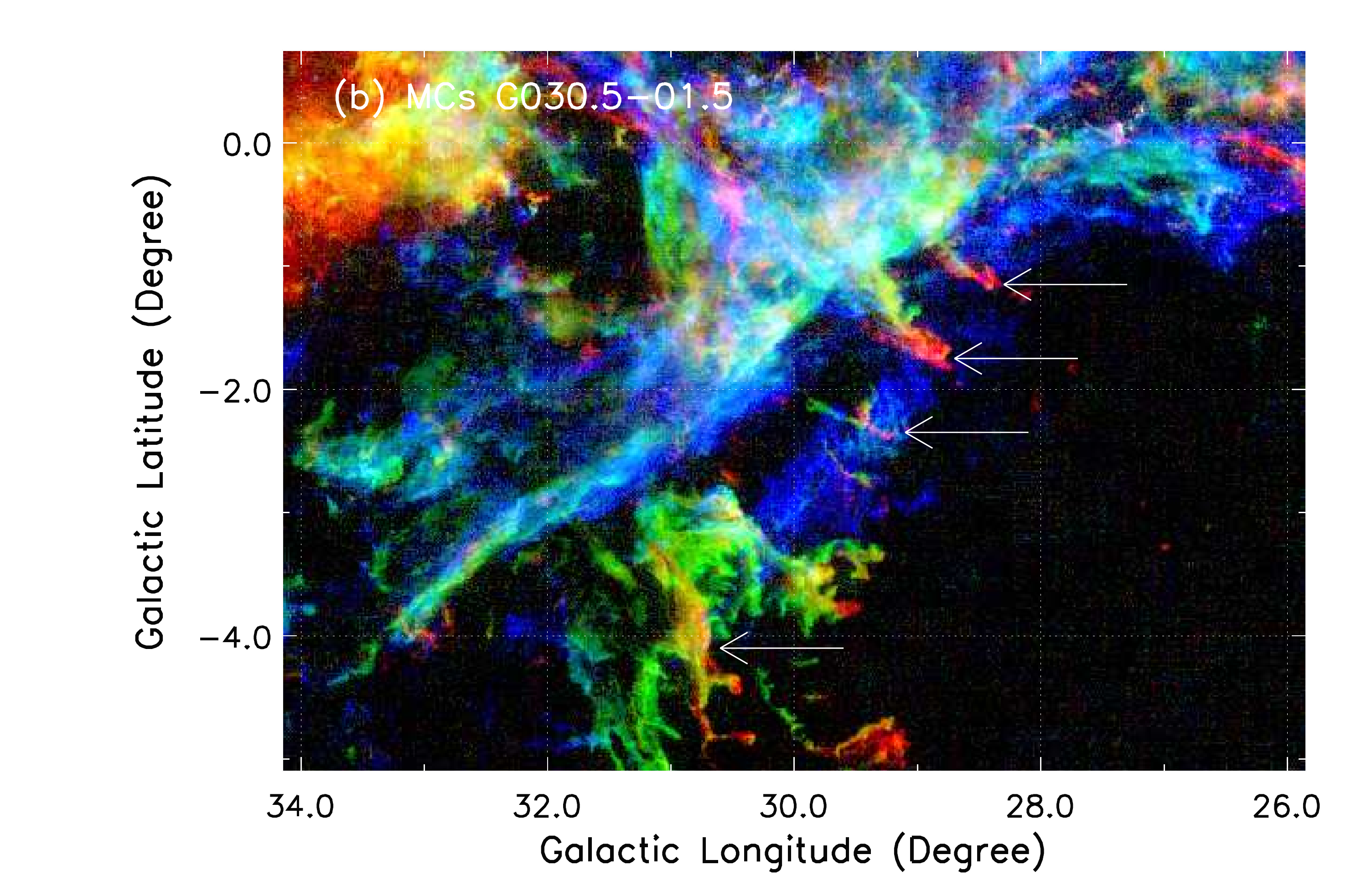}{1.\textwidth}{}
          }
\vspace{-6ex}
\caption{
Panel (a): integrated \twCO\ ($J$=1--0) emission in the interval of [1.0, 3.0]~km~s$^{-1}$ (blue),
[3.0, 5.0]~km~s$^{-1}$ (green), and [5.0, 7.0]~km~s$^{-1}$ (red) toward
MCs G029.0$-$02.0a at a distance of $\sim$~260~pc.
Panel (b): integrated \twCO\ ($J$=1--0) emission in the interval of [6.0, 8.0]~km~s$^{-1}$ (blue),
[8.0, 10.0]~km~s$^{-1}$ (green), and [10.0, 12.0]~km~s$^{-1}$ (red) toward
MCs G030.5$-$01.5 at the distance of $\sim$~400~pc.
The white arrows indicate protruding structures perpendicular to the elongated
molecular gas.
\label{aqu_17_612_rgb}}
\end{figure}

\begin{figure}
\vspace{-3ex}
\gridline{
          \hspace{-10ex}
          \fig{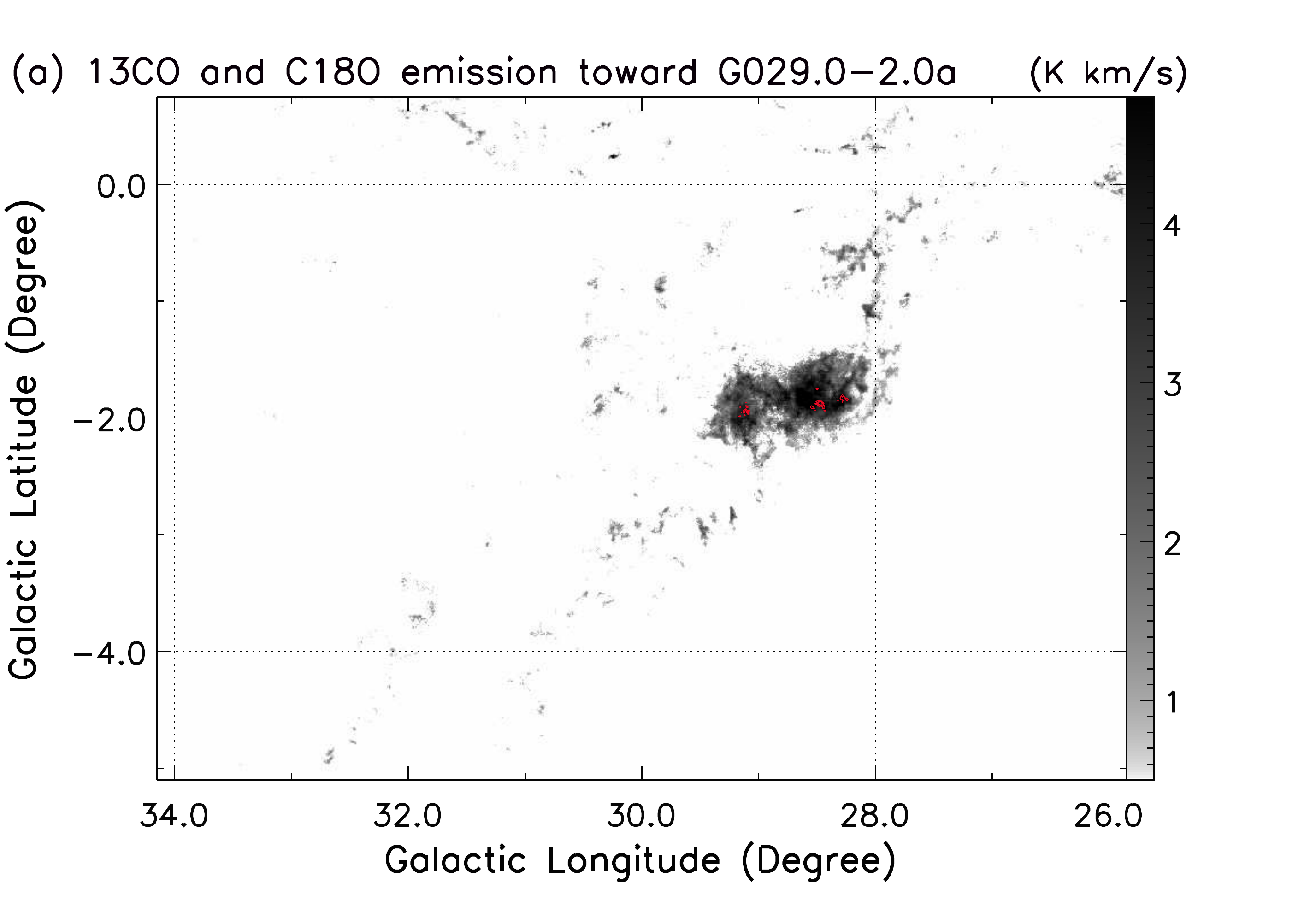}{0.6\textwidth}{}
          \hspace{-4ex}
          \fig{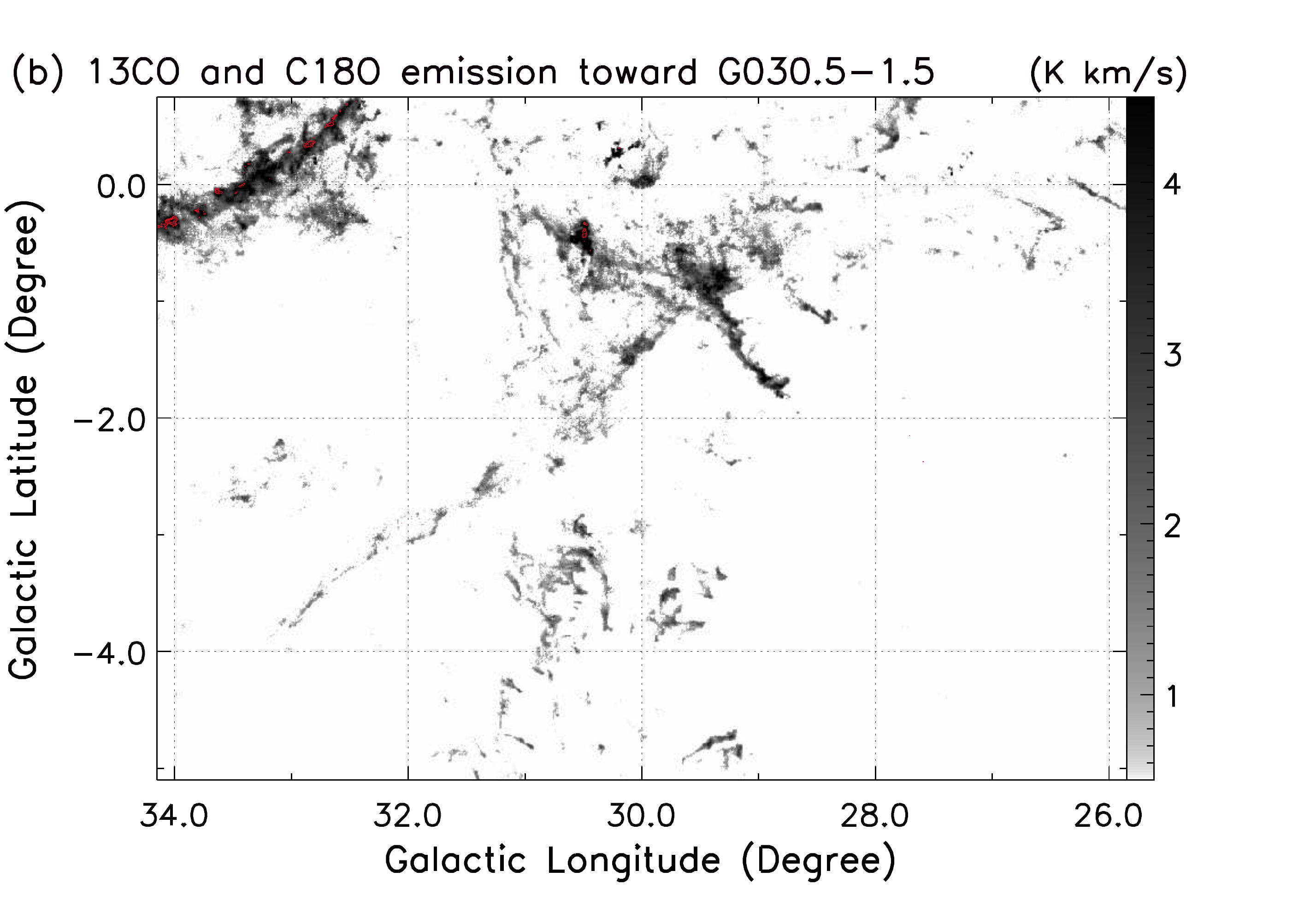}{0.6\textwidth}{}
          }
\vspace{-7ex}
\gridline{
          \hspace{-10ex}
          \fig{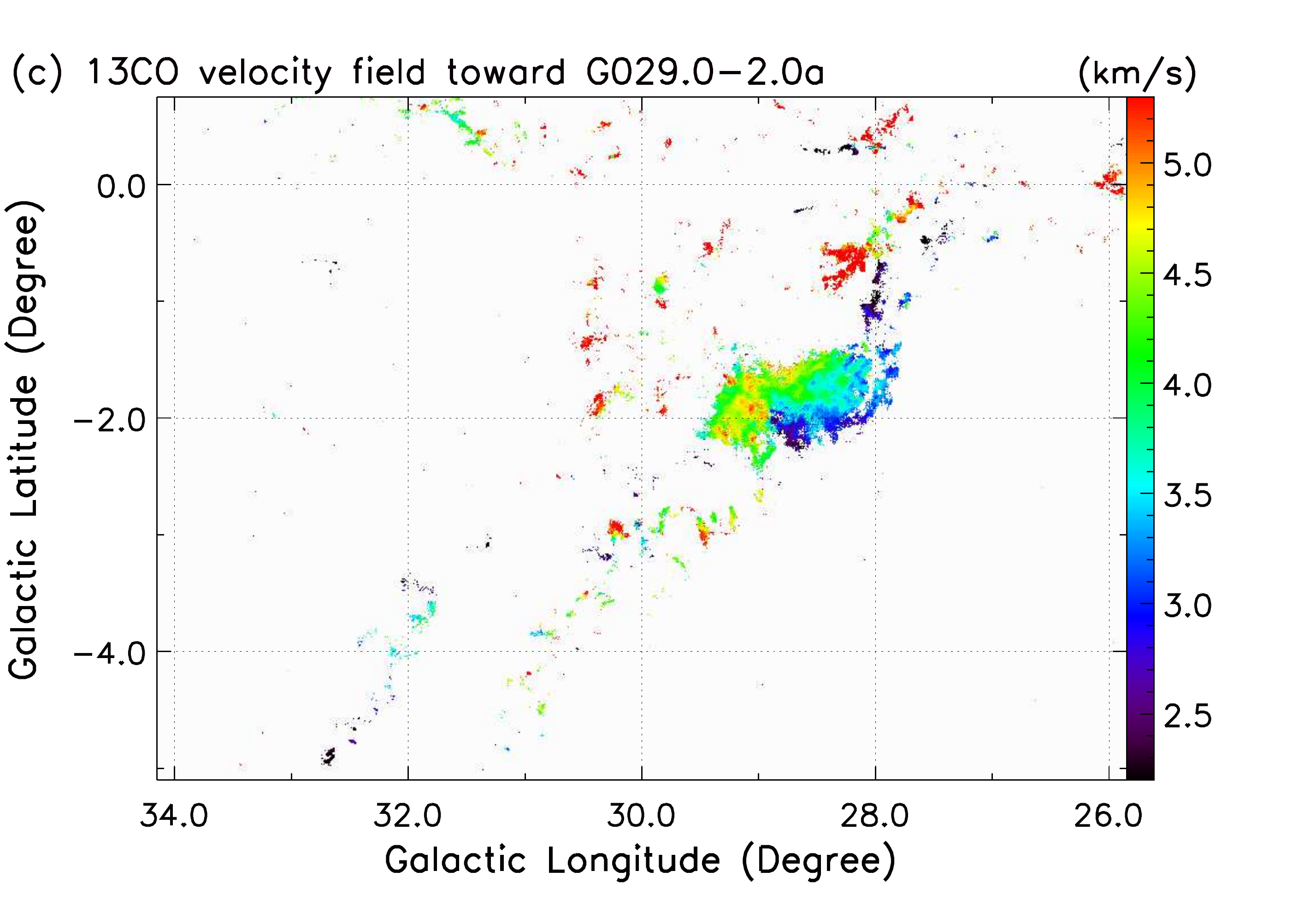}{0.6\textwidth}{}
          \hspace{-4ex}
          \fig{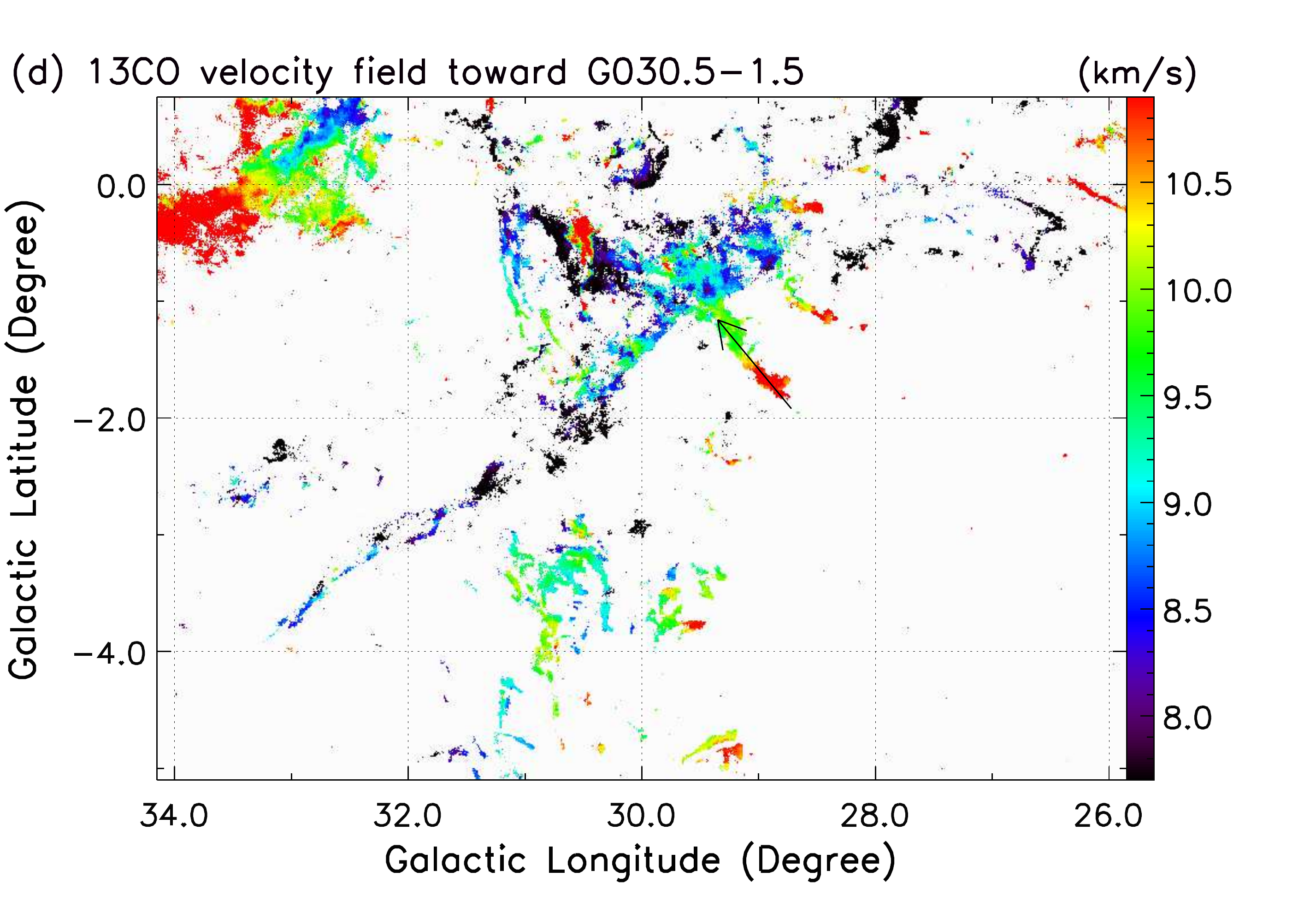}{0.6\textwidth}{}
          }
\vspace{-5ex}
\caption{
Panel (a): integrated \thCO\ ($J$=1--0) emission (gray) and C$^{18}$O ($J$=1--0) emission
(red contours,  0.4, 0.8, and 1.2~K~km~s$^{-1}$) toward MCs G029.0$-$02.0a 
in the velocity range of [1.0, 7.0]~km~s$^{-1}$.
Panel (b): same as panel (a) but for MCs G030.5$-$01.5 in the velocity range 
of [6.0, 12.0]~km~s$^{-1}$.
Panel (c): intensity-weighted \thCO\ mean velocity (first moment) map of MCs G029.0$-$02.0a.
Panel (d): same as panel (c) but for MCs G030.5$-$01.5. 
The black arrow indicates the direction of the PV diagram shown in
Figure~\ref{fingerpv}.
\label{aqu_17_612}}
\end{figure}

\begin{figure}[ht]
\includegraphics[trim=0mm 0mm 0mm 0mm,scale=0.4,angle=0]{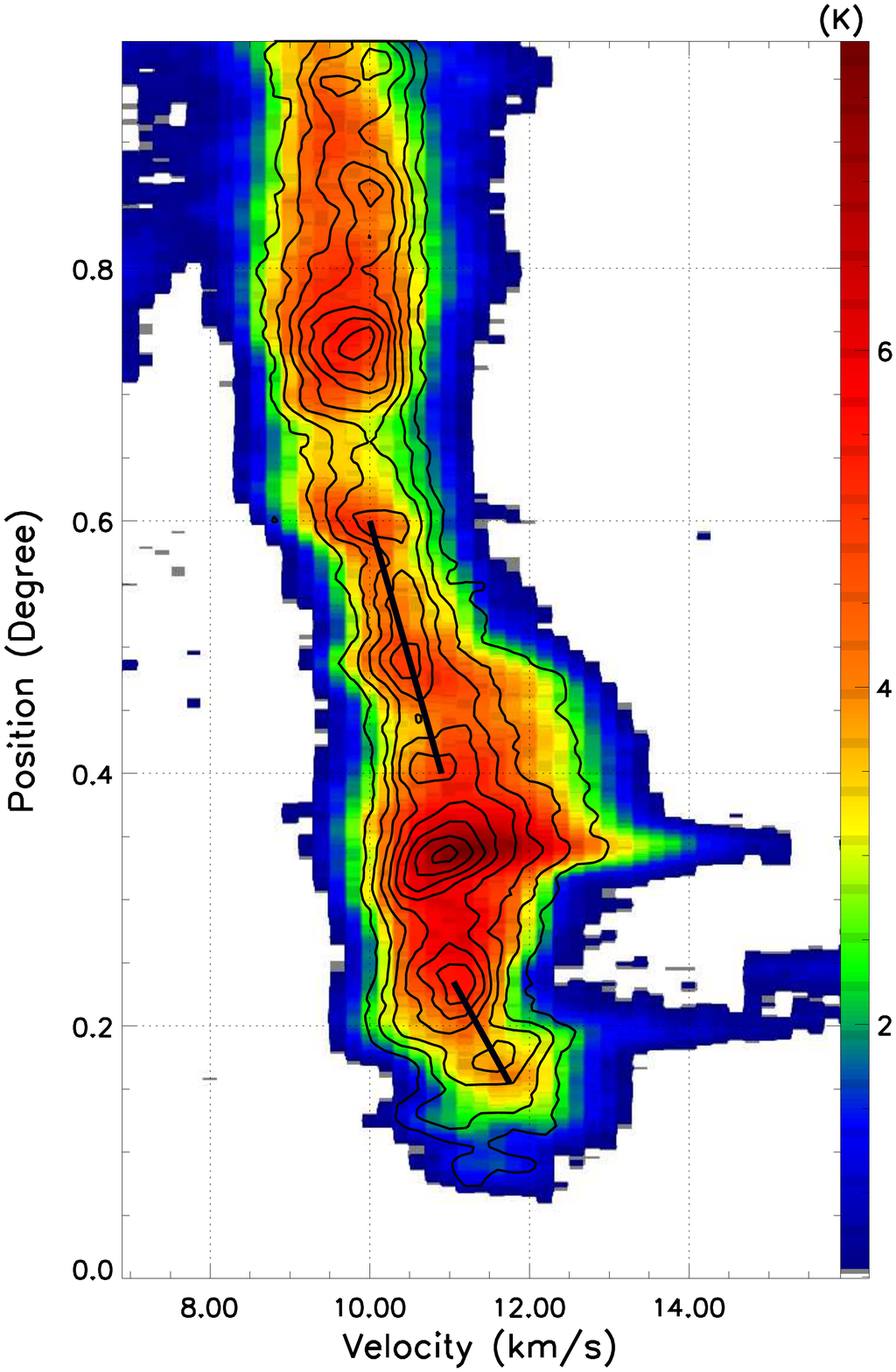}
\caption{
Position--velocity diagram of the \twCO\ (color) and \thCO\ (contours) emission
of the protruding structure labeled with the black arrow in Figure~\ref{aqu_17_612},
i.e., from ($l=$28\fdg72, $b=-$1\fdg92) to ($l=$29\fdg35, $b=-$1\fdg36).
The slice has a length of $\sim 1^{\circ}$ and a width of 0\fdg14 
from the \thCO\ emission.
The contours of \thCO\ emission start from 0.3~K with a step of 0.3~K.
The black lines indicate the velocity gradient of 
$\sim0.7-1.3\km\ps$pc$^{-1}$ for the CO gas at the distance of $\sim 400$~pc.
\label{fingerpv}}
\end{figure}

\begin{figure}[ht]
\includegraphics[trim=20mm 0mm 0mm 0mm,scale=0.6,angle=0]{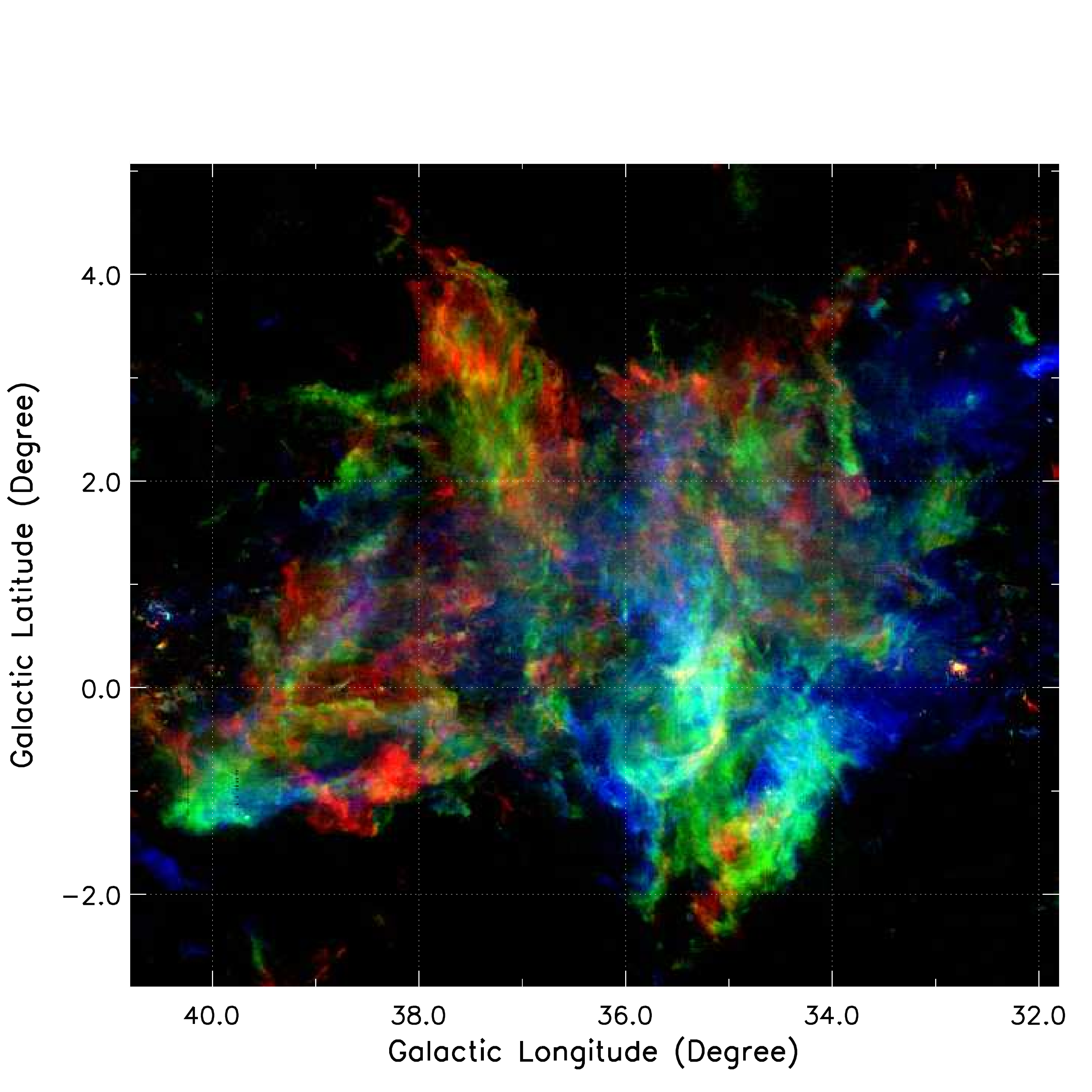}
\caption{
Integrated \twCO\ ($J$=1--0) emission in the interval of [11.0, 13.0]~km~s$^{-1}$ (blue),
[13.0, 15.0]~km~s$^{-1}$ (green), and [15.0, 17.0]~km~s$^{-1}$ (red) toward 
the GMC complex G036.0$+$01.0 at the distance of $\sim$~560--670~pc.
\label{b_1117_rgb}}
\end{figure}

\begin{figure}
\vspace{-17ex}
\gridline{
          \fig{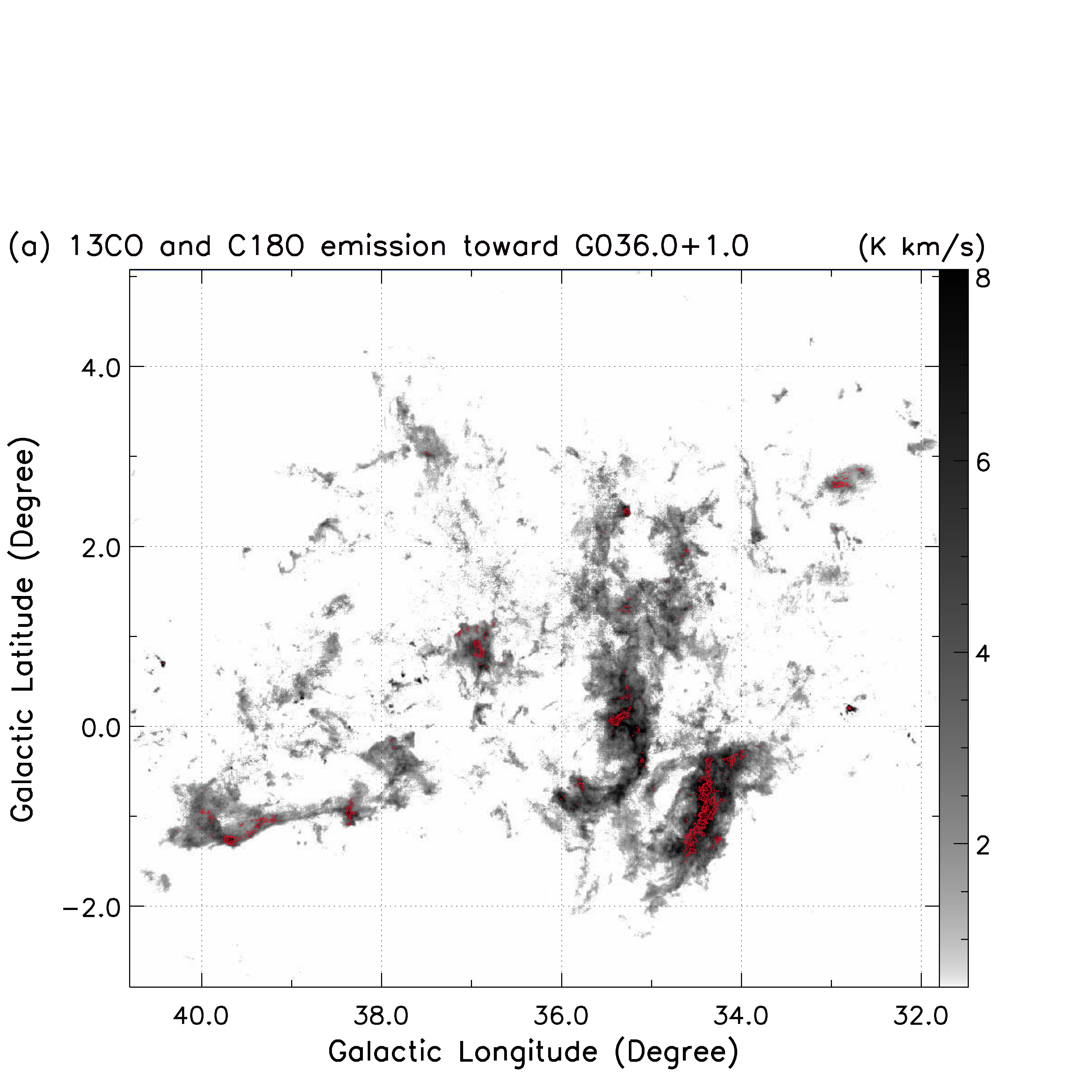}{0.9\textwidth}{}
          }
\vspace{-22ex}
\gridline{
          \fig{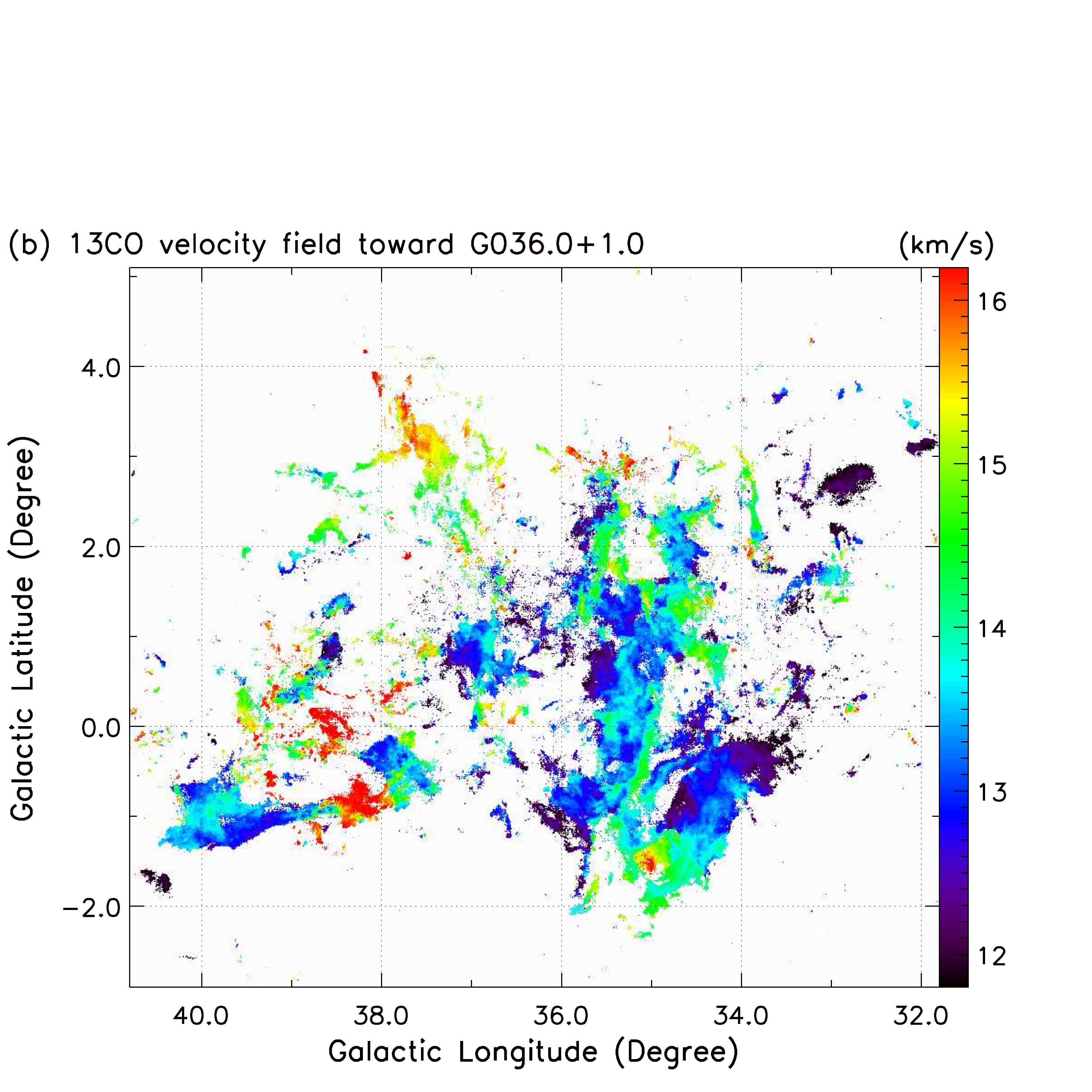}{0.9\textwidth}{}
          }
\vspace{-5ex}
\caption{
Panel (a): integrated \thCO\ ($J$=1--0) emission (gray) and C$^{18}$O ($J$=1--0) emission
(red contours, 0.4, 0.8, 1.2, 1.6, and 2.0~K~km~s$^{-1}$) toward the GMC complex 
G036.0$+$01.0 in the velocity interval of [11.5, 16.5]~km~s$^{-1}$.
Panel (b): intensity-weighted \thCO\ mean velocity (first moment) map of
the GMC complex G036.0$+$01.0 in the same velocity range.
\label{b_1117}}
\end{figure}

\begin{figure}[ht]
\includegraphics[trim=0mm 0mm 0mm 0mm,scale=0.4,angle=0]{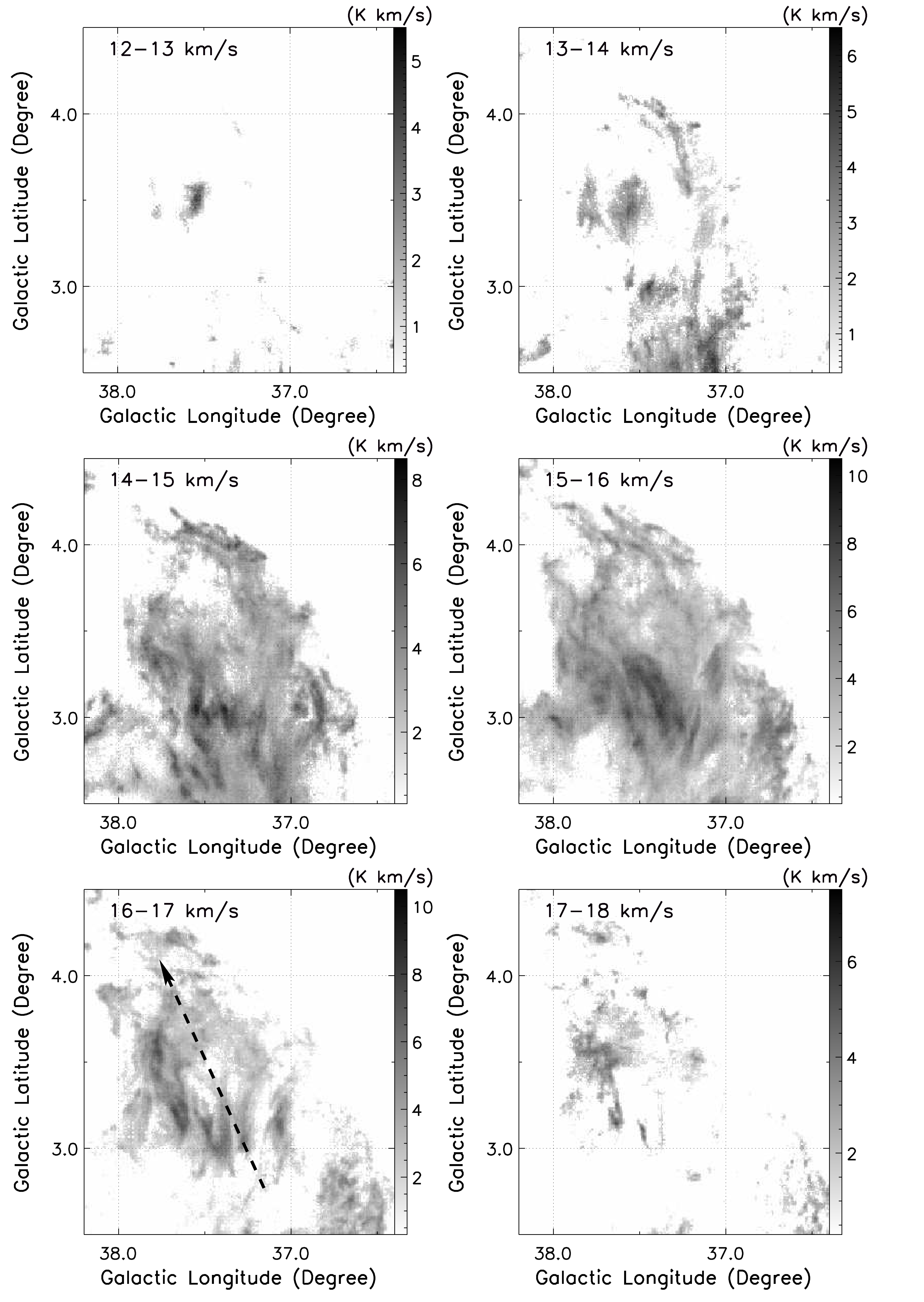}
\caption{
Channel maps of \twCO\ emission toward the northeastern part of 
the GMC complex G036.0$+$01.0. The arrow indicates the direction
of the large-scale elongated CO structure. 
\label{b12channel}}
\end{figure}

\begin{figure}[ht]
\includegraphics[trim=0mm 0mm 0mm 0mm,scale=0.4,angle=0]{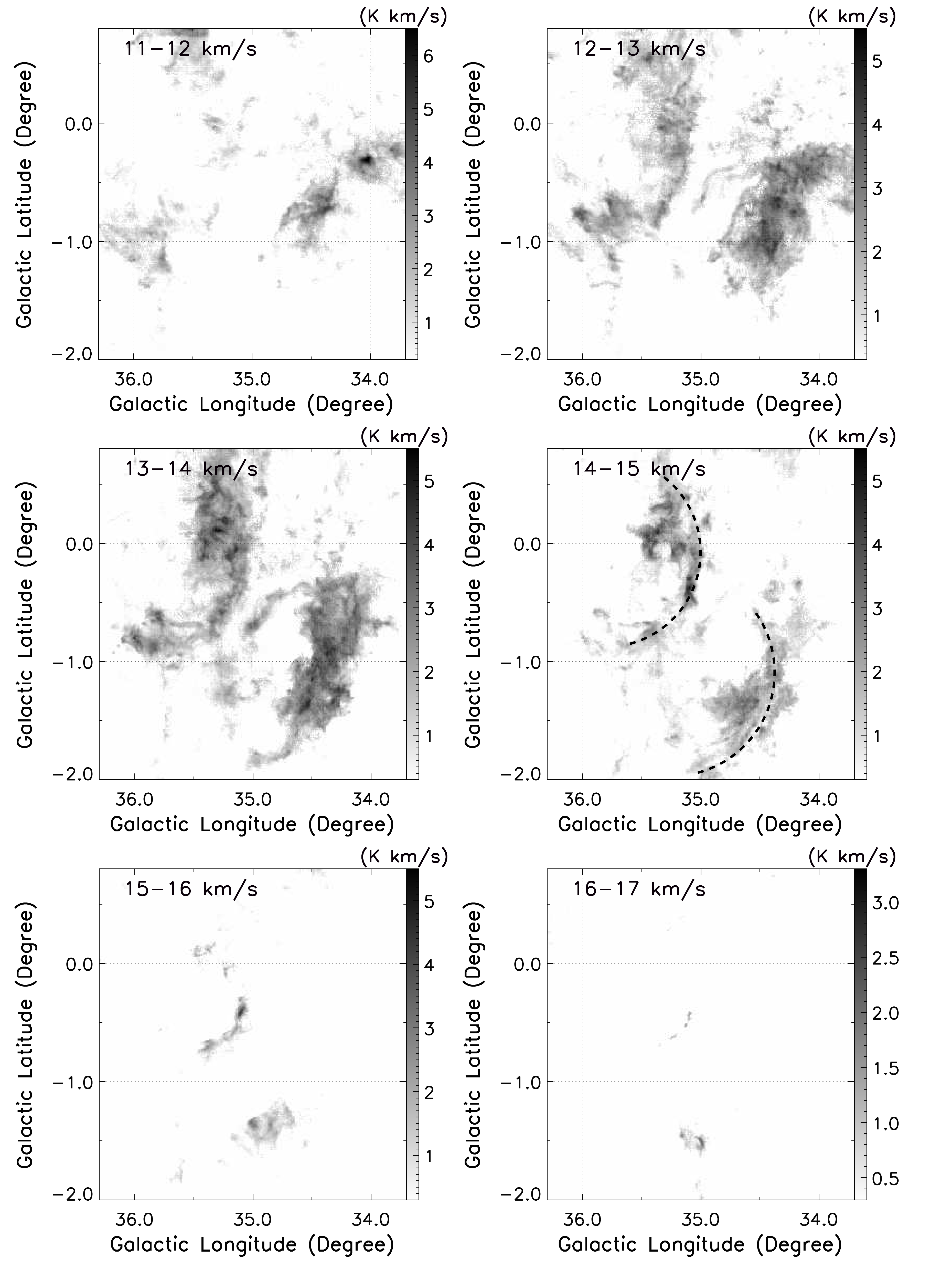}
\caption{
Channel maps of \thCO\ emission toward the southwestern part of
the GMC complex G036.0$+$01.0. The dashed lines indicate the large-scale
arc-like structures traced by \thCO\ emission.
\label{b13channel}}
\end{figure}

\begin{figure}[ht]
\includegraphics[trim=20mm 0mm 0mm 0mm,scale=0.5,angle=0]{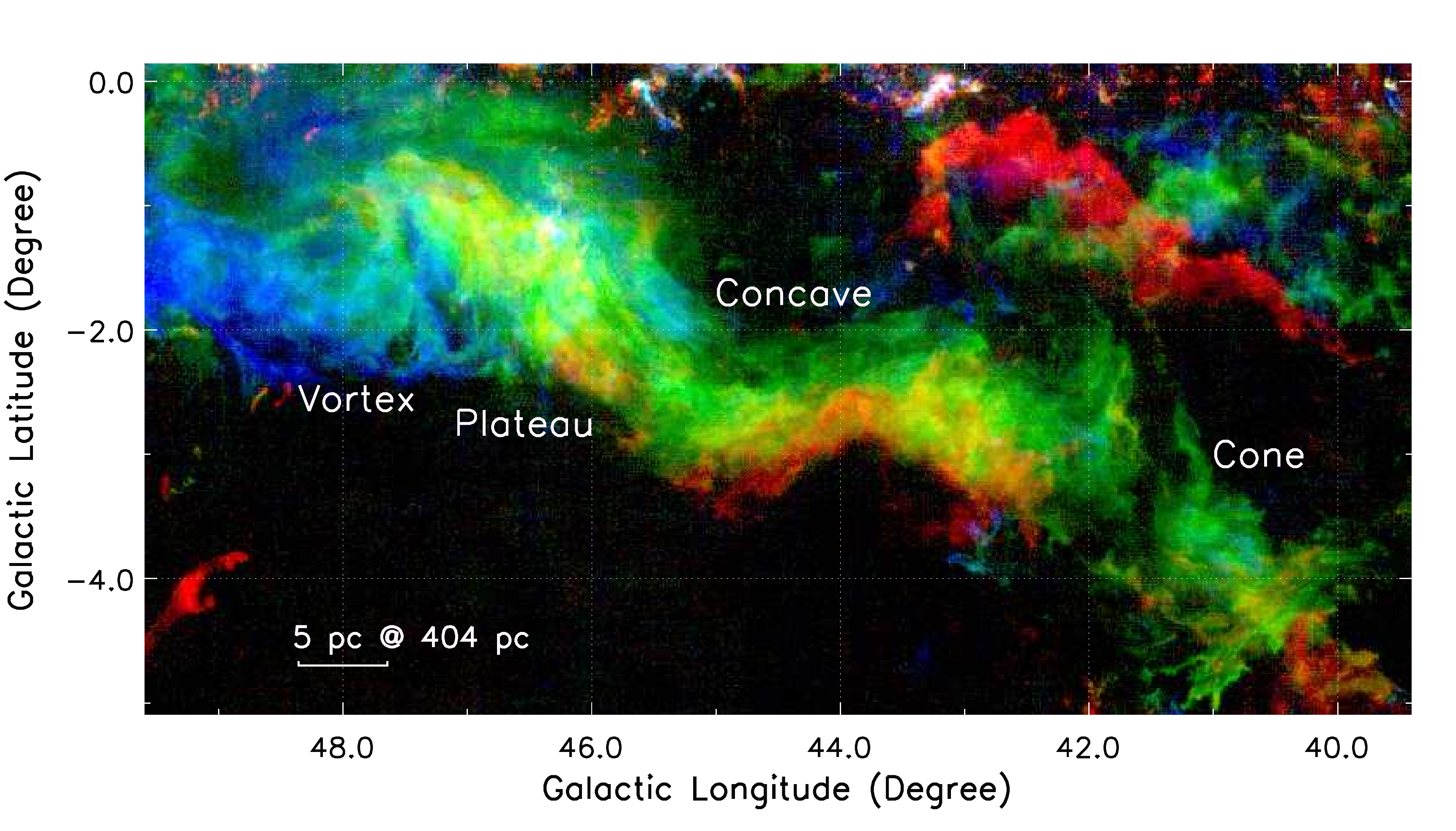}
\caption{
Integrated \twCO\ ($J$=1--0) emission in the interval of [3.0, 6.0]~km~s$^{-1}$ (blue),
[6.0, 9.0]~km~s$^{-1}$ (green), and [9.0, 12.0]~km~s$^{-1}$ (red) toward
MC G044.0$-$02.5 at the distance of $\sim$~404~pc.
The four sub-structures of the long filamentary MC are named as
Vortex, Plateau, Concave, and Cone, respectively.
\label{aqu_312_rgb}}
\end{figure}

\begin{figure}[ht]
\includegraphics[trim=20mm 0mm 0mm 0mm,scale=0.45,angle=0]{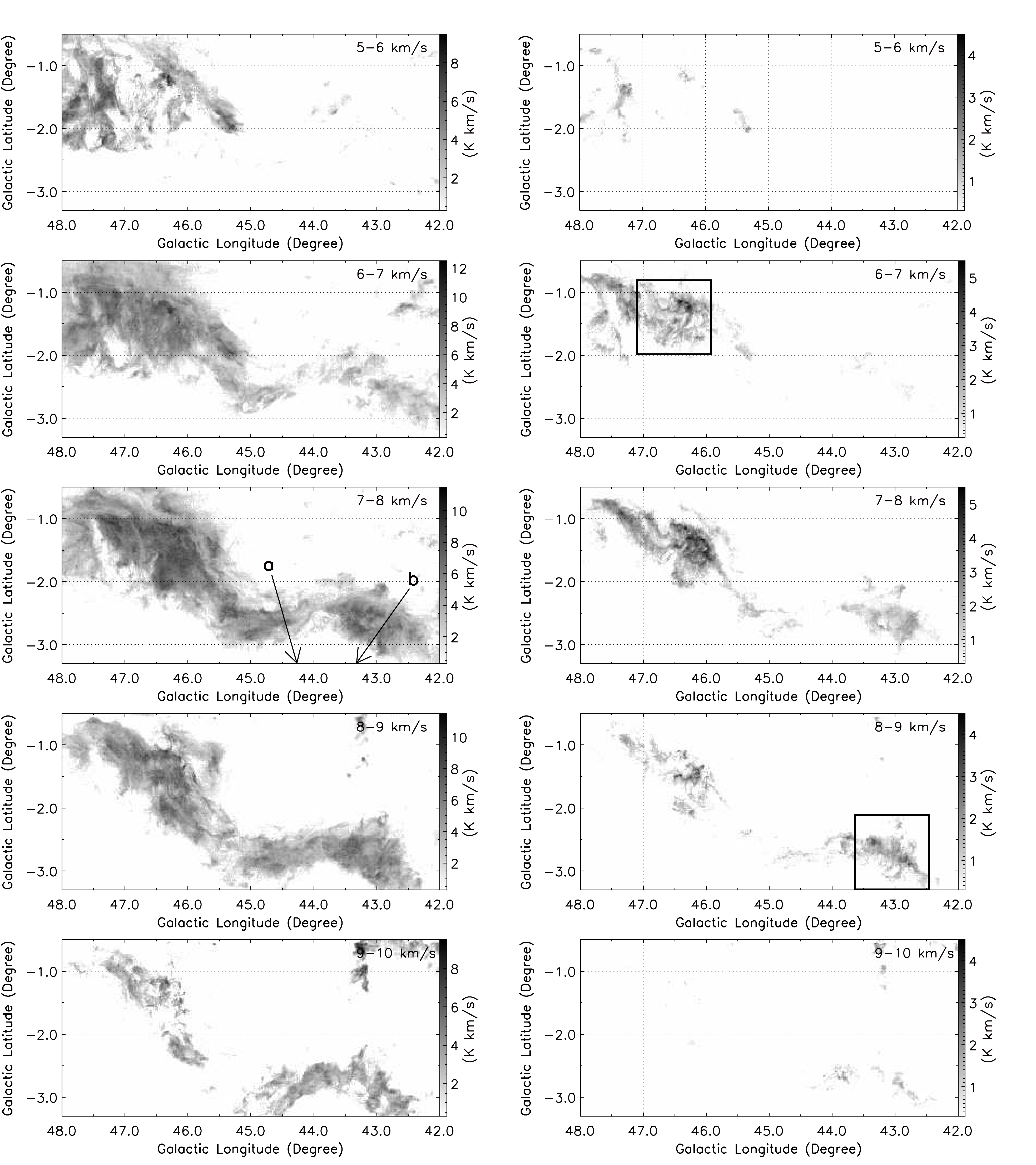}
\caption{
Channel maps of \twCO\ emission (left panels) and \thCO\ emission (right panels)
toward the zoom-in region of the filamentary MC G044.0$-$02.5.
The arrows show the direction of the velocity gradient
(Figure~\ref{conepv}) perpendicular to the long filamentary structure,
while the two boxes display the typical
regions of filamentary networks in the MC.
\label{c13channel}}
\end{figure}

\begin{figure}[ht]
\gridline{
          \hspace{-5ex}
          \fig{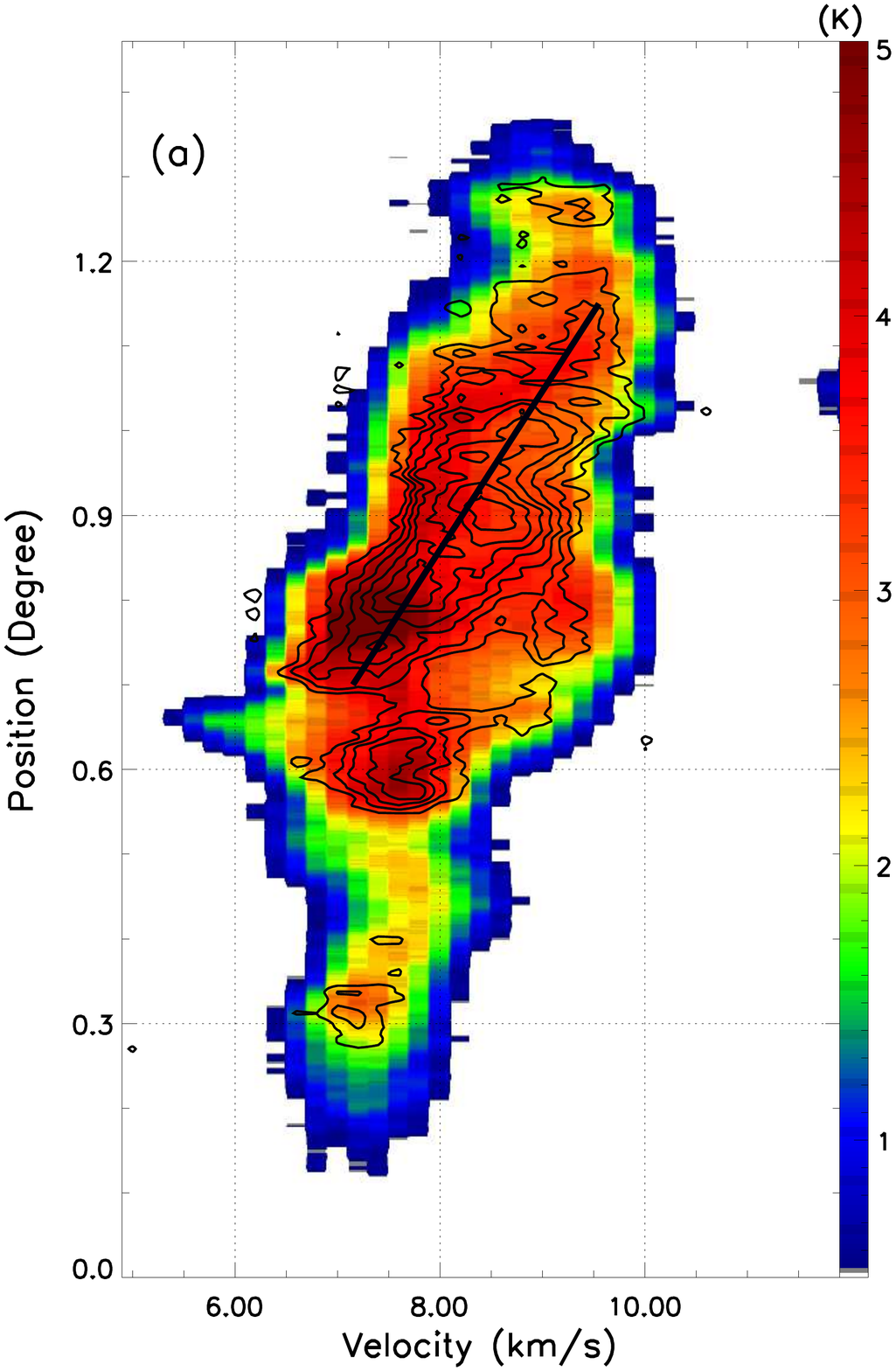}{0.5\textwidth}{}
          \hspace{-5ex}
          \fig{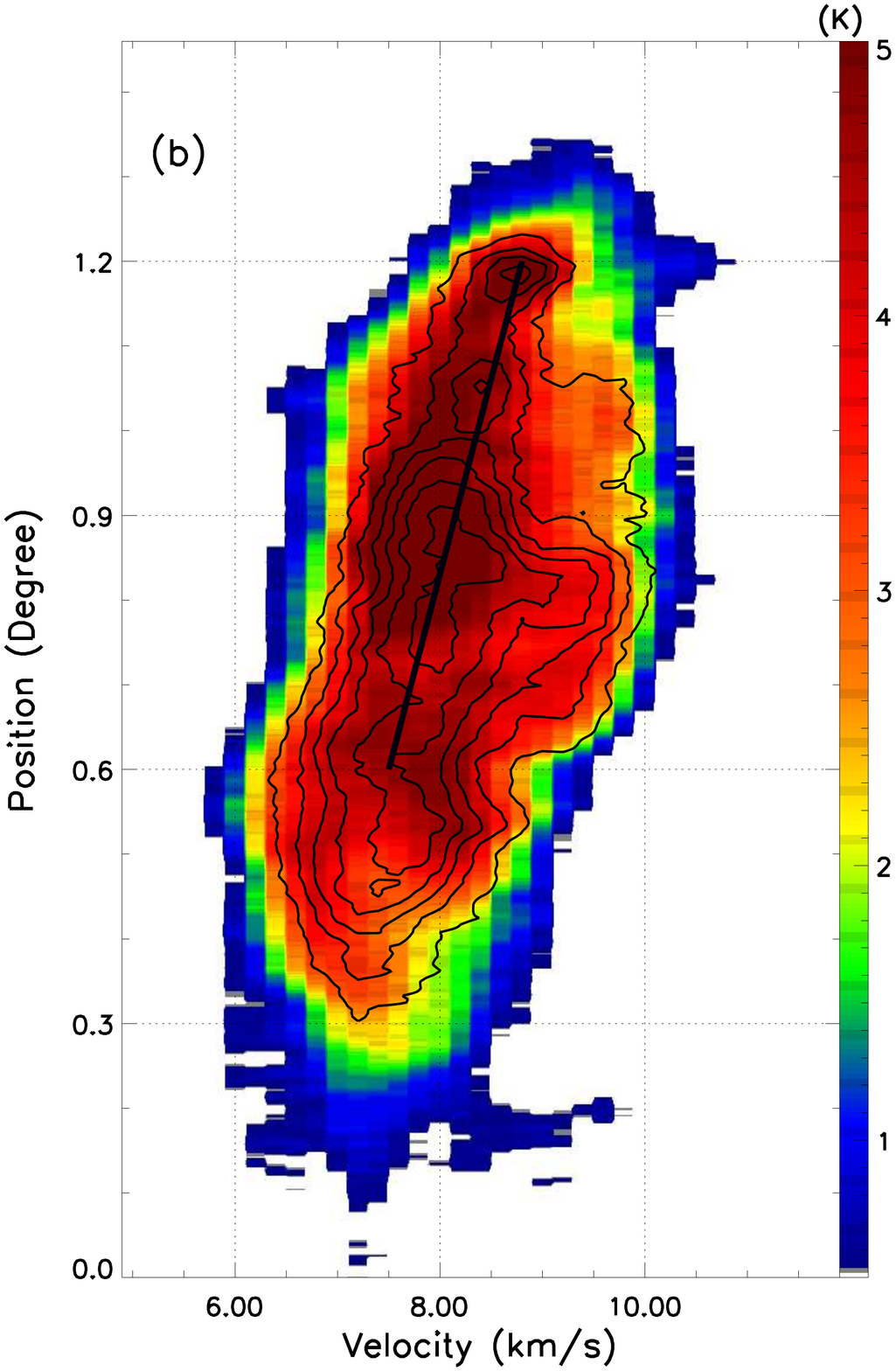}{0.5\textwidth}{}
          }
\caption{
Position--velocity diagrams of the \twCO\ (color) and \thCO\ (contours) emission
perpendicular to the long filamentary MC G044.0$-$02.5
(see arrows in Figure~\ref{c13channel}).
The slice has a length of 1\fdg46 and a width of 0\fdg225.
The contours of \thCO\ emission start from 0.3~K with a step of 0.3~K.
The black lines indicate the velocity gradient of
$\sim0.8\km\ps$pc$^{-1}$ in panel (a) and $\sim0.3\km\ps$pc$^{-1}$ 
in panel (b) for the gas at a distance of $\sim$404~pc.
\label{conepv}}
\end{figure}

\begin{figure}
\vspace{-15ex}
\gridline{
          \fig{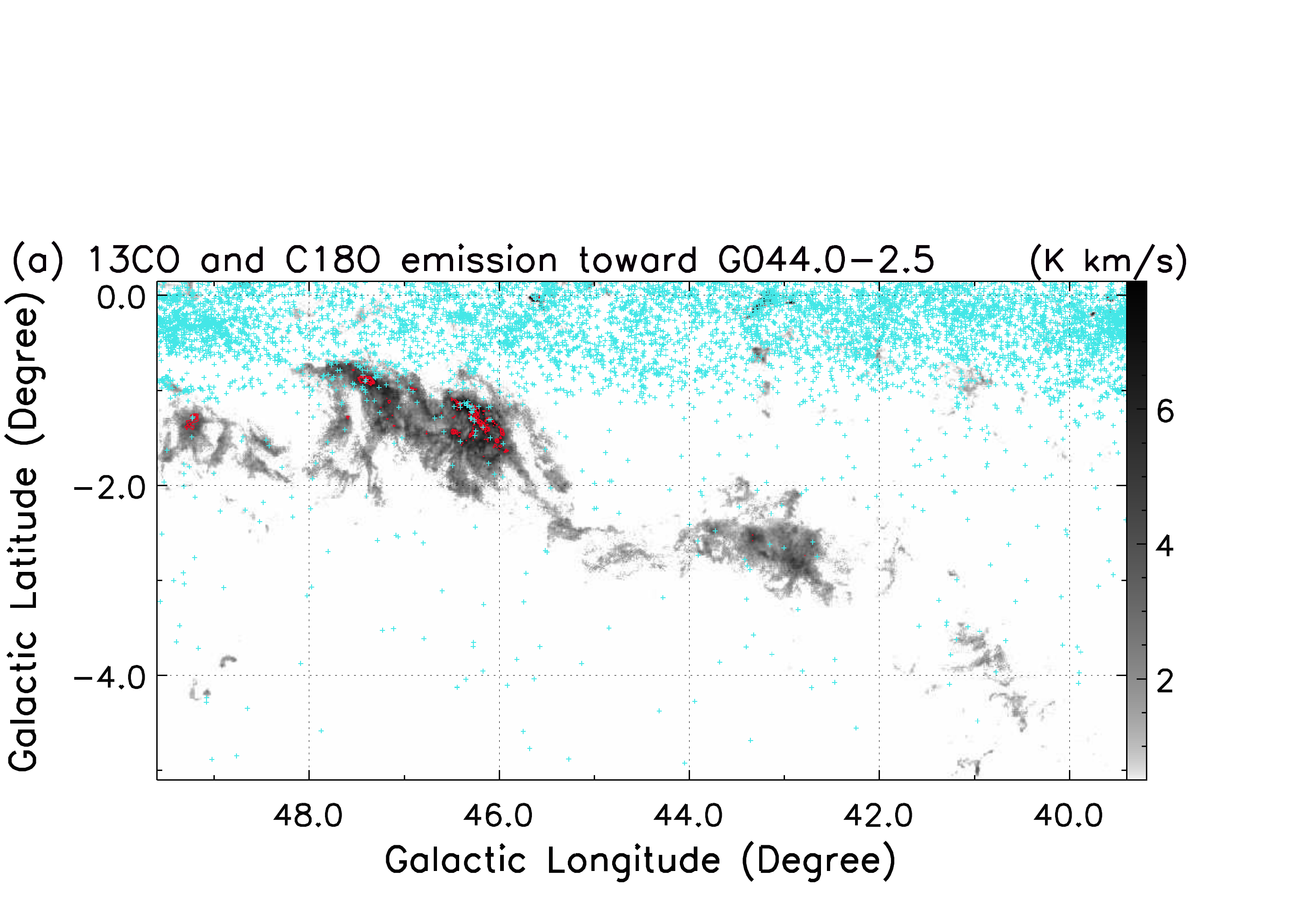}{1.2\textwidth}{}
          }
\vspace{-27ex}
\gridline{
          \fig{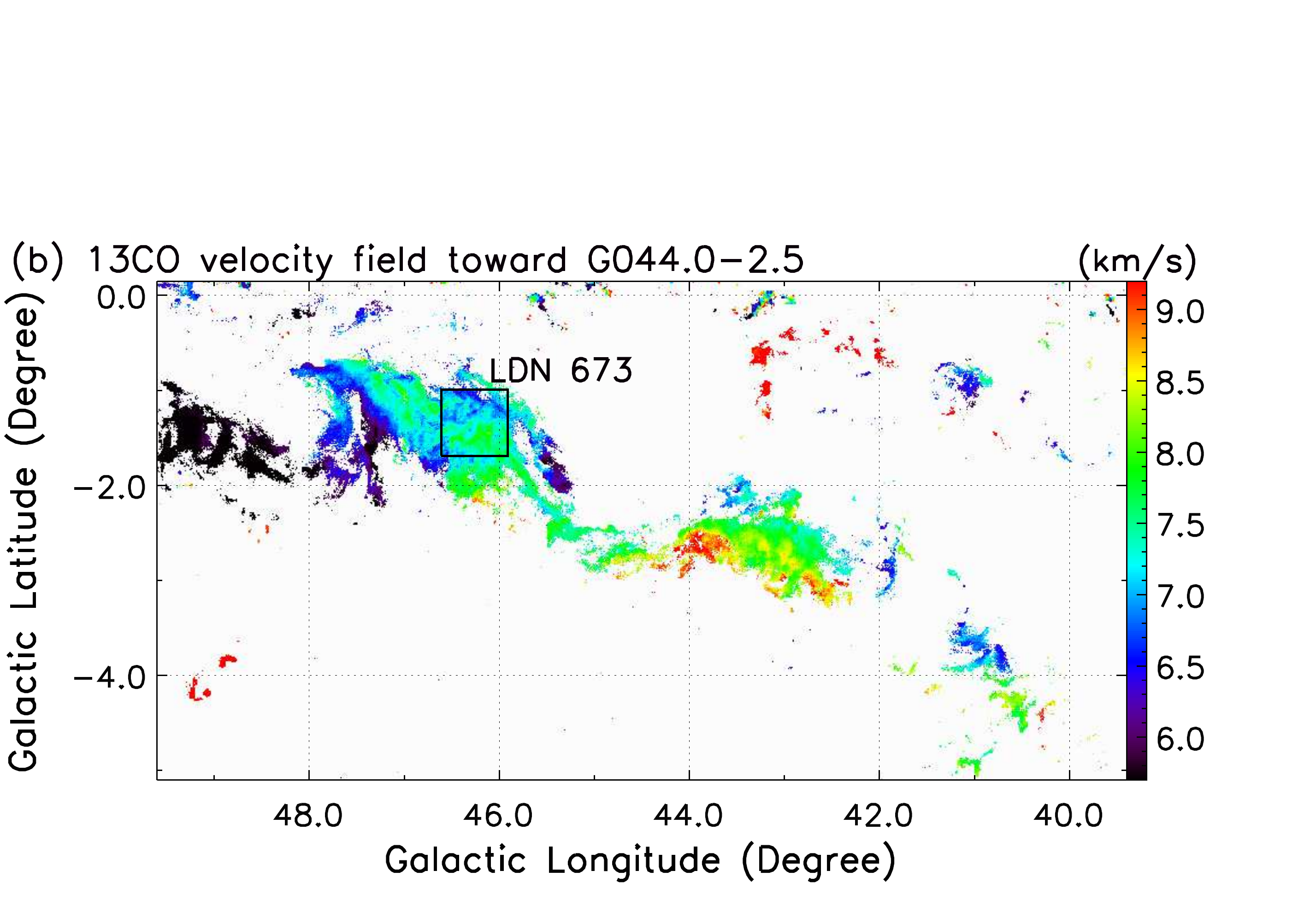}{1.2\textwidth}{}
          }
\vspace{-5ex}
\caption{
panel (a): integrated \thCO\ ($J$=1--0) emission (gray) and C$^{18}$O ($J$=1--0) 
emission (red contours, 0.4, 0.8, 1.2, and 1.6~K~km~s$^{-1}$) 
toward MC G044.0$-$02.5 in the velocity interval 
of [3.0, 12.0]~km~s$^{-1}$.
Cyan pluses indicate the YSO candidates (Class I and II sources
and possible contamination of AGB stars) identified from
the infrared multicolor criteria based on various
IR data \citep[see Table~1 and discussions in][]{2019A&A...622A..52Z}. 
The identified YSO candidates are found to be overdense
in LDN 673, in which the enhanced C$^{18}$O emission is concentrated.
Note that samples in the $|b|\lsim$~1\fdg0 region
suffer high contamination due to the complex environment 
in the Galactic plane.
Panel (b): intensity-weighted \thCO\ mean velocity (first moment) map of
MC G044.0$-$02.5 in the same velocity range. The black box shows 
the region of LDN 673.
\label{cone_312}}
\end{figure}

\begin{figure}[ht]
\includegraphics[trim=20mm 0mm 0mm 0mm,scale=0.7,angle=0]{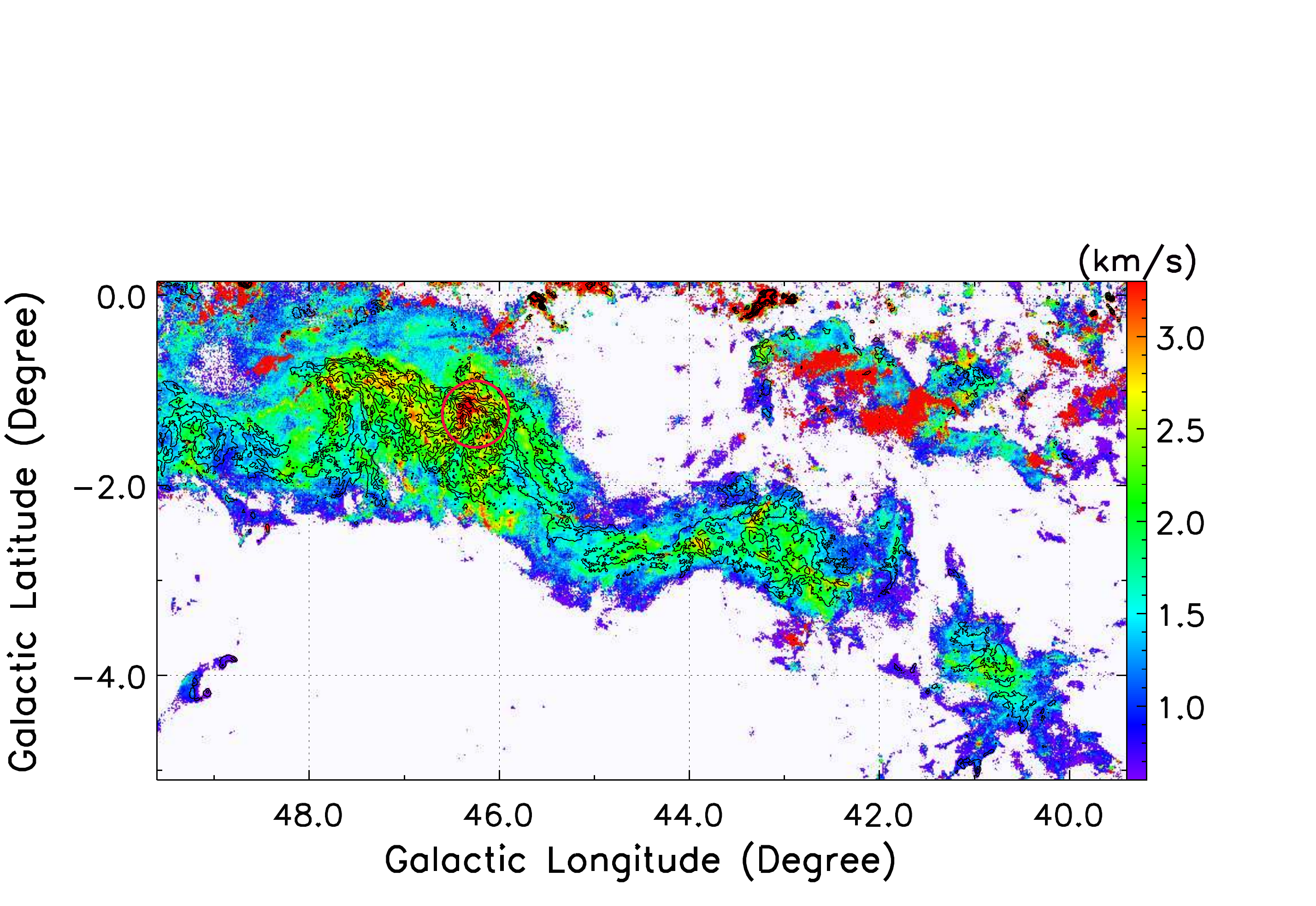}
\caption{Intensity-weighted \twCO\ line width (second moment) map 
of MC G044.0$-$02.5 in the 3.0--12.0~km~s$^{-1}$ interval. The contours 
are 0.5, 2.0, 3.5, 5.0, 6.5, and 8.0~K~km~s$^{-1}$
for \thCO\ emission integrated in the same velocity range.
The red circle indicates the hub-like region, i.e., LDN 673.
\label{cone_1213}}
\end{figure}

\begin{figure}[ht]
\includegraphics[trim=20mm 0mm 0mm 0mm,scale=0.6,angle=0]{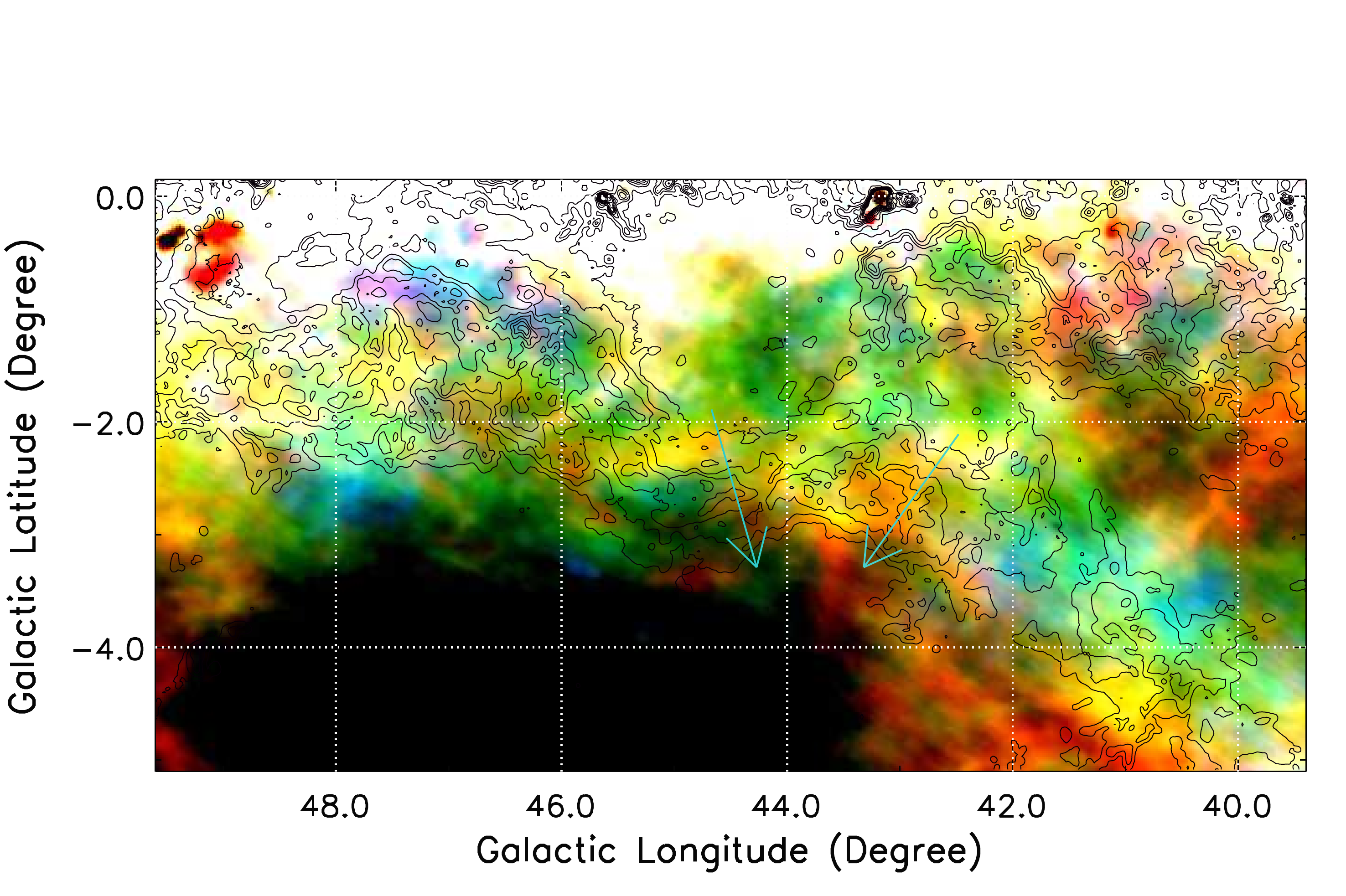}
\caption{
Integrated HI emission from Arecibo in the interval of [4.0, 6.0]~km~s$^{-1}$ (blue),
[6.0, 9.0]~km~s$^{-1}$ (green), and [9.0, 11.0]~km~s$^{-1}$ (red) toward
the MC G044.0$-$02.5. The contours, which are from 0.5~K~km~s$^{-1}$
to 32.5~K~km~s$^{-1}$ with a step of 4.0~K~km~s$^{-1}$, 
indicate the integrated \twCO\ emission in the velocity range of 
3.0--12.0~km~s$^{-1}$. The cyan arrows are the same as those in 
Figure~\ref{c13channel}.
\label{HI_aqu_411_rgb}}
\end{figure}

\begin{figure}[ht]
\includegraphics[trim=35mm 100mm 0mm 80mm,scale=1.1,angle=0]{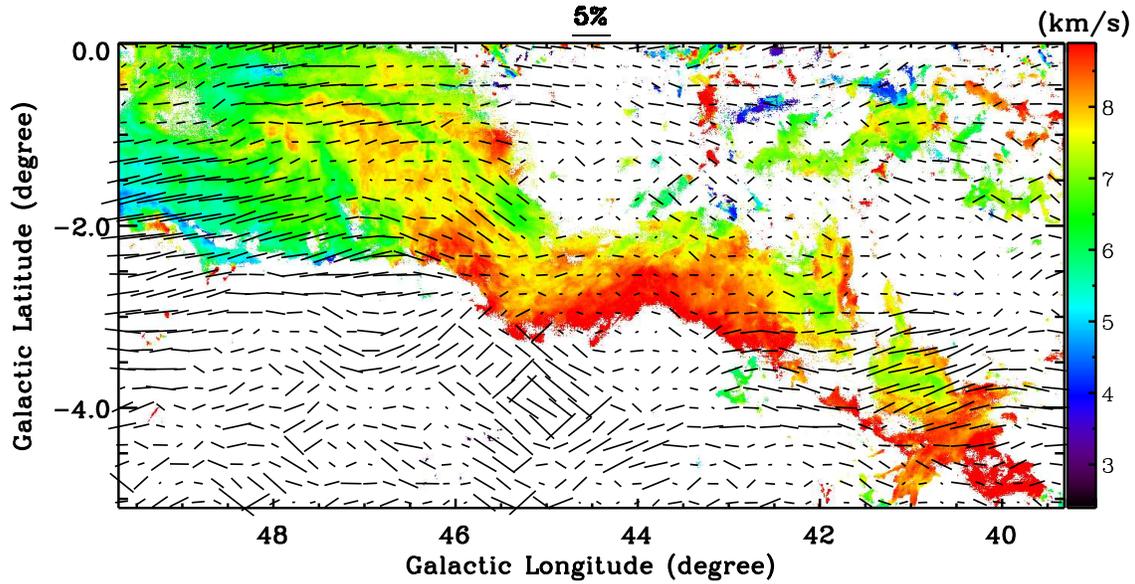}
\caption{
\twCO\ velocity field of MC G044.0$-$02.5, overlaid with the orientations
of the apparent magnetic field from the $Planck$ 353~GHz data \citep{2015A&A...576A.104P}.
The 353~GHz data are smoothed to a resolution of 12\farcm5 to improve the
signal-to-noise ratio (S/N). The length of each 
polarization segment is proportional to the polarization fraction, and the position 
angle has been rotated by $90^{\circ}$. The uncertainties of polarization fraction
and position angle are computed following the method described in Appendix B.1 of
\cite{2015A&A...576A.104P}. Note that the polarization segments with 
$P/\sigma_P <$ 3 are excluded in this plot.
\label{plank_pol}}
\end{figure}

3. Toward the Aquila Rift region, large-size MC structures are 
identified from \twCO\ emission in coherent spatial and velocity 
distributions. In combination with the MWISP CO data and the
Gaia DR2, distances of these local molecular-gas structures 
are well determined in the range of $\sim$~200--800~pc 
(Figures~\ref{aqu_03}--\ref{aqu_312} and Table~1) 
through Bayesian analysis and an MCMC algorithm.
The dominant mass of the local molecular gas is in the W40 region
and the GMC complex G036.0$+$01.0, 
both of which are located at a distance of $\gsim$~470~pc.

4. For the W40 region, two overlapping 
molecular-gas components with similar extensions 
are located at different distances 
of $\sim 235$~pc and $\sim 474$~pc, respectively.
The $\sim 474$~pc MCs are associated with its surrounding star formation
activities, i.e., Sh 2$-$62, \HII\ region W40, and Serpens NE Cluster.
Generally, in regions of $l\sim +26^{\circ}$ to $+33^{\circ}$, 
the $\lsim$~240--260~pc MCs with lower gas masses have spatial
distributions from the northwest to the southeast, 
which are similar to the results of the extended dust clouds 
at $\sim$~250~pc \citep[i.e., the Serpens Cirrus in front of 
the star-forming region of W40,][]{2019ApJ...878..111H}. 

5. The $\sim 2.0\times 10^{5} \Msun$ GMC complex G036.0$+$01.0,
which is less studied in the literature, consists of several 
molecular-gas components at a distance range of $\sim 560$--670~pc.
Along that direction, some smaller MCs at $\sim$~220--240~pc appear to be 
in front of the GMC complex (Figures~\ref{aqu_29} and \ref{aqu_1319}). 
These nearer MCs at $\sim 220$--240~pc
are widely distributed in the whole mapping region, 
although the mass of the molecular gas is generally small, 
i.e., $\sim 10^{2}-10^{3} \Msun$.

6. In regions with a larger Galactic longitude, the main MC structure 
is GMF G044.0$-$02.5, which has a mass of $\sim 2.4\times 10^{4} \Msun$
and is located at a distance of $\sim 404$~pc (Figure~\ref{aqu_312}). 
Due to a little confusion between different cloud components in the region, 
the long filamentary MC becomes a good target to study
the detailed spatial and velocity structures of the molecular
gas from several arcminutes to ten degrees
(or $\sim$~0.5~pc to tens of pc). 
The $\sim$~80~pc long MC G044.0$-$02.5 appears to be surrounded 
by more extended filamentary HI gas that has a similar spatial
structure and velocity gradient to that seen in the CO emission, 
indicating that the molecular gas is probably forming 
from large-scale atomic gas. 
Interestingly, the dense gas of the MC is often located in 
enhanced CO regions with asymmetric line profiles
or multi-velocity components.

7. These characteristics from the new CO data suggest that 
material accumulation, together with localized gas 
dynamics and various instabilities, 
is essential for the formation and evolution of MCs.
We suggest that the MCs toward the Aquila Rift region are 
likely formed by large-scale converging HI flows in the local ISM.
This mechanism can naturally explain the rapid MC formation and
simultaneous star formation in localized dense regions. 
In fact, we note that the ongoing star formation occurs in the densest part 
of the MC G044.0$-$02.5 (i.e., the overdensity of 
YSO candidates and groups of HH objects in LDN 673 at $l=$46\fdg26, $b=-$1\fdg33).

\acknowledgments
We gratefully acknowledge the staff members of the Qinghai Radio
Observing Station at Delingha for their support in obtaining the observations.
We would like to thank the anonymous referee for 
carefully reading through the manuscript and,
we appreciate the many constructive comments and suggestions 
that improved the paper.
MWISP is funded by the National Key R\&D Program of China
(2017YFA0402700) and the Key Research Program 
of Frontier Sciences of CAS (QYZDJ-SSW-SLH047).
Y.S. is supported by the NSFC grant No. 11773077.
X.C. acknowledges support by the NSFC through grant No. 11629302.

This work has made use of data from the European Space Agency (ESA) mission 
Gaia (https://www.cosmos.esa.int/gaia), processed by the Gaia Data Processing 
and Analysis Consortium (DPAC, https://www.cosmos.esa.int/web/gaia/dpac/consortium). 
Funding for the DPAC has been provided by national institutions, in particular 
the institutions participating in the Gaia Multilateral Agreement.

This publication utilizes data from the Galactic ALFA HI (GALFA HI) survey
data set obtained with the Arecibo $L$-band Feed Array (ALFA) on the
Arecibo 305~m telescope. The Arecibo Observatory is part of the National
Astronomy and Ionosphere Center, which is operated by Cornell
University under Cooperative Agreement with the U.S. National Science
Foundation. The GALFA HI surveys are funded by the NSF through grants
to Columbia University, the University of Wisconsin, and the
University of California.

\facility{PMO 13.7m}
\software{GILDAS/CLASS \citep{2005sf2a.conf..721P}} 

\bibliographystyle{aasjournal}
\bibliography{references}

\clearpage

\clearpage

\clearpage

\clearpage

\clearpage

\clearpage

\clearpage

\clearpage

\clearpage

\clearpage

\clearpage

\clearpage

\clearpage

\clearpage

\clearpage

\begin{deluxetable}{cccccccc}
\tabletypesize{\tiny}   
\tablecaption{Parameters of Local Molecular Clouds toward the Aquila Rift}
\tablehead{
\colhead{\begin{tabular}{c}
Name/Region  \\
      \\
(1)   \\
\end{tabular}} &
\colhead{\begin{tabular}{c}
$l$ range               \\
($^{\circ}$)      \\
(2)  \\
\end{tabular}} &
\colhead{\begin{tabular}{c}
$b$ range              \\
($^{\circ}$)      \\
(3)  \\
\end{tabular}} &
\colhead{\begin{tabular}{c}
$V_{\rm LSR}$      \\
($\km\ps$)         \\
(4)  \\
\end{tabular}} &
\colhead{\begin{tabular}{c}
Area $^{\mathrm {a}}$            \\
(arcmin$^2$)     \\
(5)  \\
\end{tabular}} &
\colhead{\begin{tabular}{c}
Distance $^{\mathrm {b}}$            \\
($\rm {pc}$)     \\
(6)  \\
\end{tabular}} &
\colhead{\begin{tabular}{c}
Mass  $^{\mathrm {c}}$                 \\
($\Msun$)                 \\
(7)  \\
\end{tabular}} &
\colhead{\begin{tabular}{c}
Note                            \\
                \\
(8)  \\
\end{tabular}}
}
\startdata
W40a region       & [25\fdg8, 32\fdg2]  &  [0\fdg5, 5\fdg1]       & [0, 6]    &  1.4$\times 10^4$  
                  & 235$^{+4}_{-4}$     &   2.2$\times 10^3$      & $b\gsim 0^{\circ}$ in the left 
panel of Figure~\ref{aqu_03}  \\
W40b region       & [25\fdg8, 33\fdg7]  &  [0\fdg5, 5\fdg1]       & [4, 12]   &  1.0$\times 10^5$
                  & 474$^{+6}_{-5}$     &   1.4$\times 10^5$      & Figure~\ref{W40_45115_rgb}  \\
MCs G029.0$-$02.0a & [27\fdg2, 32\fdg8]  &  [$-$5\fdg1, $-$0\fdg2] & [0, 7]    &  2.6$\times 10^4$
                  & 263$^{+8}_{-8}$     &   4.3$\times 10^3$      & $b\lsim 0^{\circ}$ in the left
panel of Figure~\ref{aqu_03}   \\
MCs G029.0$-$02.0b & [26\fdg3, 30\fdg7]  &  [$-$5\fdg1, $-$0\fdg2] & [12, 25]  &  2.1$\times 10^4$
                  & 508--550     	&   1.4$\times 10^4$ $^{\mathrm {d}}$ & $b\lsim 0^{\circ}$ 
in the right panel of Figure~\ref{aqu_03}  \\ 
MCs G030.5$-$01.5  & [27\fdg7, 33\fdg8]  &  [$-$5\fdg1, $-$0\fdg1] & [4, 12]   &  3.5$\times 10^4$
                  & 403$^{+15}_{-13}$   &   1.3$\times 10^4$      & $b\lsim 0^{\circ}$ in the middle
panel of Figure~\ref{aqu_03}  \\  
GMC G036.0$+$01.0  & [32\fdg2, 40\fdg4]  &  [$-$2\fdg4, 5\fdg1]    & [8, 20]   &  1.3$\times 10^5$
                  & 555--673            &   2.0$\times 10^5$ $^{\mathrm {e}}$ & Figures~\ref{aqu_1319} and \ref{b_1117_rgb}; the nickname of ``Phoenix cloud" \\
MC G044.0$-$02.5   & [39\fdg3, 49\fdg7]  &  [$-$5\fdg1, 0\fdg4] & [3, 12]   &  5.7$\times 10^4$
                  & 404$^{+4}_{-4}$     &   2.4$\times 10^4$      & Figures~\ref{aqu_312} and 
\ref{aqu_312_rgb}; the nickname of ``River cloud"\\
MC G041.5$-$01.0   & [40\fdg0, 43\fdg4]  &  [$-$2\fdg1, $-$0\fdg2] & [9, 13]   &  7.0$\times 10^3$
                  & 769$^{+15}_{-14}$   &   9.1$\times 10^3$      & the upper-right part in 
Figure~\ref{aqu_312}  \\
\enddata
\tablecomments{
$^{\mathrm {a}}$ The area is defined by the integrated \twCO\ emission in the 
corresponding velocity interval. 
$^{\mathrm {b}}$ The distance is estimated based on the MWISP CO data and the 
Gaia DR2. The 5\% systematic error is not included.
$^{\mathrm {c}}$ The total mass is estimated from the integrated \twCO\ emission 
using the mean CO-to-H$_2$ mass conversion factor of $2\E{20}$~cm$^{-2}$(K~km~s$^{-1})^{-1}$
(see Section 3.3).
$^{\mathrm {d}}$ Adopting a mean distance of 530 pc for calculating the total mass of the MCs.
$^{\mathrm {e}}$ Adopting a mean distance of 610 pc for calculating the total mass of the GMC 
complex.
}
\end{deluxetable}

\end{document}